\documentclass{iopart}
\usepackage{epsfig}
\usepackage{graphicx}
\usepackage{harvard}
\usepackage{citesort}
\usepackage{iopams}
\usepackage{epstopdf}
\newcommand{\bm}[1]{ \mbox{\boldmath $#1$} }

\begin{document}

\title[Comparing and contrasting nuclei and cold atomic gases]
{Comparing and contrasting nuclei and cold atomic gases}

\author{N.T. Zinner and A.S. Jensen}
\address{Department of Physics and Astronomy, University of Aarhus,
DK-8000 Aarhus C, Denmark}
\ead{zinner@phys.au.dk}

\begin{abstract}
The experimental revolution in ultracold atomic gas physics over the
past decades have brought tremendous amounts of new insight to the
world of degenerate quantum systems. Here we compare and constrast
the developments of cold atomic gases with the physics
of nuclei since many concepts, techniques, and nomenclatures are common to 
both fields. However, nuclei are finite systems with interactions
that are typically much more complicated than those of ultracold 
atomic gases. The simularities and differences
must therefore be carefully addressed for a meaningful
comparison and to facilitate fruitful crossdisciplinary activity.
We first consider condensates of bosonic and paired
systems of fermionic particles with the mean-field description 
but take great care to point out potential problems in the limit
of small particle numbers. Along the way we review some of the 
basic results of BEC and BCS theory, as well as the BCS-BEC
crossover and the Fermi gas in the unitarity limit, 
all within the context
of ultracold atoms. Subsequently, we consider the specific example
of an atomic Fermi gas from a nuclear physics perspective, comparing
degrees of freedom, interactions, and relevant length and 
energy scales of cold atoms and nuclei. 
Next we address some attempts in nuclear
physics to transfer the concepts of condensates in nuclei that 
can in principle be built from bosonic alpha-particle constituents.
We also consider Efimov physics, a prime example of nuclear physics transfered to 
cold atoms, and consider which systems are
more likely to show interesting bound state spectra. Finally, we
address some recent studies of the BCS-BEC crossover in light 
nuclei and compare them to the concepts used in ultracold atomic 
gases. While many-body concepts such as BEC or BCS states 
are applicable in both subfields,
we find that the interactions and finite particle numbers in 
nuclei can obscure the clear meaning they have in cold atoms. On 
the other hand, universal results from atomic physics should have
impact in certain limits of the nuclear domain. In particular,
with advances in the trapping of few-body atomic systems we 
expect a more direct exchange of ideas and 
results.
\end{abstract}

\date{\today}


\maketitle
\tableofcontents

\section{Introduction}\label{intro}

The last two decades have witnessed dramatic developments in terms
of realizing
Bose-Einstein condensates and degenerate Fermi gases which has been driven by the 
experimental advanced in trapping and cooling
dilute atomic gases
\cite{pet02,pit03,leg06,bloch2008,giorgini2008}. 
Typically, these system contain thousands or even millions of 
particles, but very recently samples with very small particle
numbers have been realized \cite{selim2011,zurn2011,bakr2010,sherson2010,weitenberg2011}.
This regime is particularly interesting since the physics that 
can be learned here might aid our understanding of
mesoscopic and microscopic systems in other subfields.
One such venue is nuclear physics, where particle numbers
are finite. This raises an interesting question 
of how concepts from 
the large system limit, such as 
condensation or superfludity, 
transfer to small systems. In the 
current presentation we will address similarities and 
differences between the physics of nuclei and 
ultracold atomic gases.

There are a number of immediate contrasts that must be 
highlighted in a comparison of nuclear and cold atoms. 
First, the notion of scattering length and effective
range expansion \cite{lan81} is very efficient for
cold gases since they are valid for very low energy and 
the typical collision energies approach the zero-energy limit. 
In more traditional atomic collision physics
and in charged Fermi liquids in metals the long-range Coulomb
interaction complicates this simple description of the two-body
interaction. Secondly, nuclei are self-bound system implying
that repulsive and attractive parts of the nuclear interaction must
balance. In the atomic gas with external confinement there is 
no such requirement and the system can be studied in a wider range
of regimes in principle. The relatively small particle numbers and the 
self-bound structure can have important consequences to the 
conceptual transfer between the subfields and will be a prime
concern in this presentation. 
Furthermore, nuclear methods
deal with structures resembling those of present
interest in cold atomic gas investigations, i.e. (i) two inherently
different but similar components, spin-up and down of neutrons and protons, as in mixed
condensates, (ii) ``unbound'' boson pairs of nucleons in the
superfluid BCS low-energy regime, (iii) boson expansions in terms of pairs
of fermionic nucleon as molecular condensates, 
(iv) bosonic $\alpha$-particle cluster structure as in
one-component Bose condensates, (v) $\alpha$-clusters and nucleons in
mixed boson-fermion structures.

In broad terms, nuclear and atomic-molecular physics have a lot of
concepts, phenomena, and techniques in common.  The reasons are to be found
in the short-range character of the different interactions and the
finite spatial extension of systems with a finite number of particles.
Specifically characteristic units of length $L_0$ (interparticle spacing) 
and mass $m$ can be
used to construct a unit of energy $E_0 = \hbar^2/(m L_0^2)$.  Using
$L_0$ and $E_0$ as length and energy units provide dimensionless
quantities which remove superficial large differences, emphasize
similarities and allow focus on the interesting principal
scale-independent differences \cite{amorim1997,riisager2000}. We will
carefully compare relevant dimensionless parameters for both fields.
An important dimensionless quantity is the density $n$ multiplied
by the third power of the scattering length $a$, i.e. n$a^3$, which is
related to the classical mean free path $\lambda$ by $na^3\sim a/4\pi\lambda$. 
In general the size of $\lambda$ indicates which structure is preferred,
e.g. mean field structure when $\lambda \gg L_0$ and strong
correlations when $\lambda \simeq L_0$ \cite{boh69,sie87}. 
This corresponds roughly respectively to
$na^3\ll 1$ and $na^3 \simeq 0.1-10$ for atoms where $a/L_0\sim 10-100$ and
nuclei where $a/L_0\sim 1$ (away from resonance).

The nucleon-nucleon interaction range  in
finite nuclei is comparable to the nuclear radius and the mean free
path at low excitation energy \cite{boh69,sie87}. These nucleon-nucleon properties are not (necessarily) the same
as in free space.  The nuclear system is dense, and more than
$s$-waves contribute.  Both mean-field
(single-particle or quasi-particle motion) and collective 
(rotational and vibrational motion of spatial, spin, and pairing) degrees of freedom are
important in realistic descriptions \cite{boh69}.  For atoms and molecules it is intuitively
clear that the important degrees of freedom must be those related to
the spatial orientation, the intrinsic excitations and
the relative motion of the individual atoms
\cite{molecule}. For both nuclei and atoms or molecules macroscopic as well as
microscopic properties are therefore necessary ingredients in the
descriptions.

The self-bound, leptodermous (thin skin) nuclear systems  \cite{myers1969} emphasize the importance
of surface properties as for mesoscopic condensed matter systems \cite{mesoscopic}.
Similar deviations from homogeneous bulk properties appear as
correlations in trap- or optical lattice confined atomic systems \cite{bloch2008,giorgini2008}.  The
common theoretical quantitative techniques include the mean-field and BCS pairing
approximations, linear response (including random phase approximations), ab initio variational
minimization, shell model diagonalization in restricted Hilbert
spaces, and density functional methods.

The present paper is devoted to a comparison of 
common concepts and structures in nuclei and 
cold atomic gases. Striking similarities are 
seen in many discussions that employ the notion 
of condensation and paired states in both 
subfields. However, a careful comparison is 
demanded when concepts that traditionally 
refer to infinite systems are applied to finite
systems such as nuclei or trapped atomic gases
containing only very few atoms. 
Our aim is
to facilitate and clarify exchange of important ideas, basic concepts
and essential techniques between the two broad fields of physics. 
For a comprehensive review of experimental and theoretical developments in 
cold atomic gases we refer to \cite{bloch2008,giorgini2008}.

Our starting point will be mean-field theory and we define
the concepts of condensate and paired state in this context. 
Particular care is taken to address the problems with finite particle
number and we therefore discuss at length the notion of 
condensate and condensate fraction. Paired states of fermionic
particles in both nuclear and cold atomic gases are discussed 
mainly within the BCS theory which is widely used in both arenas. 
We also discuss the interesting physics of the transition from weakly-coupled
paired states to a condensate of strongly bound two-body molecules as
it is described within the BCS formulation (BCS-BEC crossover), 
and we 
address strongly-interacting Fermi gases in the unitarity limit. 
Along the 
way we take care to distinguish between generic features
and model-dependent details that originate from certain
convenient choices of model-space, effective interactions, etc.

An important topic is the comparison of different scales in 
cold atomic gases and in nuclei. In particular, the isolation of 
relevant dimensionless quantities is essential. We undertake such
a discussion next, explaining the degrees of freedom of both fields, 
and pointing out similarities and differences.
An extended discussion of the cold atomic two-component fermionic 
gas is given from a nuclear physics perspective within the 
traditional shell model picture that should be helpful for
applying nuclear physics techniques and concepts directly
to cold atomic gases. As an example, we discuss few-body 
fermionic trapped atomic gases and compare them to
light nuclei.

We then turn to some specific examples of bosonic structures and 
condensates in few-body nuclear systems and also in trapped
few-body atomic gases that have been studied theoretically in 
recent years. We employ a concept of quantum localization which can predict
zero-temperature phases of quantum matter, and show how it is
connected to more common measures of quantum degeneracy. The case of
finite systems of bosons is subsequently discussed with emphasis on
the competition between localization of particles and
mean-field states. For finite systems localization is unfavourable 
to the formation of a condensate. In nuclear physics, we apply these
concepts to light nuclei that are believed to have cluster structures, 
and for which condensation of $\alpha$-particles have been proposed
as a potential structure of the system.

Efimov physics is discussed next. Originating in nuclear physics, the
concepts of an infinite number of bound three-body states around the threshold
for two-body binding which has recently been experimentally 
studied with cold atomic gases. This constitutes a beautiful example
of crossdisciplinary fertilization. We discuss the concept
of universal three-body bound states and give criterions for which systems
are best suited for its observation. Finally, we discuss neutron-rich
matter within the context of the BCS-BEC crossover. Again, we
focus on few-body aspects and consider light halo nuclei that have
a low-density surface occupied mostly by neutrons. Here we address 
which features related to the crossover can be described as universal and 
which are model-dependent. 

The current presentation is organized in sections with slightly different
aims. The discussion of condensates in Section \ref{condens} is directed from
atomic towards nuclear physics, implying that the content is well
known in atomic and has to be defined properly in nuclear physics.
The properties of pair correlations in Section \ref{pairing} are common to
nuclear and atomic physics but there are subtle issues with choices
of model-space and residual pairing interactions in nuclear physics 
that we point out.
Section \ref{atomgas} contains a description of atomic gas properties such
as length and energy scales, degrees of freedom, etc.,
translated into nuclear physics language. 
Section \ref{fewbody}
discusses few-body applications of the concepts of atomic boson gases 
in nuclear physics and in
Section \ref{efimov} we specialize to the topical 
Efimov effect in both nuclei and atomic gases. 
Section \ref{nucbcs} is devoted to a discussion of BCS-BEC
crossover physics in neutron-rich matter with particular
emphasis on light nuclei and particular choices
in the models that describe the nuclear interactions.
Finally, section \ref{conc} contain a brief
summary and conclusions.

\section{Condensates}\label{condens} 
Bose-Einstein condensates were initially
predicted as the result of a phase transition of non-interacting bosons
below a critical temperature. This was
at first considered a pathology of no interaction, but 
after the discovery of superfluidity in bosonic $^{4}$He 
in 1938 by Kapitza \cite{kapitza1938}, Allen and Misener \cite{allen1938}, Fritz London had the insight
that superfluidity might in fact be a sign of condensation \cite{london1938}. A few years
later Landau explained superfluidity in terms of the linear spectrum
of the low-energy excitations in the system \cite{landau1941}. He also introduced a two-fluid model 
with a superfluid and a normal component, the former of which is related
to the condensed particles.
Superfluidity is connected with the excitation spectrum, whereas condensation
is a property of the 'natural' stationary state of the system as given
by the one- or two-body density matrix introduced below. From the 
point of view of atomic and nuclear physics, the latter concept is the 
more important and hence our focus in the following. 

As a large number 
of identical bosons are allowed simultaneously in the
same single-particle quantum state it can be
macroscopically occupied. 
An important aspect of condensed matter physics 
and the theory of phase transitions in three-dimensional space is that of long-range order. 
As we will
discuss below, this concept is also useful in the case
of finite systems where it is connected to the coherent state that a Bose system
can condense into. The properties of boson condensates
are extensively investigated in laboratories under a variety of
conditions defined by ingenious combinations of external
electromagnetic fields \cite{pet02,bloch2008,optlattice}.  
In contrast, for fermions only one particle
is allowed in a given quantum state. However, an entity consisting of
an even number of fermions exhibits boson symmetry
properties, and in particular these composite particles can occupy the
same quantum state.
The Cooper pairs of fermions in the BCS theory
are examples arising due to an attractive two-body interaction \cite{cop56,bcs57,leg06}.
Superconductivity and superfluidity are in this sense the same phenomena which can
be seen in Landau's theory of phase transitions for these systems \cite{lan58,pit80}. 
For a summary of the concepts and historical development in superfluidity
we refer to Leggett \cite{leggett1999}.

In nuclei, prominent examples of bosonic substructures are $\alpha$-particles which from the
early days were suggested as building blocks of nuclei \cite{wig37,whe37,wef37,bri66}, and
pairs of nucleons within the BCS theory \cite{bmp58,boh69}.
Numerous examples exist of both fundamentally bosonic atoms and molecules 
and also of binary bound fermionic molecules \cite{molecule}.  When
composite boson structures begin to get close to each other and
interact, the ideal boson structure is no longer maintained.  The
intrinsic degrees of freedom become important and true microscopic
fermionic nature is revealed. We shall here first discuss the
definition and structure of boson condensates and in the next
subsection the connection to bosons formed by two fermions via
BCS-pairing.

\subsection{Structure of Bose condensates}\label{condens1}
The ideal Bose condensate (BEC) is defined through
quantum statistics for bosonic
particles \cite{fet71,pet02,pit03}. The prediction is that below a certain temperature these
systems will have a macroscopic number of particles, $N_0$, in the
lowest energy state. Here macroscopic occupation refers to the
thermodynamic limit where $\lim_{N\rightarrow\infty}\,(N_0/N)$ has to
be finite when $N$ is the total number of particles.  The ideal Bose
condensate with $N=N_0$ is then by definition non-interacting since
the inevitable correlations otherwise would prevent some particles from
being in the same quantum state. This depletion of the condensate due to 
interactions depends on $na^3$ where $a>0$ is the scattering length
characterizing the two-body interaction at low energy and $n$ is the particle density. 
When $a<0$, the bosons attract
and only a limited number of particle can condense without collapsing as we will
discusse below.
However, to achieve condensation in
an experiment the particles must be confined inside a trap. The
presence of such an outer one-body trapping potential alters the particle
spectrum but BEC can still occur for dilute gases in spatially
extended traps \cite{pet02,pit03}.

To be accurate in discussions of BEC we need to decide on an
appropriate and robust definition. The most widely accepted basic
definition, revolving around the density matrix for the system in
question, was given early on in \cite{pen51,lan58,pen56,yang62}. We
shall follow \cite{yang62} and specify properties of the $N$-body
density matrix arising from the wave function $\Psi$ for the many-body
system of $N$ particles.  We first define the one-body density matrix,
$\rho_1$, by
\begin{eqnarray}  \label{e410}  
 && \rho_1(\vec{r},\vec{r}^{\prime}) 
 \equiv Tr|\Psi(\vec{r})\rangle\langle\Psi(\vec{r}')|  \\  \nonumber
  &=& \int d\vec r_2 d\vec r_3 ...d\vec r_N 
 \Psi^{*}(\vec{r},\vec{r}_2, ... ,\vec{r}_N)
 \Psi(\vec{r}^{\prime},\vec{r}_2, ... ,\vec{r}_N) \; ,
\end{eqnarray}
where $\vec{r}_i,i=1,..,N$, are the particle coordinates. The choice of
integration omitting $i=1$ is arbitrary as $\Psi$ is symmetric under
any particle interchange.  We assume that $\Psi$ is normalized to $N$
after integration over all coordinates of $|\Psi|^2$.

The positive definite matrix $\rho_1$ with two continuous indices is
hermitian and can thus be diagonalized by a unitary transformation,
i.e. we can decompose this density matrix as
\begin{eqnarray} \label{spec}
\rho_1(\vec{r},\vec{r}')=\sum_i \lambda_i \phi^*_i(\vec{r}) 
\phi_i(\vec{r}') \; ,
\end{eqnarray}
where $\lambda_i$ are the eigenvalues (which are positive) and $\phi_i$ the
corresponding single-particle eigenfunctions each normalized to
unity. We order the eigenvalues to decrease in size with $i$.
Integrating the diagonal part of the density matrix we get $N$, i.e.
\begin{eqnarray} \label{e420}
  \int d\vec r \rho_1(\vec{r},\vec{r})= \sum_i \lambda_i = N \; .
\end{eqnarray}
We define the system as a simple BEC when one and only one of the
eigenvalues $\lambda_1$ remains of the order of magnitude $N$ for
increasing $N$, i.e. one large eigenvalue in the thermodynamic limit.
Thus the wave functions for a BEC is dominated by a wave function that
is a simple (symmetric) product of the same single-particle
wave function $\phi_1$, i.e.  $\Psi \approx \Pi_i \phi_1(\vec r_i)$.
Notice that this holds for condensates that are close to the ideal 
non-interacting case. A counterexample is atomic $^4$He which has very strong
correlations and thus only 10\% of a given sample will be in the condensed
state.
This definition inherently compares structures for different (large)
numbers of particles.  More than one eigenvalue may be large even in
the limit of large $N$.  Such states are called fragmented BEC
and
they have recently received increasing attention, particularly in
connection with gases of more than one species of atoms and also with
respect to condensation of asymmetric (different number of particles)
two-component systems \cite{pet02,pit03,mueller2006}.

The concept of off-diagonal long-range order is closely related to the
behavior of the one-body density matrix in the limit where $|\vec r
-\vec r^{\prime}|$ becomes large compared to the system size \cite{yang62}.  Let us then consider a possible
limiting form of $\rho_1$, i.e. 
\begin{equation} \label{odlro}
\lim_{|\vec{r}-\vec{r}'|\rightarrow\infty} 
 \rho_1(\vec{r},\vec{r}')= F^*(\vec{r})F(\vec{r}')
 +\rho_1'(\vec{r},\vec{r}')\;,
\end{equation}
where the first term has the product form obtained by diagonalization.
If the function $\rho_1'$ tends to zero in this limit one would say
that the system is a BEC with order parameter $F$ which depends on the
coordinate. Then $F$ is interpreted as the condensate wave function, $F
\rightarrow \phi_0 \sqrt{N}$, containing $N$ particles and obtained in
the mean-field approximation.  Thus Eq.~(\ref{odlro}) separates
$\rho_1$ into the product part related to the largest eigenvalue and
the rest of the sum in Eq.~(\ref{spec}) denoted by $\rho_1'$.  One can
show under rather general conditions that $\rho_1'$ vanishes for
homogeneous 3-dimensional Bose systems \cite{leg06} and also 
in the case of trapped condensates \cite{naraschewski1999}. Thus this
alternative definition of BEC is equivalent to that of one
large eigenvalue in the expansion in Eq.~(\ref{spec}). 

The one-body 
density matrix for a BEC in a trap was experimentally measured
about a decade ago \cite{bloch2000} and the expected coherence was
confirmed. This corresponds to what is known as first-order coherence \cite{naraschewski1999}.
Second-order coherence (related to the two-body density matrix)
is connected to the famous Hanbury-Brown and Twiss effect \cite{hbt1956,baym1969,baym1998}. The 
corresponding bunching effect has been observed in optics \cite{bachor2004} and in cold bosonic atoms \cite{yas96,fol05,sch05}. 
The analogous anti-bunching in electronic systems \cite{hen99,oli99,kie02}, for neutrons \cite{ian06}, and in fermionic cold atoms \cite{rom06}. Recent 
experiments have even begun to study third-order coherence (large eigenvalues in the third-order correlation function) and the nucleation of a $^4$He BEC \cite{truscott2010}.

\subsection{Center-of-mass motion}\label{condens2}

There is one possible flaw in this equivalence between the two
definitions of BEC above.  Unfortunately that is related to trapping in
external one-body potentials which is employed in all laboratory
experiments.  The problem is that the center-of-mass motion is not
decoupled as for a self-supported system. The center-of-mass of the
$N$ particles do not coincide with the center of origin for the
external field.  Then the particles can for example move coherently in rotational
states without changing their relative motion.  This would show up as
simultaneously existing BEC-like structures with eigenvalues of
comparable magnitude even when the number of particles increases.
This was pointed out in \cite{pet00} and an alternative definition of
BEC was suggested to remove the effect of the center-of-mass motion,
i.e. using the internal one-body density matrix where only relative
coordinates enter. Then different rotational motion build on the same
intrinsic BEC would all lead to one large eigenvalue of the internal
one-body density matrix.

\begin{figure}[ht]
\centering
  \epsfig{file=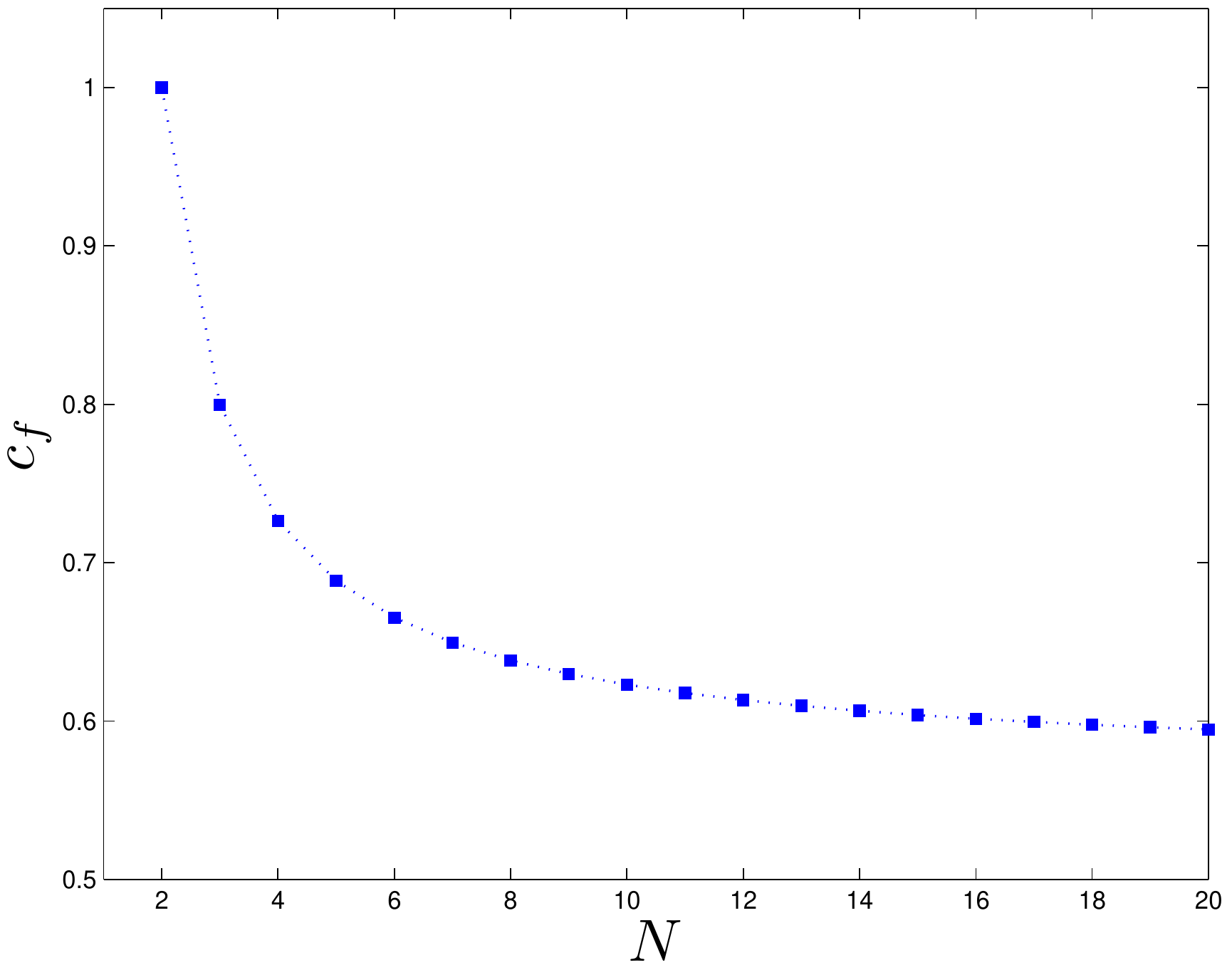,clip=true,scale=0.5}
  \caption{  \label{condfrac}
  Condensate fraction, $c_f$, as function of the 
  number of particles, $N$, for a gaussian mean-field wave function
  when the center-of-mass has been separated out as discussed in the 
  text.}
\end{figure}

We can illustrate by a gaussian mean-field wave function of range $b$
where the center-of-mass coordinate separates out leaving only
relative coordinates in the internal wave function, i.e.
\begin{eqnarray} \label{e35}
 \Psi_{int}(\{\bm{r}_i\}) = (b\sqrt{\pi})^{-3(N-1)/2} \exp(-\rho^2/(2b^2)) 
  \;,\\  \rho^2 \equiv  \sum_{i=1}^N  q_{i}^2   \;\;,\;\;
   \bm{q_{i}} \equiv \bm{r_{i}} - \bm{R}  \;\;,\;\;
 \bm{R} \equiv \frac{1}{N} \sum_{i=1}^N   \bm{r_{i}} \; , \label{e37}
\end{eqnarray}
where the coordinates $\bm{q_{i}}$ are measured from the common
center-of-mass $\bm{R}$.  This wave function is invariant under
rotations around $\bm{R}$.  Following \cite{pet00} the internal
one-body density matrix $\rho(\bm{q}_1,\bm{q}'_1)$ is now found
by inserting $\bm{q_{N}} = - \sum_{i=1}^{N-1}
\bm{q_{i}}$ in Eq.~(\ref{e35}), i.e.
\begin{eqnarray} 
&& \rho(\bm{q}_1,\bm{q}'_1) \propto \int d^3\bm{q}_2
d^3\bm{q}_3...d^3\bm{q}_{N-1} |\Psi_{int}|^2 \nonumber \\ &\propto&
\exp\bigg(-\frac{\bm{q}^2 + \bm{q'}^2}{b^2} +
\frac{(N-2)(\bm{q'}+\bm{q})^2}{(N-1) 4 b^2}\bigg) \ \label{e53}\;.
\end{eqnarray}
The condensate fraction obtained through the largest eigenvalue is
then \cite{gaj06} $c_f = 8/(1+\sqrt{2-2/N})^3$ which decreases with
$N$ from $1$ for $N=2$ towards about $0.57$ for large $N$. This 
is illustrated in figure \ref{condfrac}.
However,
the choice of relative coordinates is arbitrary \cite{gaj06} and we
could as well choose $\bm{q_{1}}$ supplemented by a set of $N-1$ {\em
independent} Jacobi coordinates.  Then the density matrix
corresponding to Eq.~(\ref{e35}) would factorize and give
$c_f=1$. This fact has also been addressed in \cite{yamada2008,yamada2009}. 
Clearly these discussions are only relevant for finite
systems where the center-of-mass can move.  Thus the distinctions
become less and less important with increasing particle number, but at
the same time interesting for a smaller number of particles in
BEC-like states.

\subsection{Mean-field description}\label{condens3}
The initial definition of BEC as a number of particles in the same
single-particle state immediately points to a mean-field approximation
or independent particle model.  The only
assumption is that the many-body wave function is a symmetric product
of single-particle wave functions.  With this restriction the
variational principle gives the lowest energy solution for any
Hamiltonian.  This implies that the average effect of the interactions
between the particles are included in this approximation.  Obviously
any interaction would prefer to correlate the particles either to
avoid or to exploit the interaction.  This cannot be expressed in the
mean-field product wave function which is the ideal BEC structure.
Thus any deviations beyond the mean-field must reduce the BEC content.
In other words interactions introduce correlations which attempt to
destroy the BEC structure.  

\begin{figure}[ht]
\centering
  \epsfig{file=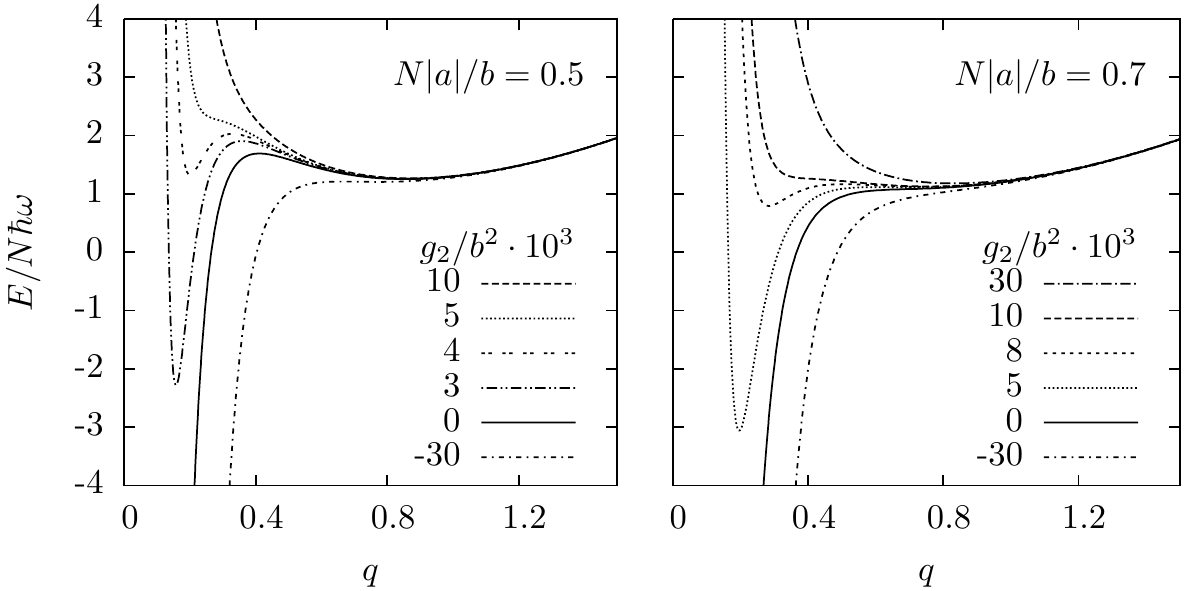,clip=true,scale=1}
  \caption{\label{varene}
    Energy of a BEC with fixed $N|a|/b$ as function
    of the variational parameter $q$, i.e. the size of the
    condensate (proportional to $b$).
    The higher-order interaction term $g_2=a^3/3-ar_{e}/2$ 
    includes the effective-range correction through $r_e$. It
    modifies the
    height and shape of the barrier through which macroscopic
    quantum tunneling occurs. The left panel has 
    $N|a|/b=0.5<0.65$ and the right panel shows results for
    $N|a|/b=0.7>0.65$, i.e. on both sides of the 
    stable regime when $g_2$ can be neglected. The right panel demonstrates
    that higher-order corrections can stabilize an attractively interacting
    condensate. Taken from \protect\cite{tho09a}.}
\end{figure}

For cold atoms the zero-range approximation for bosons has been used
successfully in the form of the self-consistent
mean-field Gross-Pitaevskii equation \cite{gro61,pit61,pet02,pit03}.  
It is demonstrated for dilute systems to be very efficient in
reproducing data for interacting condensates with inter particle repulsion.
For attractive interactions ($a<0$), the
system collapses unless some measures are taken to avoid this part of
the Hilbert space.  However, quasi-stable solutions exist of finite
size and energy if $N|a|/b<0.65$ where $b$ is the trap length of
the external harmonic potential \cite{ruprecht95,kagan98,pit03,pet02,boh98,sor03}.
Figure \ref{varene} shows the effective potential barrier of an 
attractively interacting condensate based on a gaussian variational 
calculation including also higher-order effective range corrections \cite{tho09a}.
The variational wave function is 
\begin{equation}
\Psi(r)=\frac{\sqrt{N}}{\pi^{3/4}\sqrt{(qb)^3}}\textrm{exp}\left(-\frac{r^2}{(qb)^2}\right),
\end{equation}
where $b=\sqrt{\hbar/m\omega}$ is the oscillator length, $N$ is the particle number, and 
$q$ is the dimensionless variational parameter.
The quasi-stable solution are located in the local minimum around 
$q\sim 1$. These solutions decay into deeper lying (collapsed) states of the
many-body system, e.g. non-condensate like states where a number of
binary or many-body bound states are present perhaps in some mixture
with dilute atomic states confined by the external field. When $N|a|/b$ 
is much lower than the critical value the lifetimes of the condensate is
much longer than the experimental time.

The mean-field treatment provides the optimal solution where the
BEC product wave function is maintained.  The most sensible definition
of correlations in many-body physics is in terms of deviations from
the simple mean-field product wave function.  Therefore when we define
BEC through the one-body density and its eigenvalues we have an
intuitive access to understand the effects of correlations on a
BEC. Expanding the full correlated many-body wave function on a basis
of different mean-field product wave functions, a number of non-zero
eigenvalues ($\lambda_i$) would appear with corresponding
single-particle wave functions.  The
condensate fraction is then given by the largest eigenvalue, $\lambda^*$, as $\lambda^*/N$.  
Imagine that we turn on
interactions slowly in an ideal BEC.  The condensate fraction then
decreases from unity and many eigenvalues of order $1$ would appear.
Approaching the thermodynamic limit by increasing $N$ while
maintaining the same interactions would again increase the largest
eigenvalue and the condensate fraction towards unity.  Even rather
strong two-body interactions would not destroy the BEC structure
reached for large $N$ \cite{mueller2006,tho07}. This does not necessarily mean
that the BEC structure is unaltered by the interactions and the
additional particles.  It means that the best approximation by a
product wave function, depending on interactions and $N$, is
approached \cite{lieb2001}.

\subsection{Condensates of Pairs of Fermions}\label{condens4}
Combining two fermions into a, possibly bound, entity produces a
system with boson characteristics. Redefining the degrees of freedom
from single-particle to two-particle center-of-mass and relative
coordinates, and integration over the relative coordinates in the
density matrix, allow use of the general definition of boson
condensates.  However, if the intrinsic two-body structure cannot be
frozen, we need a generalization of the BEC concept to systems of
fermions.  Due to the Pauli exclusion principle any one-body density
matrix will always have eigenvalues that are less than or equal to
one, so using this to look for coherence makes little sense. However,
as Yang \cite{yang62} shows, the two-body density matrix, $\rho_{2}$,
for fermions should be used instead of the one-body density matrix
$\rho_{1}$.  The definition of $\rho_{2}$ is
\begin{eqnarray}  \label{e412}  
 && \rho_2(\vec{r}_1,\vec{r}_2,\vec{r}^{\prime}_1,\vec{r}^{\prime}_2)
  \\ \nonumber &=& \int d\vec r_3 ...d\vec r_N
 \Psi^{\dagger}(\vec{r}_1,\vec{r}_2, ... ,\vec{r}_N)
 \Psi(\vec{r}^{\prime}_1,\vec{r}^{\prime}_2, ... ,\vec{r}_N) \; ,
\end{eqnarray}
which may have eigenvalues, $\lambda_i$, of order $N$, i.e.
\begin{equation} \label{dens2}
\rho_2(\vec{r}_1,\vec{r}_2,\vec{r}^{\prime}_1,\vec{r}^{\prime}_2) =
\sum \lambda_i \phi_i^*(\vec{r}_1,\vec{r}_2) 
\phi_i(\vec{r}^{\prime}_1,\vec{r}^{\prime}_2) \;.
\end{equation}
It is then natural to find these eigenvalues and define a fermion
condensate as the structure where one of these eigenvalues is large as
for BEC.  The connection to a boson condensate of composite particles
is then readily seen by change to relative and center-of-mass
coordinates $\vec{r} = \vec{r}_1 -\vec{r}_2$, $\vec{R} = (\vec{r}_1
+\vec{r}_2)/2$, and subsequent integration over the ``intrinsic''
$\vec{r}$-coordinates, i.e.
\begin{equation} \label{densbos2}
\rho_1(\vec{R},\vec{R}^{\prime}) =
 \int  \rho_2(\vec{r}_1,\vec{r}_2,\vec{r}^{\prime}_1,\vec{r}^{\prime}_2)
 d\vec r d\vec{r}^{\prime} \;.
\end{equation}
If $ \phi_i(\vec{r}_1,\vec{r}_2)$ factorizes in the
$(\vec{R},\vec{r})$ coordinates the fermion condensate reduces to a
boson condensate of pairs of fermions. This two-body density matrix determines
the second-order coherence of the system for both fermions and 
bosons as was mentioned above.

In many cold atomic gas experiments with fermions the system has two internal
states, similar to electrons in condensed-matter systems.
The fermions carry an effective half integer quantum number (the physical 
origin of this will be discussed in later sections). We have suppressed this 
index here. When internal states are present, 
one can think of the coordinates in Eq.~\ref{e412} as containing both spatial
coordinate and internal state information. 

Some very simple, although
pathological, limits of very few particles have  
been discussed in nuclei. For two
identical bosons the one-body density matrix is automatically of the
form in Eq.~(\ref{spec}) with only one term in the summation which
means maximum eigenvalue and ideal Bose condensate. Thus when the
center-of-mass wave function is controlled by an external field or
included as in \cite{pet00} this definition implies that 
two bosons always form a condensate. Two fermions with identical 
orbital wave function and antisymmetric
relative spin wave function automatically factorize as in
Eq.~(\ref{dens2}) with one term. This implies that two identical
fermions in a state of lowest energy are in a fermion
condensate. Furthermore, the composite system of two fermions is
by definition in a boson condensate consisting of one particle. 
These rather pathological 
limits demonstrate that composites of few particles with properly
chosen definitions can fulfil the criteria for a condensate.
Of course one must still determine whether this has 
any additional physical significance in the relevant system at hand. 
We will discuss these questions in a few simple cases in the second part of
the review.

\section{Pair correlations}\label{pairing}
Before we discuss pairing correlations in two-component Fermi systems, we 
first address an important question related to the choice of model space
and interactions. This is necessary since there are several choices involved
before one arrives at some properly defined model to which pair correlations
and a particular method of solving pairing Hamiltonians can be applied. This is
the case for both finite and infinite systems, and it is important to 
keep track of the effective interactions and model spaces that are applied 
when addressing generic features or universal behavior.

In the spirit of density functional computations, a sufficiently well
chosen density functional would provide any desired accuracy of the
true energy \cite{koh65,DFT}.  However, correlations are entirely missing in the
one-body structures of the wave functions.  To investigate effects of
correlations it is therefore necessary to use a Hilbert space that goes
beyond one-body product states.
As correlations are driven by interactions it is also necessary to
know precisely which interaction to use in the extended space.  These
connections and the practical implementations in nuclear and atomic
physics are completely different.

A natural starting point in both subfields is some form of self-consistent
mean-field model. For fermions the most common choice is 
the Hartree-Fock approximation \cite{fet71,negele1982,sie87,pet02}.  For
identical bosons (fermions) the Hilbert space is (anti)symmetrized
products of single-particle wave functions. Several 
extensions of the Hartree-Fock method are in use. In nuclear physics, a 
prominent example is the Hartree-Fock-Brueckner scheme \cite{fet71,sie87} which
takes medium effects of Pauli blocking into account in a self-consistent
manner. For cold atomic gases similar medium effects have been taken into account 
in various other ways \cite{heiselberg2000,stoof09}. For the present 
discussion we keep things simple and stay at first within the framework of
the standard Hartree-Fock approximation. 
Later on we comment on some important features of going beyond this approximation 
in connection with polarization effects, induced interactions and 
the renormalized zero-range BCS model.

A mean-field Hartree-Fock calculation requires
suitable two-body interactions as input.  
In atomic gases the diluteness of the system makes the zero-range 
approximation extremely accurate even around broad Feshbach resonances
where the effective range remains small \cite{kohler2006} 
(a more detailed discussion of Feshbach resonances is given in Section \ref{atomgas1}).
However, 
it is remarkable that
nuclear physics mean-field calculations employ
zero-range interactions since nuclei are not dilute as
the range of the nucleon-nucleon potential is comparable to the 
interparticle distance in the nucleus. Cold atomic gases are usually dilute
in this sense and zero-range interactions are more appropriate (at least away from
interaction resonances as we will discuss below). An interesting comparison can be 
made to fermionic liquid $^3$He which is a dense system.

In general, it is crucial to relate the
chosen interaction with the allocated Hilbert space 
\cite{bar80,suzuki1,suzuki2,brandow67,poves81}. This must be done whether one 
uses zero-range or some other form as the model interaction, although using
a finite-range potential can often help avoid divergences. 
The physical
interaction must be transformed to apply in a restricted Hilbert
space, and the operators for other observables than
the energy must be correspondingly transformed.  Properly done all physics is
then correctly maintained.  The connection between basic and effective
(transformed) interactions can easily be complicated or downright
impossible to trace. It is for example rigorously necessary to work
with up to $N$-body interactions, not only the simple initial two-body
interactions \cite{suzuki1,suzuki2}.  In any case, nuclear and atomic physics differ
tremendously in both effective interactions and Hilbert space 
even when identical techniques are employed \cite{bar80}.

The nuclear force is at low energy expanded in terms of the relative
momentum between the nucleons, and only terms up to second order is
maintained \cite{sie87}.  This Skyrme interaction corresponds to zero-range in
coordinate space \cite{sky56}. The inevitable collapse for attractive potentials in
Hartree-Fock applications is avoided by a completely phenomenological
(still zero-range) density dependence, or alternatively multi-body
forces \cite{neg72,neg75}.  The strengths of the different terms are then adjusted to
reproduce observables like energies and sizes of a series of nuclei.
The connection to the basic nucleon-nucleon interaction is lost.

In atomic physics the zero-range interaction in mean-field
calculations leads for identical bosons to the Gross-Pitaevskii
equation \cite{gro61,pit61,pet02,pit03}. The total wavefunction for 
$N$ bosons in factorized form, $\Psi=\Pi_{i=1}^{N} \phi(\vec r_i)$,
leads to the stationary Gross-Pitaevskii equation for $\phi$ which reads
\begin{equation}\label{gp-eq}
-\frac{\hbar^2}{2m}\nabla^2\phi(\vec r)+V_{ext}(\vec r)\phi(\vec r)
+\frac{4\pi \hbar^2 a}{m}|\phi(\vec r)|^2\phi(\vec r)=E\phi(\vec r),
\end{equation}
where $m$ is the boson mass and $E$ the energy per particle. 
The non-linear interaction term, which 
gives rise to many interesting physical effects in condensates, 
are well-described by this equation \cite{pit03}.
The strength of the
interaction is chosen to reproduce large distance, or equivalently
low-energy scattering properties within a confining external
field. This amounts to reproduce the physical atom-atom scattering
length by the Born approximation of the potential \cite{pet02,pit03}.
The intuitive explanation is that Hartree-Fock product wave functions
in the one-body external potential corresponds to free uncorrelated
solutions for which the Born approximation is sufficient.  For
repulsion the solution can then immediately be obtained. The collapse
for attraction is avoided by restriction of the Hilbert space to
larger distances or, equivalently to, lower energies.  The connection
to low-energy two-body scattering properties is maintained, and the
strength related to the scattering length in this limit.

\subsection{The BCS-approximation}\label{pairing2}
Correlations in general and pair correlations in particular arise only
by going beyond the Hartree-Fock or Gross-Pitaevskii approximation.
The neglected residual interaction must be defined relative to the
main part accounted for in the mean-field approximation. In
atoms the residual interaction is rather well-determined as connected
to the basic physical interaction.  However, for nuclei this
immediately presents the problem that the procedure is not unique
since the starting point is phenomenological. The residual and mean-field
interactions for nuclei must then be related through the same phenomenology.
Conclusions about physics beyond the region where the parameters are
adjusted can be very uncertain. Recent advances in deriving low-momentum interactions
for use in few- and many-body nuclei promise to improve on this situation
by decoupling the troublesome high-momentum parts of the typical nucleon-nucleon
interaction while preserving the correct scattering phase shifts \cite{epel09,bog10}. 
These developments may be relevant for the interchange between nuclei and cold
atoms. However, as these technical issues are not necessary for the purpose of 
discussing pairing correlations in the BCS approximation, 
we focus here on the more phenomenological approach traditionally 
employed for nuclei.

Pair correlations can be incorporated on an equal foot with the
single-particle degrees of freedom by extension of the Hartree-Fock to the
Hartree-Fock-Bogoliubov (or Bogoliubov-de Gennes in the atomic
literature) approximation \cite{deGennes99,boh69,ring80,doba2012}.
To grasp the main idea, the simplest
extension is sufficient which is to supplement the mean-field treatment by the
BCS approximation.  For simplicity we assume identical energy spectra, $\epsilon$, for the 
different internal states of the fermions we have in mind. In traditional superconductivity, 
the two internal states of the electrons are of course the spin degrees of freedom, whereas
in a two-component ultracold atomic Fermi gas the internal states are typically two 
states with different hyperfine projections (this will be discussed in greater detail below).
We will also assume that the
same number of particles occupy each internal state forming pairs. The latter condition 
means that we consider here only such balanced systems where the original implementation 
of the BCS theory applies. When the system is imbalanced, a number of exotic states like
the FFLO pairing state (characterized by a oscillating pairing gap parameter) have been 
predicted (see \cite{casal2004} for a recent review).
In nuclei, the different internal states are
given by time-reversal symmetry \cite{bmp58,boh69,ring80,sie87}, 
and in cold
atomic gases by the two decoupled hyperfine states with controlled frozen
occupation (see Section \ref{atomgas}). The role of time-reversal symmetry was
already discussed in the context of superconductors by Anderson \cite{anderson59}.
The Hamiltonian, after subtraction of
the Lagrange multiplier term $\mu N$, is
\begin{eqnarray} \label{e103}
 H_{\mu} &=& H_{mf} - \mu N + H_R  \\ \nonumber &\equiv&
 \sum_{i}  (\epsilon_i - \mu)  
(a_{i}^{\dagger} a_{i} + a_{\bar{i}}^{\dagger} a_{\bar{i}})
-  \sum_{i,j} G_{ ij} a_{i}^{\dagger} a_{\bar{i}}^{\dagger} a_{\bar{j}} a_j \;,
\end{eqnarray}
where the single-particle energies $\epsilon_i $, the creation
$a_{i}^{\dagger}$, and annihilation $a_{i}$ operators refer to the
single-particle mean-field states $|i\rangle$.  The internal states, time-reversed or hyperfine,
are denoted by a bar and we assume $\epsilon_i=\epsilon_{\bar{i}}$.
The single-particle energies are measured with
respect to the chemical potential $\mu$ multiplying the number
operator $N$. If the spectra of each component are not the same, 
we need a sum of two terms each related
to the two types of internal states.  The matrix elements $G_{ij}$ of the
residual interaction $H_R$, beyond the mean-field $H_{mf}$ then denotes
the two-body interaction between pairs of particles in time-reversed
(hyperfine) mean-field states. This piece of the interaction is called
the pairing interaction, and it should not be confused with neither
the full two-body interaction employed in the mean-field potential nor
any other residual contribution to the matrix elements, e.g. related to other
correlations like vibrational or rotational excitations in 
the RPA treatment.

The BCS approximation then consists in
\begin{eqnarray} \label{e143}
 H_{\mu} \approx   H_{mf} -\mu N
 - \sum_{i}  \Delta_i (a_{i}^{\dagger} a_{\bar{i}}^{\dagger} 
 + a_{\bar{i}} a_i)  \;\;,
\end{eqnarray}
which is diagonalized exactly by the Bogoliubov transformation
\begin{eqnarray} \label{e114}
a_{i}^{\dagger} &=& u_i \alpha_{i}^{\dagger} -  v_i \alpha_{\bar{i}} \;,\; 
a_{\bar{i}}^{\dagger} = u_i \alpha_{\bar{i}}^{\dagger} +  v_i \alpha_{{i}} \;, 
\end{eqnarray}
where the new quasi-particle creation $\alpha_{i}^{\dagger}$ and
annihilation $\alpha_{i}$ operators also obey fermion anti commutation
rules.  The result is
\begin{eqnarray} \label{e107}
H_{\mu} &\approx& U_0 + \sum_i E_i
(\alpha_{i}^{\dagger} \alpha_i + \alpha_{\bar{i}}^{\dagger} \alpha_{\bar{i}})
 \; , \\ \label{e116}
U_0 &=& 2 \sum_i \epsilon_i v_i^2 -2 \sum_i \Delta_i u_i v_i \;, 
\end{eqnarray}
The occupation numbers $v_i^2$ and quasi-particle excitation energies
$E_i$ are
\begin{eqnarray} \label{e120}
 v_i^2 &=& 1- u_i^2 =  \frac{1}{2}(1 - \frac{\epsilon_i -\mu}{E_i}) \;, \\
 E_i &=& \sqrt{(\epsilon_i-\mu)^2 + \Delta_i^2} \;\;  . \label{e122} 
\end{eqnarray}
The chemical potential $\mu$ and the gaps $\Delta_i$ are determined
for a given average number of particles $N_0=\langle N\rangle$ from the equations
\begin{eqnarray}  \label{e135}
  N_0 = 2 \sum_i v_i^2 \;\; , \;\; 
  2 \Delta_j  = \sum_i G_{ij} \frac{\Delta_i}{E_i} \; .
\end{eqnarray}
The new BCS ground state wave function related to the
$\alpha_{i}^{\dagger}$ quasi-particle operators is expressed by
\begin{eqnarray}  \label{e157}
 |BCS>  = \Pi_{i} (u_i + v_i a_{i}^{\dagger} a_{\bar{i}}^{\dagger}) |0> \;, 
\end{eqnarray}
where $|0>$ is the ``vacuum'' related to the $a_{i}$ operators. It is also
possible to write an analogue of the BCS theory presented here that conserves 
particle number by using projection \cite{pra1964,ado1992,leg06} although
these can become considerably more complicated. Schematic and exactly 
solvable pairing models have also been studied in great detail \cite{dukelsky2004,pan1999}.

The pair correlations obtained in the BCS-approximation are derived
from the residual interaction expressed in terms of the two-body
matrix elements, $G_{ij}$, of the interaction between pairs in
time-reversed states, see Eq.~(\ref{e103})   An approximation of constant pairing
matrix-elements, $G$, independent of $i$ and $j$ is very instructive
and often also reflects correct average pairing properties. This is
an approximation when the single-particle
states are different from plane waves even for a zero-range interaction.
Then
the gap $\Delta_i=\Delta$ also becomes state-independent and the
occupation numbers $v_i^2$ in Eq.~(\ref{e120}) are seen only to deviate
from zero or unity around the Fermi surface where the least bound
particles reside.  This indicates a tendency of pairing as a surface
phenomenon for finite systems \cite{bmp58,boh69}.

A zero-range residual interaction leads to constant matrix-elements
for plane wave single-particle states.  However, the summations for constant $G$, e.g. in
Eq.~(\ref{e135}), produce an unphysical divergence which is removed by
a cut-off in the sums at a finite value of $i$.  The price is
an adjustment, depending on the cut-off, of the value of $G$ to a
physical observable \cite{hu87,pet02}. The mean-field solutions usually deviate from
plane waves arising in free space with no external confinement. 
However, the gap equation can in fact be solved for arbitrary mean-field
states with constant matrix-elements, as discovered early on by Richardson \cite{rich1963}.
Any residual pairing interaction,
including zero-range, will have state-dependent matrix-elements in general. It is then as
convenient to use a realistic residual interaction of finite range
which automatically removes the divergence. Numerical treatment is
then necessary but nowadays only marginally more difficult.

The new ground state in Eq.~(\ref{e157}) describes a system of (Cooper)
pairs of particles with opposite momenta and zero total angular
momentum. This result applies to continuous mean-field spectra for any
attractive short-range residual interaction, and for discrete (finite)
systems above a critical strength, $G_c$ \cite{bmp58,fet71,sie87}.  
This ground state is
sometimes called the pair condensate to emphasize the analogy with
boson condensates.  The first excited state is, for constant matrix
elements, at $2\Delta$ corresponding to the energy required to break
one pair.  The gain in energy from the normal state with a Slater determinant wave function 
(a filled Fermi sea) to the
superfluid state is $g(\mu) \Delta^2/2$ where $g(\mu)$ is the
single-particle level density at the Fermi energy $\mu$ \cite{fet71}. Here
we ignore low-energy collective modes which appear at zero excitation energy
in the infinite system limit but which are pushed up to finite energy (i.e. they 
have an excitation gap) in finite systems
as well \cite{lan58}.

These
conclusions are quite remarkable when considering that in three
dimensions it takes a finite attraction to even produce a bound state.
Here (almost) any small attraction produces a qualitatively different
solution. In the original treatment of the effect by Cooper \cite{cop56} 
this occurs since the pairing is assumed to occur only near the Fermi surface.
The problem then becomes effectively two-dimensional or, equivalently, a
one-dimensional problem with a linear spectrum around the Fermi energy.
Here it is 
well-known that an arbitrarily weak attraction leads to a bound state \cite{lan81,simon76,vol2011}.
The presence of the Fermi sea, however, implies that 
we are dealing with a true many-body effect. In the cold atomic gases the zero-range
approximation is accurate and should be used also in the Cooper pair problem. However,
the pairing interaction is now constant everywhere in the Fermi sea and this
introduces the need for renormalization of the divergent bound-state equation. 
The bound-state remains
but its energy changes. In nuclear physics, the problem is usually 
treated by employing a cut-off either by
hand or by using a finite-range interaction that gives vanishing matrix elements above 
some energy scale. Thus, we see that different methods are in fact needed to extract the
same physics in different fields. We also caution that one should be careful to distinguish 
between the Cooper pair energy and the BCS gap parameter which have similar expressions as 
function of the pair interaction strength (or scattering length) but different exponents
\cite{deGennes99,leg06}.
All this should serve as a warning of the trouble that one faces when extending these
concepts to small systems with few particles both in nuclear and 
cold atomic systems.

To summarize for nuclei, both mean-field and residual interactions are
often chosen phenomenologically.
It is then of less importance
whether parameterizations are for a pairing residual potential, for the
matrix elements themselves, or if less details are required, for the
constant matrix elements combined with a cut-off.  For atoms the
mean-field potential is to a large extent determined by the external
one-body potential although the physical two-body interaction also may
contribute.  The contributing matrix elements in mean-field and
residual pairing interaction are not the same although for atoms both
are obtained directly from the low-energy scattering properties of the
physical two-body interaction.  It should be emphasized once again that
the two distinct states that have the paired structure 
are time-reversed orbits in nuclei but they are states of different 
hyperfine projection in the case of cold atoms 
(we give more details in Section \ref{atomgas3}).

For atoms the procedure for the BCS-calculations could be to choose a
potential of for example gaussian shape and adjust range and strength
to reproduce the desired scattering length for particles in paired
states.  The range is very small compared to the dimension of the
systems and the simplifications of a zero-range interaction is
therefore tempting.  However, this immediately introduces a divergence
in the gap equations (Eq.~(\ref{e135}) with $G_{ij}=G$), and a renormalization procedure
becomes necessary.  The method usually adopted is based on the observation
that the scattering length expression from the Lippmann-Schwinger
equation formally resembles the gap-equation, Eq.~(\ref{e135}), with
the same type of divergence (similar to the procedures used 
for Fermi gases with repulsive hardcore interaction \cite{fet71,pit03}).  
Subtraction of these two divergent
equations replaces the pairing matrix element by the physical
scattering length for particles in paired states, i.e.
\begin{eqnarray}  \label{e167}
 \frac{ m}{2\pi \hbar^2a} =  \sum _{\bf k}  \frac{1}{\epsilon_{\bf k}} -
 \sum _{\bf k}  \frac{1}{ E_{\bf k}}  \;,
\end{eqnarray}
which combined with the number equation in Eq.~(\ref{e135}) provide
gap, $\Delta$, and chemical potential, $\mu$ \cite{eag69,leg80,engel97,pap99,chen05}.  
This derivation is
based on the assumption that the mean-field wave functions are plane
waves leading to constant pairing matrix elements for the zero-range
residual interaction.  Thus the approximation is exact for a
non-interacting system in an infinite volume.  To be valid the system
must therefore be dilute and spatially extended. If we start with a finite-range
potential then the limit of zero-range must be approached after the infinite
volume limit.

In the discussion above we have ignored the effect on pairing from polarization of the 
background Fermi gas. This effect was considered shortly after the 
introduction of BCS theory 
by Gorkov and Melik-Barkhudarov \cite{gorkov1961} (often refered to as the GMB correction)
who calculated a reduction by a factor of $(4e)^{-1/3}\sim 0.4514$ 
on the transition temperature from paired to 
normal phases in a dilute Fermi gas. The zero-temperature value of the gap discussed above 
is reduced by the same factor. This effect was later discussed in the specific context of 
ultracold atomic Fermi gases \cite{heiselberg2000} and similar effects have been pointed out 
for Bose-Fermi and Bose-Bose mixtures \cite{pet02}. Such medium effects are relevant 
for many condensed-matter systems such as for instance liquid $^3$He where the 
equilibrium ground-state is determined from a manifold of degenerate possiblities
by medium effects \cite{voll90}. In the context of nuclear physics, this 
discussion also has a long history which is intimately connected with 
Hartree-Fock-Bruckner theory and 
the independent pair approximation \cite{fet71,sie87,dickhoff05}. 
For nuclear matter, the inclusion of medium polarization reduces 
the gap similar to the GMB corrections \cite{chen1993,wambach1993,schulze1996}.

\subsection{Pertinent properties of the BCS solution}\label{pairing3}
The BCS solution reveals
two-particle correlations through the so-called pair
wave function $\Psi_{pair}$, i.e. the amplitude for finding one
particle in $\vec{r}_1$ and another in $\vec{r}_2$.  We have
\begin{equation} \label{e305}
\Psi_{pair}(\vec{r}_1,\vec{r}_2) \propto \langle BCS | 
\psi^\dagger(\vec{r}_1)\psi^\dagger(\vec{r}_2)  | BCS \rangle  \; ,
\end{equation}
where $\psi^\dagger$ is the spatial creation operator which is
expressed by the mean-field wave functions $\phi_i$ in the external
field as
\begin{equation} \label{e310}
 \psi^\dagger(\vec{r}) = \sum_i (\phi_i(\vec{r}) a^{\dagger}_{i}
 + \phi_{\bar i}(\vec{r}) a^{\dagger}_{\bar i} )  \; .
\end{equation}
In terms of the $u$ and $v$ coefficients of Eq.(\ref{e114}) \cite{leg06}, one obtains
\begin{equation} \label{e320}
\Psi_{pair}(\vec{r}_1,\vec{r}_2) \propto  \sum_i  u_i v_i (\phi_i(\vec{r}_1)  
 \phi_{\bar i}(\vec{r}_2) + \phi_i(\vec{r}_2)  
 \phi_{\bar i}(\vec{r}_1) )  \; ,
\end{equation}
which is a symmetrized product wave function of the spatially identical
wave functions $\phi_i$ and $\phi_{\bar i}$. We have omitted the
antisymmetric internal wave function, spin singlet for nuclei and
antisymmetrized product of hyperfine states for atoms.

The $uv$-factor, $u_i v_i=\Delta_i/2E_i$, 
implies that the states within an energy interval of
$\Delta$ around the Fermi level are the main contributors to
$\Psi_{pair}$.  When the mean-field wave functions are plane waves as
for a homogeneous system the pair wave function in Eq.~(\ref{e305})
becomes the Fourier transform of the $uv$-product with respect to the
relative coordinate, and as such it depends only on the difference
between $\vec{r}_1$ and $\vec{r}_2$.  For finite systems like nuclei
the center-of-mass motion of the two particles are equally important
in $\Psi_{pair}$ as suggested by the relation $2(\vec{r}_1^2
+\vec{r}_2^2) = (\vec{r}_1- \vec{r}_2)^2 + (\vec{r}_1+ \vec{r}_2)^2$.
The wave function separates exactly into an independent center-of-mass factor
for a harmonic oscillator potential, and approximately for other
potentials.

The pair wave function, $\Psi_{pair}$, carries information about the
two-body correlation for particles around the Fermi level.  It
can be used to compute properties related to the pairs as
if it is an ordinary wave function.  For instance, the relative mean
square radius is obtained by
\begin{equation} \label{e330}
\langle r^2\rangle =  \int d^3\vec{r}_1  d^3\vec{r}_2 
 (\vec{r}_1 - \vec{r}_2)^2  |\Psi_{pair}(\vec{r}_1,\vec{r}_2)|^2  \; ,
\end{equation}
where the center-of-mass coordinate in $\Psi_{pair}$ can be removed completely for a
homogeneous system and decoupled for a harmonic oscillator external
mean-field. The expectation value in Eq.~(\ref{e330}) is often called
the mean square radius of a Cooper pair \cite{leg06}.

\begin{figure}[ht]
\centering
  \epsfig{file=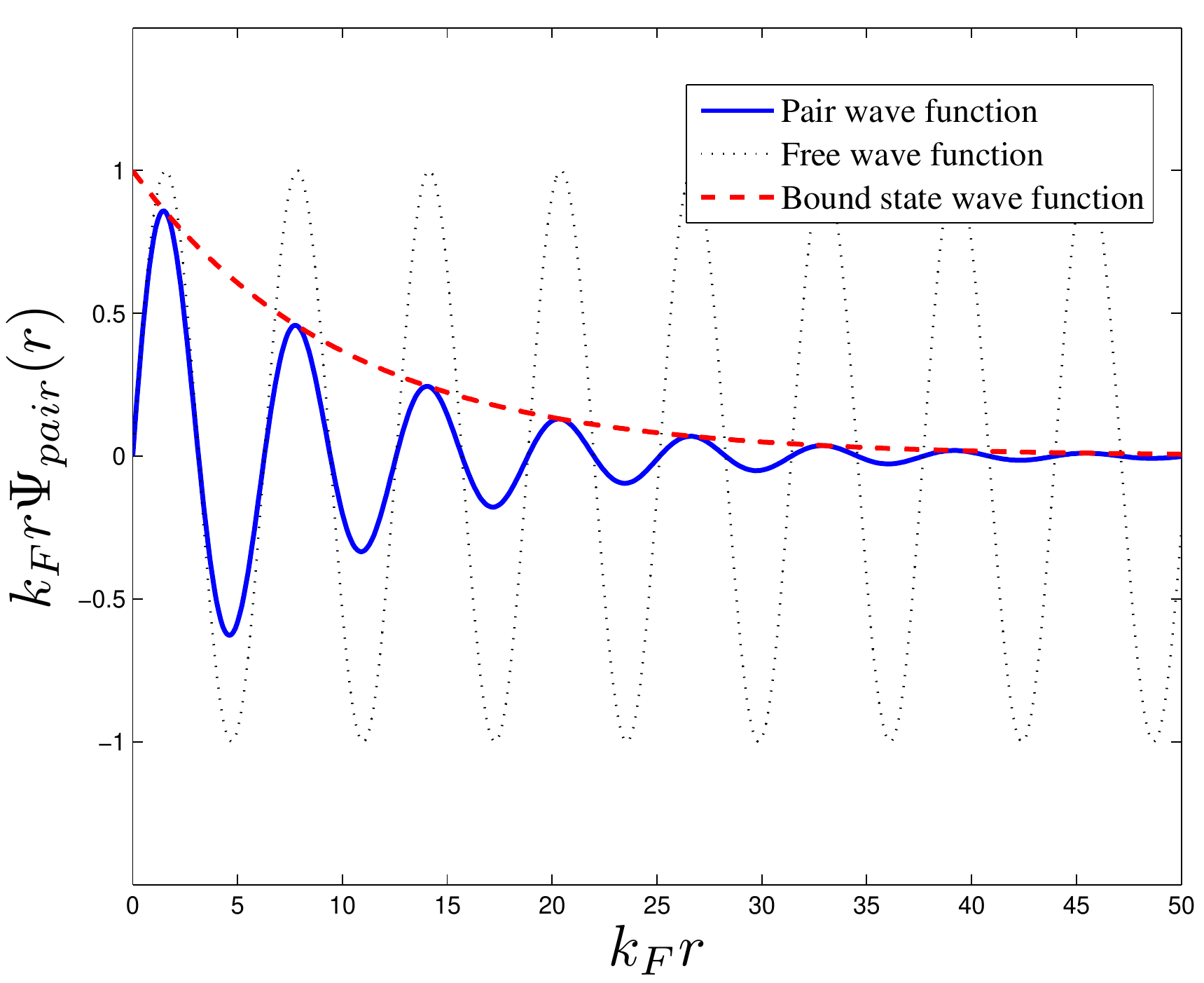,clip=true,scale=0.6}
  \caption{\label{pairwave} Plot of the pair wave function, Eq.~\eref{legwave}, multiplied 
  by $k_Fr$ as function of $k_Fr$ (solid (blue) line) for $k_F\xi'=\sqrt{200}$ (weak-coupling
  BCS limit).
  For comparison, we also show the
  non-interacting wave function with momentum $k_F$ (dotted (black) line) and the 
  bound state wave wavefunction corresponding to the exponential
  damping in Eq.~\eref{legwave} (dahsed (red) line).}
\end{figure}

For a homogeneous system of fermions with two internal states 
and a zero-range two-body pairing
potential, $\Psi_{pair}(\vec{r})$ is
for large values of $r=|\vec{r}_1-\vec{r}_2|$ given by \cite{leg06}
\begin{equation} \label{legwave}
\Psi_{pair}(r)\propto \frac{\sin(k_F r)}{k_F r}e^{-\sqrt{2}r/\xi'},
\end{equation}
where $k_F$ is the wavenumber at the Fermi energy and $\xi'=\hbar^2
k_F/(m\Delta)$ with $\Delta$ as the pairing gap at the Fermi
surface. 
Thus $\xi'$ is
seen as the characteristic length of the pair correlation
(historically one considers $\xi=\xi'/\pi$ which is the so-called
Pippard coherence length \cite{fet71}). 
The constant of proportionality in Eq.~(\ref{legwave})
depends on $\Delta$ and the density of state at $k_F$. The 
generic form of the pair wave function is shown in figure \ref{pairwave}.
From Eq.(\ref{legwave}) we
clearly see that the value of $k_F$ determines the density of nodes of
the pair wave function. These oscillations are often used as a
signature of BCS behavior. However, in finite systems they can arise
from other mechanism as we shall see in the nuclear case in Section \ref{nucbcs}.
Except for the exponential factor, the expression in Eq.~\eref{legwave} 
is exactly that of a
free particle (or pair in relative coordinates) having an energy
corresponding to $k_F$ (also shown in figure \ref{pairwave}). 
It is therefore apparent that the true
many-body character of the BCS solution is found in the exponential
decay of the pair correlation. However, in nuclei $\xi'$ is large
and therefore the exponential region is not entered before finite-size effects
set in. We will return to this point when we consider BCS in finite nuclei 
in Section~\ref{nucbcs}.

For finite systems large $\xi'$ complicates the signature of BCS from the
pair-wave function since there will always be an exponential suppression of the
wave function for radii that are outside
the potential range.  Reduction of the BCS many-body wave function to a
two-body pair wave function yields Eq.~(\ref{e320}) which has the
largest amplitude when the $uv$-factor is maximum, i.e. around the
Fermi level where $u^2=v^2=1/2$.  This largest component ($k=k_F$) is
responsible for the oscillating term in the approximation of
$\Psi_{pair}$ in Eq.~(\ref{legwave}).  The exponential damping
on the other hand becomes pronounced at large distances and grows
with $\xi^{-1}\propto\Delta$. Since the condensate fraction (of Cooper pairs)
is also propertional to $\Delta$ \cite{leg06}, we conclude that this 
damping is caused by the presence of 
the condensate at low energy (corresponding to large distance).

\subsection{The BCS-BEC crossover}\label{pairing4}
The physics of the BCS-BEC crossover 
originates from considerations of strongly coupled
superconductors by Eagles \cite{eag69} and atomic Fermi gases by
Leggett \cite{leg80}. The central observation is that by varying the
two-particle interactions between the electrons/alkali atoms from weak
to strong, one should observe a transition from the BCS paired state to a 
molecular BEC
state where the relative two-body bound state is the relevant degree of freedom,
hence the name BCS-BEC crossover. Intermediate stages would presumably
have both pairs, molecules, and single fermions. It is important to 
emphasize that this is not a phase transition in the usual sense 
of Landau's theory of symmetry breaking transition but rather 
a smooth transition.
The BCS-BEC crossover is the focus
of intensive studies currently, both in condensed-matter, atomic, and
nuclear physics. In particular, the use of optical lattices in
ultracold atomic gas physics hold great promise for achieving insights 
into models that are typically applied in condensed-matter systems and
are likely a key to a better
understanding of high-$T_c$ superconductivity \cite{chen05,leg06,bloch2008,giorgini2008}.

The physics associated with crossover, i.e. the transition from 
extended pairs to tightly-bound molecules, is found already in the
relatively simple Leggett model \cite{leg06}, which is expressed in
the homogeneous version of the gap and number equations in
Eqs.~(\ref{e135}) and (\ref{e167}).  These equations yield gap,
$\Delta$, and chemical potential, $\mu$, for any given density $n =
k_{F}^{3}/(3\pi^2)$, where $k_{F}$ is the Fermi momentum related to
the Fermi energy $\epsilon_F = \hbar^2 k_{F}^{2}/(2m)$.  By
dimensional analysis, $\Delta=\epsilon_F f(k_F a)$ and $\mu=\epsilon_F
g(k_F a)$, where $f$ and $g$ can be determined numerically. We also have
$(k_F a)^3\propto n a^3$ (typical values are given in Table~(\ref{tab:overview})). 
In the ultracold gases one fixes the density and
then vary $a$ through a Feshbach resonance \cite{bloch2008}. 
Then we can move from the
BCS regime ($k_Fa\to 0^-$), where $\Delta\propto \exp(-1/k_F|a|)$ and $\mu\propto
\epsilon_F$ \cite{leg06}, through unitarity at the threshold for 
two-body binding, where $|a|^{-1}\rightarrow 0$, and
into the deep BEC regime ($k_Fa\to 0^+$), where $\mu\propto -\hbar^2/2ma^2$. The latter 
shows that the energy per particle is half the binding energy
of the molecules in the deep BEC limit as it should be. Also, one can show that 
in the BEC limit, $\Delta\propto \epsilon_F / (na^3)^{1/3}$. For the 
pair wave function of (\ref{legwave}), this implies that $\xi'\propto a$. 
The bound state wave function in figure \ref{pairwave} is then exactly 
of the expected form, $e^{-r/a}/r$, and becomes a delta function for 
very tightly bound pairs ($a\to 0$).
Importantly, keeping $a$ constant and varying $n$ is another way that
one can explore the crossover. This can be seen by realizing that in the 
simple Leggett model, the dimensionless control parameter is $k_Fa\propto an^{1/3}$.
This observation will be important for
the application discussed in Section \ref{nucbcs}.

The crossover is then the sequence of structures occurring as the
interaction is varied from weak attraction describable in BCS-theory,
via stronger attraction barely binding pairs, to strong attraction
with substantial binding energy of pairs.  When pairs can bind it is
energetically advantageous to place all bound pairs in the same lowest
mean-field level for pairs, thus forming a BEC of these bosons made of two
fermions that exists only in the confinement of the external field.
The transitions in this sequence is driven by adiabatic or sudden
changes of the interaction or equivalently of $a$, resulting in
corresponding dynamic evolution of the structure.  This can either be
evolution towards an equilibrium or by establishing quasi-stationary
coherent oscillations between the initial state of paired atoms to
the structure of a molecular BEC \cite{pet02,bloch2008,giorgini2008}.

The unitary region is of considerable interest since $|a|\rightarrow
\infty$ means that the system effectively loses a scale, and
the divergent scattering length can thus not be used as expansion
parameter when calculating the energy of the system 
\cite{bertsch98,giorgini2008}. In this situation, the structure has to 
be independent of $a$, and hence is referred to as the universal limit.
For a Fermi
system, we thus expect the energy to be proportional to the only scale
in the problem: $E/N=\frac{3}{5}(1+\beta)\frac{\hbar^2 k_{F}^{2}}{m}$, where
$\beta$ (sometimes given as $\eta=1+\beta$) is a parameter that must be determined
numerically. Monte Carlo approaches \cite{carl03,ast04,cha04} give a value of $\beta=-0.58\pm 0.01$, 
which indicates considerable more binding than the BCS mean-field result of $\beta=-0.41$.
Analogous universal relations can be defined also in the imbalanced case \cite{giorgini2008} 
where the two internal states of the fermions are unequally occupied.
The bottom line is that any Fermi
system is expected to exhibit similar universal behavior provided $a$ is large compared to
other length scales.

In atomic physics, the simple homogeneous version of the gap equation
in Eq.~(\ref{e167}) is most commonly used.  The assumptions are low
density, negligible effects from the trap, and consequently plane
wave states.  In this dilute limit distances are large, momenta small, and
only binary encounters are expected to contribute to the physics.
These conditions are met in a typical atomic physics setup, where the
density, $n$, is fixed and the scattering length, $a$, is varied.  

Low density and plane waves are usually not applicable in
finite nuclear systems, where the basic interaction in principle is fixed by
the nucleon-nucleon potential. The bare (physical) neutron-neutron
scattering length ($\sim -18$ fm) is an order of
magnitude larger than the range of the nuclear force ($\sim 1.4$ fm), see Tab.~(\ref{tab:overview}). Thus, there is
hope that a system of low density neutrons would show universal behavior and be
described by something resembling a BCS-BEC crossover model. The
approximately infinite neutron matter that one might expect to find in
a neutron star therefore seems like an obvious candidate for applying
crossover theory to nuclear systems \cite{stars}.

However, the need for effective nucleon-nucleon potentials complicates
the problem.  This leads us back to the problem of separating
mean-field and residual interactions for nuclei. To illustrate, we
note that if one uses BCS theory in infinite nuclear matter with a
realistic bare neutron-neutron potential, then one finds a value that
is much smaller than both the measured gap and the value from the
otherwise successful semi-empirical mass formula \cite{fet71}.  Of
course, finite systems yield different results, but for large
nuclei one should expect fair agreement with this limit.  An effective
neutron-neutron interaction can be used to give a better gap, but this
is then only applicable in a stricly defined limited context in a 
local area of the nuclear chart and within a particular approximation 
scheme (such as BCS theory).

The unitary Fermi gas and the BCS-BEC crossover dynamics that can 
be studied with cold atomic Fermi gases also plays an important 
role in elucidating the nature of the so-called 'pseudogap' regime
which has been proposed in connection with high-temperature 
superconductivity \cite{chen2005,chen2009}. Theoretical 
studies find a pseudogap phase above the critical temperature
for superfluidity due to preformed 
pairs \cite{randeria98,janko97,yanase99,perali02,bruun06,barnea08}. However, these preformed
pairs are not condensed due to large fluctuations and the underlying
Fermi statistics of the particles. These predictions have been 
confirmed by ab-initio Monte Carlo calculations \cite{magier09,magier11} and 
can also be obtained using quantum cluster expansions \cite{hu10}. A number 
of recent experiments \cite{stewart08,gaebler10,perali11} provide strong support for preformed
pairs and for the pseudogap theory \cite{randeria10}.

\section{The Atom Gas from a Nuclear Perspective}\label{atomgas}
In order to meaningfully discuss the similarities and
differences between the atomic and nuclear physics systems it is
imperative to understand the relevant quantum states involved in the
two situations. Therefore we will now discuss the much studied
two-component Fermi gas from atomic gas physics. Bosonic states in
nuclei will be addressed later in relation to $\alpha$-particles.  
Then we will compare and contrast the interaction in the
atomic physics systems with that typically employed in nuclear physics
in the framework of the (nuclear) shell model.  We shall continue to
emphasize corresponding degrees of freedom in nuclei and cold atomic
gases.  Also for convenience we have collected in Table~(\ref{tab:overview}) 
a number of typical values for characteristic key quantities in the two systems.

\subsection{Two-component Atomic Fermi Gases}\label{atomgas1}
By now the production of degenerate Fermi gases has been achieved by
many experimental groups around the world \cite{bloch2008,giorgini2008}.  However, this was
initially a considerable challenge since the evaporative cooling of a
single species of fermions is hindered by the Pauli exclusion
principle since they are basically non-interacting. 
This means that the gas cannot equilibrate and lower its
temperature by expelling the fastest moving atoms. If one instead
prepares a gas with a population of two different species of fermions,
then low-energy $s$-wave interactions are no longer forbidden and degeneracy can be
reached \cite{ketterle2008}. Typically the Fermi temperature, $T_F=\hbar^2k_{F}^{2}/2mk_B$, 
is around $T_F\sim 1\mu$K,
whereas experiments can cool the gas to $T\lesssim 200$nK, thus $T/T_F\lesssim
0.2$ so the degenerate regime is truly reached. The
typical density of atoms in the trap is around $10^{13}$
cm$^{-3}$. This translates to an average interatomic distance of about 500 nm, which is
much bigger than atomic sizes (of order 0.1 nm) and inter-atomic potential ranges of order 1-10 nm. In contrast, nucleons in
nuclei have average interparticle distances of the order $\sim 1$ fm which is the same order as the nucleon size
(see Table~(\ref{tab:overview})) \cite{boh69}, and nuclei are never dilute in the sense of atoms in an
ultracold trapped gas. The criterion for diluteness in the atomic case is usually
$na^3\ll 1$ (away from Feshbach resonances). In nuclei, $na^3$ is of order one or larger. 
This is one reason to be careful in comparing the two systems.

The atoms must be confined by an external field, a trap, and two
distinctly different degrees of freedom are present, i.e. internal ones
related to the atom itself and external ones related to the atoms in the
trap.  Correspondingly, two different energy scales emerge, i.e. the
spacing of (external) trap eigenstates and the energy difference of
the (internal) atomic states.  We will now discuss these in detail for
the relevant experimental setup.

The internal atomic quantum numbers are those of the standard
hyperfine interaction, obtained from the coupling of the angular momentum 
(orbital, $\bm L$, and spin, $\bm S$) of the
electrons and the nuclear spin, $\bm I$, into total spin $\bm F$ and projection $m_F$. 
The complete set of quantum numbers is $(\bm J,\bm I,\bm F,m_F)$. In
a magnetic field only $m_F$ is conserved.  In the limits of high and
low field strengths there are good asymptotic quantum numbers
\cite{pet02}.  Systems with two components corresponding to population
of the lowest two hyperfine states (of two different given projections $m_F$) have been
extensively studied \cite{ketterle2008}, and recently also three-component systems 
(three different $m_F$ values) have
been realized in experiments \cite{ott08,huc09,naka2010}.  The
hyperfine energy splitting for different $m_F$ due to the magnetic
field is of order $10^{-6}$ eV for the alkali atoms typically used in
experiment \cite{pet02}. This is much smaller than any electronic transition in
the atoms.

The external quantum numbers depend on the trap which often is
harmonic and we shall use those of the harmonic oscillator, which for
the isotropic case is $(nlm)$.  Typically, the oscillator (trap) length
$b=\sqrt{\hbar/A m_N\omega}$ (where $m_N$ is the nucleon mass, $\omega$ the trap
angular frequency, and $A$ the mass number of the trapped species) 
is of the order of $1 \mu$m in a large trap.  This can be
translated into an energy $\hbar\omega=4.18\cdot
10^{-11}A^{-1}(1\mu\textrm{m}/b)^2$ eV.
We immediately see that the trap level
spacing $\hbar\omega$ is much smaller than the hyperfine splitting of order $\mu$eV.
In total, the two-component Fermi gas in a trap therefore has two
different internal states, $m_F$ and $m'_F$, and each atomic state is in
addition to the $m_F$-value associated with external quantum numbers,
$(nlm)$.

If we assume a distribution of identical fermions occupying the
(lowest two) hyperfine states, the lowest oscillator levels would be
occupied for each $m_F$ depending somewhat on the temperature.
Inelastic collisions can in principle change the hyperfine projections
of a pair of atoms. These
processes are, however, suppressed and the two-component Fermi gas is
therefore in practice stable with a fixed number of atoms in each 
hyperfine state \cite{ketterle2008}.  The
hyperfine states can be considered as frozen degrees of freedom when
we describe the system. The internal and external degrees of freedom
are uncoupled, and no spontaneous process exist that changes the internal
quantum numbers on the timescale of up to seconds of a typical experiment
\cite{ketterle2008}. This decoupling is what allows one to treat
separately these two sets of degrees of freedom and use product wave functions.

Beside the one-body external field also two-body interactions between
the atoms are present, thereby introducing another energy and length
scale into the problem.  The central part of the two-body interactions
between the
atoms is of the van der Waals type with characteristic length scale of 10-100 Bohr radii (0.5-5 nm) and
typical low-energy $s$-wave scattering lengths can be 100-1000 Bohr
radii and of both signs (ignoring for now the interesting case of Feshbach resonances).
The average distance (500 nm) in the cloud estimated above is
larger than these interaction lengths.  This would imply that two-body
interactions are weak. The ratio of the interaction energy, $E_{I}$
to the energy of the trap, $E_{HO}$ can be estimated for $N$ atoms in
a harmonic oscillator mean-field potential of length $b$ and a
$\delta$-interaction of strength given by the scattering length $a$,
i.e.  $E_{I}/E_{HO}\sim N^{1/6}a/b$ \cite{pet02}.  If we take
typical numbers, $a\sim 10-100$nm and $b\sim 1-10\mu$m, we therefore
see that with $a/b\sim 0.1-0.001$ the ratio will be small even for
millions of particles. The effect of interactions on the ground state
away from resonance is therefore usually small.

\begin{figure}
\centering
\epsfig{file=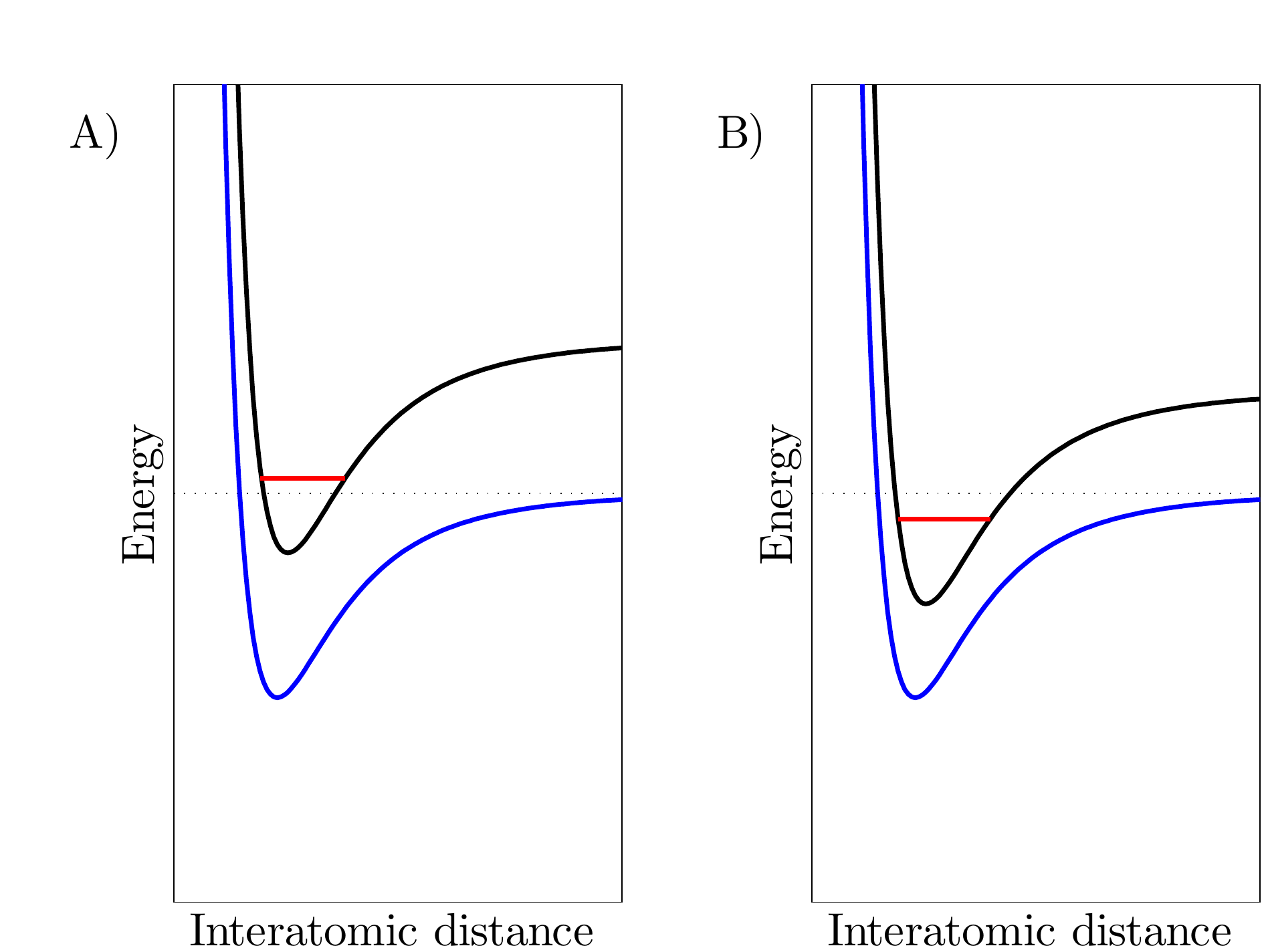,clip=true,scale=0.55,angle=0}
\caption{\label{feshbach} Illustrative plot of the physics of a Feshbach resonance. The lower (blue) potential curve
indicates the open (in-coming) channel of two atoms in a given spin state, 
while the upper (black) curve is the closed channel that 
is in a different spin state from the open channel. The dotted horizontal line indicates the 
relative energy of the incoming atoms. The horizontal (red) line in the closed channel indicates the presence
of a bound state. By tuning an externally applied magnetic field the open and closed channel potential curves can 
be moved relative to each other. In A) the bound state is above the relative energy, while it is goes below 
in B). The characteristic divergence of the scattering length, $a$, happens when the bound state energy coincides
with the reletive energy of the incoming atoms, and $a$ changes sign when going from situation A) to B).}
\end{figure}

The two-body interaction may, however, be varied experimentally by utilizing Feshbach
resonances resulting in anything from attraction to repulsion
corresponding to scattering lengths in the interval $[-\infty;\infty]$ \cite{chin2010}.
The Feshbach resonance technique exploits two internal atomic two-body states of 
different magnetic moments, which means that the relative energy of the two states can 
be controlled by an externally applied magnetic field. We illustrate the situation in 
figure~\ref{feshbach}.
When the scattering energy in 
the in-coming (open) channel is resonant with a bound state in a closed channel of 
different magnetic moment, the scattering length blows up and passes through infinity as
the magnetic field is varied. One thus has laboratory control of the effective interaction
of the atoms in the (open) scattering channel.
This situation is connected with the number of bound states of the two-body interaction.
The point at which the scattering length diverges is where another 
bound state appears (or disappears) in the potential. Note that Feshbach
resonances which are properties of the two-body potentials occur for
all atomic systems, irrespective of whether they are bosonic or fermionic
species. However, the $s$-wave interaction can of course only occur between
two bosons, a boson and a fermion, or two different internal states of 
fermions. For a single-component Fermi gas (fully polarized) the dominant
interactions are usually in the $p$-wave channel. We also note here that
there is a slight abuse of common language in scattering theory where the
word 'resonance' usually denotes a feature (typically a peak) in the 
scattering cross section. In the realm of ultracold atomic gases, a 
Feshbach 'resonance' most often refers to the divergence in the 
scattering length as a certain external (usally magnetic) field is varied.
The cause of this behavior of the scattering length is of course associated 
with a resonance in the traditional sense between an open and a 
closed channel as explained above. 

A common parametrisation of the $s$-wave scattering length variation as 
the magnetic field is tuned around a Feshbach resonance
takes the form
\begin{equation}\label{afesh}
a(B)=a_{bg}\left(1-\frac{\Delta B}{B-B_0}\right),
\end{equation}
where $B_0$ is the magnetic field position corresponding to a resonance at
zero energy, $\Delta B$ is called the resonance
width, and $a_{bg}$ is the background value of the scattering length
away from the resonance. $\Delta B$ and $a_{bg}$ can be of both signs
in general. Values and parameters for Feshbach resonances in 
different atomic systems can be found in \cite{chin2010}.
Correlations may increase tremendously when the scattering length diverges as the resonance
is approached.  Different features like BCS-BEC
crossover and universality as in Efimov physics (Section~\ref{efimov}) 
then arise with a demand to describe the
correlations. This can be compared with the nuclear physics case where
the predetermined strong nuclear interactions causes a highly correlated ground
state configuration.  Mean-field effects are still prominent but
correlations often contribute significantly.

We now estimate the ratio of interaction energy to mean-field energy in 
a nucleus in order to compare to the atomic physics ratio found above. 
We use a crude
harmonic oscillator value for the mean-field part and a
phenomenological estimate for the residual pairing interaction.  The
mean-field energy is $N\epsilon_F \sim N\hbar \omega (6N)^{1/3}$,
where $\epsilon_F$ is the Fermi energy, and we have $\hbar\omega=41/\sqrt{N}$ MeV 
(see Table.~(\ref{tab:overview})).
The energy gain from the
pairing interaction is proportional to the single-particle level
density $g_F\approx N/10$~MeV$^{-1}$ at the Fermi energy times the
square of the pairing gap $\Delta \approx 12/\sqrt{N}$ MeV \cite{boh69}. Thus the ratio
is about $1/(5N^{5/6})$ which is small and further decreases with the number of
nucleons. In terms of this estimate, the nuclear and atomic system are 
only comparable in the weak-coupling limit where $|a|\to 0$. When
the atomic scattering lengths are large, 
the nuclear and atomic systems are basically incomparable, 
i.e. the fundamental interactions are
completely different implying that the physics also is different.
We note that the weak-coupling limit is not very interesting in nuclei since
there is a critical interaction strength for superfluidity in finite 
Fermi systems.
However, this does not imply that the methods and techniques cannot be
carried from nuclei to atomic gases. Advanced nuclear models can
deal with correlations in both fields.

In studies of trapped (two-component) atomic Fermi gases one can often
replace the discrete one-body oscillator levels by the homogeneous spectrum,
essentially switching to continuous momentum space.  This is
because the oscillator level spacing is very small and the spectrum
can be well approximated as a continuum.  This also holds for
strong interactions, where local density or Thomas-Fermi approximations 
often are applied \cite{pet02,pit03,giorgini2008}.

The temperature scale of ultracold atomic gas experiments is
in the range $T\sim 10-100$ nK which is
extremely small but nevertheless non-zero.  The
corresponding energy $k_B T\sim 10^{-12}-10^{-11}$ eV is larger
than the typical atomic trap spacing. Thus, the spectrum is approximately
continuous and sums can be converted to appropriately weighted
integrals in thermodynamic considerations. If the trap
becomes tighter (small $b$) this approximation is no longer expected to be accurate as
the level spacing of the trap states goes as $\hbar \omega \propto
b^{-2}$. Then one must explicitly use the discrete spectrum and the
quantum numbers of the harmonic trap.  In this respect, tight traps
with very few particles is the case most closely related to nuclei
\cite{selim2011,zurn2011}. Alternatively, a very deep periodic optical lattice potential
can have single sites that are approximately harmonic and contain 
only a few particles \cite{bloch2008}, again closer to the situation in 
small nuclei.
The
notion of temperature is, however, somewhat ill-defined for isolated
systems like self-bound nuclei where the energy is conserved.  The
temperature is only useful in connection with average quantities in
sufficiently excited nuclei where the level spacing is relatively
small \cite{boh69,sie87}.

The tunability of atomic interactions through Feshbach resonances is 
an important feature of experiments that enable studies of physical
system over broad regions of parameter space. In particular, one 
is forced to think more in terms of generally applicable theories. In 
nuclear physics a number of scales are intrinsic to the problem and
do not change, even when considering stellar environments such as
neutron stars. Typical values are listed in Tab.~(\ref{tab:overview}).
A prominent example is the neutron-neutron scattering length 
which is $a\sim -18$ fm. This is an order of magnitude larger than the
typical interaction range in nuclei of $r_0\sim 1.4$ fm. This 
would suggest that neutron matter is in a universal regime, i.e. that 
the physics does not depend on the specific details of the inter-neutron
potential only on the scattering length and the density 
of the neutron matter. The dimensionless measure of the strength 
of the interaction in this situation is $k_Fa$ with  
$k_F\propto n^{1/3}$ the Fermi wave vector for a homogeneous 
system.
This is similar to 
a two-component atomic Fermi gas close to a 
Feshbach resonance where $a$ diverges. However,
the correction to the scattering length description which is 
given by the effective range is intrinsically large in nuclei, 
$r_e\sim 2.7$ fm \cite{sie87}. So when $k_Fa$ is large so is $k_Fr_e$.
A theory that takes the effective range correction into account was 
discussed in \cite{schwenk05} while a recent numerical study
can be found in \cite{gezerlis2010}.
Problems with non-universal corrections also arise in 
connection with bosonic few-body systems, when the range of the potential, $r_0$,
is neglected and only the scattering length is kept, i.e. $r_0/a$
or $r_e/a$ are small, as is often done in effective
field theory calculations of few-body properties \cite{deltuva10}.

In the cold atomic gases there is, however, a handle also on the relation
between $a$ and $r_e$ near a Feshbach resonance. This stems from the
fact that the effective range can be related to the width of the Feshbach resonance, $\Delta B$
\cite{chin2010}. In fact, in zero-range models, $r_e(B)$ on resonance ($|a|=\infty$) is given
by
\begin{equation}\label{reff}
r_{e0}=-\frac{2\hbar^2}{m a_{bg}\Delta B\Delta\mu},
\end{equation}
with $m$ the atomic mass and $\Delta\mu$ the difference in magnetic moment between
the coupled channels that cause the resonance \cite{petrov2004,jonsell2004}. The
effective range can thus be reduced by working with broad
resonances. Many cold atomic Fermi gas experiments have aimed for the 
universal physics as $a$ diverges and have worked with broad resonances where
background scales are small \cite{bloch2008,ketterle2008}. Also 
the wide resonances are easier to work with from an experimental point of view as 
they require less resolution on the applied magnetic field. However, 
very recent experiments have now started utilizing
narrow resonances \cite{gross2010}.

An aspect of recent interest for experiments is the ability to tune the scattering
length to zero at $B=B_0+\Delta B$ in the parametrisation of Eq.~\eref{afesh}
\cite{roati2007,pollack2009}. This was used
to reduce interactions in atom interferometry \cite{fattori08a} and to probe magnetic dipolar interactions
\cite{fattori08b}. In zero-range models of Feshbach resonances, the effective
range is a function of the magnetic field, $r_{e}(B)$, and in fact diverges when $a\to 0$ \cite{zinner2009}, 
a fact that also 
holds for a model potential like
the attractive $1/r^6$ potential which is the outer part of the atomic 
van der Waals interaction \cite{gao98}. The only remaining scales are 
then $a_{bg}$ and $r_{e0}$ from Eq.~\eref{reff}. For a bosonic system 
this can lead to strong modification of the stability properties 
around the resonance \cite{tho09a,tho09b,zin2012}. For atomic Fermi gases
the regime where $a\to 0$ remains largely unexplored which is due to the 
fact that the experimental use of broad resonances dominates the field and 
makes this non-interacting limit less interesting at the moment. As 
experiments become more sofisticated it is, however, clear that one 
should include the non-universal corrections from a non-zero effective range
in the description of experimental findings.

\begin{table}
\caption{\label{tab:overview} Estimates of typical parameters and ratios of interest in nuclear and cold atomic gas systems in natural units
of MeV and fm for nuclei, and eV and nm for atoms. The bosons in the table under nuclear physics are effective constituents. For cold atomic gases we merely cite a few alkali species of recent interest. $n$ is the particle density. $r_0$ is the characteristic distance of the interparticle force, which is the van der Waals length for cold atomic gases. The scattering length is given for both neutron-neutron (nn) and neutron-proton (np) for nuclei. $N$ is the number of atoms (with two spin states for the two-component gas considered in the table) and $A$ is the number of nucleons. $b$ is the harmonic trap size for atoms and the system size for nuclei. $\hbar\omega$ is the trap level spacing of the outer confinement for atomic gases, whereas for nuclei it is the level spacing in a mean-field oscillator model \protect\cite{fet71}. $E_{HO}\sim N\hbar\omega (6N)^{1/3}$ is the total mean-field energy for atoms and $E_{HO}\sim N\hbar\omega(3A/2)^{1/3}$ for nuclei. $\epsilon_K=\hbar^2/mr_{0}^{2}$ is the characteristic kinetic energy of the interparticle potential and $\epsilon_{int}$ is the energy difference between the internal degrees of freedom. For atoms this is the hyperfine scale $\Delta E_{hf}$, whereas for a single nucleon system with equal spin up and down components it is rouhgly the mean-field splitting $\hbar\omega$. $E_I$ is the average value of a typical residual interaction for nuclei and for atoms we use $E_I/E_{HO}\sim N^{1/6}a/b$ as in the text. $\Delta$ is the typical BCS pairing gap. The numbers for atoms given in the latter part of the table are for $N=10^6$, expect for the Fermi energy, $E_F$, which
is given by $E_F=(3\pi^2)^{2/3}\hbar^2n^{2/3}/2m$ where $m$ has been taken as the $^{6}$Li mass for atoms and the nucleon mass for nuclei.}
  \begin{indented}\item[]
    \begin{tabular}{lll}
			\br
      Quantity
			& Nuclear Physics 
			& Cold Atomic Gases\\
			\mr
       Fermions
			& neutron, proton
			& $^{6}$Li, $^{40}$K, etc.\\
       Bosons
			& deuteron (np), $\alpha$
			& $^{7}$Li, $^{85}$Rb, etc.\\
       $n$
			& 0.10-0.15 fm$^{-3}$
			& $10^{10}-10^{13}$ cm$^{-3}$ \\
       $d=n^{-1/3}$
			& $\sim 2$ fm 
			& 500-5000 nm \\
       $r_0$
			& $\sim 1.4$ fm
			& 0.5-5 nm \\
       $a$ (non-resonant) 
      & -18 fm (nn) 5 fm (np)
      & ($\pm$)10-100 nm \\
       $n|a|^3$ 
      & $10^{1}-10^{3}$
      & $10^{-6}-10^{-2}$ \\
       $A/N$
      & 10-100
      & $10^{3}-10^{6}$ \\
       $b$ 
      & 1.2$A^{1/3}$fm
      & 1-10 $\mu$m \\
       $\hbar\omega$  
      & $41/\sqrt{A}$ MeV
      & $\lesssim 10^{-11}$ eV \\
       $E_{HO}$ 
			& 47$A^{5/6}$ MeV
			& $\lesssim 10^{-3}$ eV \\
       $\epsilon_K$
			& 10 MeV
			& $10^{-7}-10^{-5}$ eV \\
       $\epsilon_{int}$
			& $\sim\hbar\omega$
			& $\Delta E_{hf}\sim 10^{-6}$ eV \\
       $E_I$
      & 5-15 MeV
      & $\lesssim 10^{-4}$ eV \\
       $E_F$
      & $\sim 43-56$ MeV
      & $10^{-12}-10^{-10}$ eV \\
       $\Delta$
      & 0.5-2 MeV
      & $\lesssim 0.5\epsilon_{F}$ \\
       $E_I/N\epsilon_{K}$
      & 0.02-0.2
      & 0.01-1 \\
     	 $E_I/E_{HO}$
     	& 0.006-0.04 
     	& 0.1-0.001 \\
     	 $\Delta/\epsilon_{F}$
     	& $<0.05$
     	& $\leq 0.5$ 
          \end{tabular}
  \end{indented}
\end{table}

\subsection{Interaction in a Shell Model Picture}\label{atomgas2}
We now consider the interaction in the atomic gases and compare the 
language
to the treatment of interactions in the standard nuclear shell model
through two-body matrix elements. This will also lead us to a
discussion of how to map the two-component Fermi gas with and without
trap into the nuclear system of a single nucleon species. We also
comment on the possibility of using the isospin formalism to
describe multi-species Fermi systems.

As mentioned, 
the interactions of typical two-body terms in atomic gases and
in nuclei are vastly different (see Tab.~(\ref{tab:overview})). 
The atomic interactions are of ranges
that are several orders of magnitude smaller than the average distance
in dilute gases. For nuclei it is quite the opposite. Here the range
is a few Fermis which is similar to the radius of the nucleus. 
The observation of pairing gaps and
superfluidity in atomic gases demonstrates 
that interactions are very important, so in
this respect atomic gases are similar to nuclei. However, we will now address a very
fundamental difference that can cause confusion.

In the atomic gases, the two-body interaction originates from the van
der Waals force and is a real physical interaction
\cite{weiner1999,kohler2006}. The Hamiltonian is divided into an
external one-body piece given by the trap and the two-body interaction from the
atom-atom collisions governed by the van der Waals interaction. In the
nuclear case, this kind of splitting is not provided a priori as the
nucleon-nucleon interaction is all there is. However, in practice most
methods find it convenient to split the interaction into a mean-field and a residual
two-body interaction. The mean-field is chosen to reproduce average
bulk properties such as energies and shell closures, whereas
the residual interaction should reproduce features beyond mean-field
such as collective states, superfluid properties, etc. 

In practice, this splitting for nuclei is rather arbitrary and
context dependent, because the basic as well as the effective nucleon-nucleon
interaction is phenomenologically adjusted to selected observables.
With a chosen interaction the split is seemingly optimized with the
self-consistent mean-field \cite{ring80}.  However, the very choice of
effective interaction depends on both the selected observables and the
chosen method of computation e.g. resulting in a limited Hilbert
space \cite{sie87}.  The residual interaction is then whatever is left
over from any given mean-field computation.  These are essential problems
of nuclear physics that have plagued the field.

In the following we will point out some of the problems
that different choices of mean-field and residual interactions can
cause. In the atomic case the choice is in this sense much more
pure. At most it can be debated where to put mean-field
contributions from the two-body interaction.  Great care must be
exercised to provide a useful and meaningful comparison which can be
used to transfer techniques and concepts.  

Let us now discuss the atomic interactions from a nuclear physics
perspective.  First we argue that $s$-waves of the relative motion of
two scattering atoms are the dominant two-body term for the
two-component gas. To distinguish from the
external trap quantum number $l$ we name the two-body relative orbital
angular momentum $l_r$.  Recall that the gas is extremely dilute with
average atom-atom distance much larger than the interaction range.
The two-body wave function approaches zero as $r^{2l_r}$, where $r$ is
the relative distance. For $l_r\neq 0$ there is a very small
probability of finding the particles at small $r$-values, whereas for
$l_r=0$ the probability is non-zero. For  very short-range interactions, 
it is therefore a good
approximation to restrict interactions to the $s$-wave channel.

The $l_r=0$ two-body states are symmetric, and the Pauli principle can
only be satisfied by using an antisymmetric combination of the other
parts of the wave function. If the system is fully polarized with
all atoms in a single hyperfine state, then $s$-waves would
be Pauli forbidden. The atoms therefore essentially do not interact,
only very weakly through odd parity Pauli allowed relative states,
i.e.  dominantly by $p$-waves.  On the other hand, two atoms in
different hyperfine states can interact much stronger through the
allowed relative $s$-wave interaction.

In nuclei the situation is quite different as the nucleon-nucleon
interaction has a hard-core part which generates terms of all
multipolarities where the strength diminish slowly for increasing $l_r$
\cite{boh69,sie87}. Furthermore the spin-dependence is quite
complicated and spin-dependent terms are strong in nuclei.
This is why a full analogue with dilute gases of atoms
is only appropriate for low density nuclear matter (ideally
low-density neutron matter to avoid Coulomb interaction) or perhaps in
very neutron-rich or halo nuclei. We will discuss some examples in the
following sections.

\subsection{Mapping atomic and nuclear degrees of freedom}\label{atomgas3}
The shell model has been extremely successful in nuclear (structure)
physics. First a mean-field is generated, self-consistently and often
phenomenologically adjusted, as a basis for a more detailed
description.  The preferred non-self-consistent phenomenological choice
is the so-called Woods-Saxon potential \cite{boh69}, which provides
bound state single-particle properties similar to that of a harmonic
oscillator. For nucleons in the mean-field, 
a sizable spin-orbit coupling is present with opposite sign compared
to the spin-orbit term for electrons in atoms.
Due to the strong spin-orbit
coupling in nuclear physics, it is customary to work in the coupled
representation where orbital angular momentum $l$ and nucleon spin
$s=1/2$ is coupled to $j=l\pm 1/2$ \cite{fet71}.

For atomic physics applications of nuclear methods this coupled basis
is not the most convenient choice.  To spell out the precise analogies
for a two-component atomic system we relate the external atomic trap
degrees of freedom with the nuclear orbital motion, and the internal
atomic hyperfine states with the two spin projections of the
nucleon. Since the external and internal atomic degrees of freedom are
effectively decoupled, an uncoupled product basis of nuclear orbital
and spin motion is the closest analogy.  This essentially corresponds
to a harmonic oscillator without spin-orbit coupling. 

To be specific we consider a two-body matrix element in the
interacting nuclear shell-model with the interaction relevant for the
two-component atomic gas.  First we have to specify the mapping of
atomic states to nuclear states. For this we choose to work with a
{\it single} species of a spin $1/2$ nucleon. The obvious thing is to
map the two hyperfine states of the two-component atomic gas to the
two spin projections of the nucleon. The nuclear two-body spin
wave functions can be either singlet or triplet (total spin of $0$ or
$1$) which corresponds to atomic antisymmetric or symmetric (internal)
hyperfine combinations, respectively.

The atomic two-body interaction acts in the $s$-wave (external)
channel on opposite hyperfine (internal) states only.  This maps to
$s$-wave nuclear interactions between opposite nucleon spin
projections only.  Therefore, the $s$-wave atomic interaction in nuclear physics
notation should be expressed as $V_{atom}(r)=P_{S=0} V_{2}(r)P_{S=0}$,
where $V_2(r)$ is the spatial part of the interaction in terms of the
relative coordinate $r$ and
$P_{S=0}=(1-\vec{\sigma}_1\cdot\vec{\sigma}_2)/4$ is the projection
onto the spin singlet (antisymmetric) component of the two-body state.
One can now use the standard methods of angular momentum coupling to
transform between coupled and uncoupled representations and thus
calculate any matrix element needed. An explicit calculation and
transformation for the zero-range interaction can be found in
\cite{cem2009}.

An important point in the mapping above
is that the two hyperfine atomic states in a two-component Fermi gas 
are mapped to a single species of
nucleon with each nucleon spin direction mapped to one of the two hyperfine
states. Of course there are both protons and neutrons in nuclei and
one could ask whether this has interesting analogies in atomic
physics. The first obvious system to look for is that of trapped gases
with two species of fermionic atoms, each with two internal hyperfine
states. In this case the $s$-wave interactions between identical atoms
in different hyperfine states would remain but between unlike atoms
all combinations of hyperfine states would have finite interactions.  
This could
resemble the situation in nuclear physics where neutron-neutron,
neutron-proton, and proton-proton $s$-wave interactions are all
allowed including extensively studied pairing interactions of
different types \cite{dean2003}.  However, the neutron and proton
interactions should then not be related as in nuclear physics but
translated into independent atomic interactions each connected to
their own individual hyperfine states. Recently such systems have been
seriously explored in alkaline-earth gases with the purpose of use for
quantum computation \cite{daley2008,gorshkov2009a}, and for testing
$SU(N)$-models for large $N$ \cite{gorshkov2009b,xu2009}. 
Experiments are also on-going \cite{fukuhara2009,de2009}.
Nuclear physicists often use isospin as a quantum number representing
the similarity of neutrons and protons in strong interactions
\cite{sie87}.  Isospin is a measure of the symmetry of the wave function
resulting from the permutation symmetry of neutrons and protons in the
Hamiltonian.  Therefore, it is not obvious that isospin has an
analogue in the atomic systems. This is because the scattering
properties of atoms depend sensitively on species and hyperfine states
and would therefore most likely strongly break this symmetry. The 
alkaline-earth gases above are, however, an example where
the spin of the atomic nucleus can be mapped onto isospin \cite{gorshkov2009b}.

Another possibility is a two-component system of bosons where
particles in the same hyperfine state are allowed.  The three
interactions between pairs in the same and in different hyperfine
states should then be similar to an isospin $1$ system.  Also
a three-component fermionic atomic system has three possible $s$-wave
interactions \cite{ott08,wen09} and could resemble an isospin 1 system.
However, it must be kept in mind that generally the scattering lengths
are different between different hyperfine components. Treating them 
as equal can be a good approximation at times and may help understand
the symmetries of the system. Of course, a full quantitative analogy requires that
one takes the different scattering lengths into account directly.

A concrete example of a fruitful interaction between nuclear
physics methods and the physics of cold atomic gases is the
pairing gap \cite{boh69}.  
This can be defined independently of any particular
theory, as 
\begin{equation}\label{gapsize}
\Delta(N)=\frac{E(N+1)-2E(N)+E(N-1)}{2} 
\end{equation}
with $E(N)$ the ground-state energy of the $N$-body system. Extrapolation 
of the pairing gap in nuclear physics from binding energy differences as in (\ref{gapsize})
presupposes that the underlying structure is unchanged by adding or 
subtracting one nucleon. This assumption is not always well fulfilled as for 
example deformation may change rapidly when closed shell configurations are
approached. A way to remove this problem to second order in such smoothly 
varying effects (such as deformation) is to use instead the double 
difference \cite{jensen84}
\begin{equation}
\frac{1}{4}\left[E(N-2)-3E(N-1)+3E(N)-E(N+1)\right]
\end{equation}
This is usually an improvement although a slightly larger range of 
nuclei are employed. For cold atomic gases such a procedure is typically
not necessary as the expression in (\ref{gapsize}) is sufficiently 
accurate.

The gap is a measurable
quantity, both for nuclei and for ultracold Fermi systems. In the
nuclear setup, there are many advanced many-body methods that go
considerably beyond BCS theory to calculate the gap.  As ultracold
gases become smaller and finite size and small particle numbers begin
to be important, these methods can be directly transfered from nuclear
physics by simple re-ordering and re-interpretation of models and
interactions as we have discussed above.

\subsection{Fermionic few-body systems}\label{atomgas4}
A particular promising venue where nuclear physics and ultracold atoms
can have large overlap is in the study of fermionic systems with 
a limited number of particles. This is a natural condition for the 
physics of light nuclei with intrinsically fermionic nucleon constituent.
Usually the number of atoms in a cold atomic gas experiment has been 
of order $10^3-10^6$ and a theoretical approach based on few-body 
methods is obviously doomed. However, recently microtraps have been 
built that are capable of trapping very few atoms ($1-10$) at 
very low temperatures \cite{selim2011,zurn2011}. Furthermore, it is now
possible to address single lattice sites in an optical lattice and 
produce and probe only a few atoms on such a 
site \cite{bakr2010,sherson2010,weitenberg2011}. These developments
provide hope that aspects of nuclear physics can be simulated with
cold atomic gas setups in a very direct manner.

In the meantime, a question of immense theoretical interest for 
fermionic two-component systems was posed back in 1999 by George 
Bertsch and concerns the nature of the ground-state of a system
of fermions with two internal states when the interaction between
fermions in different internal states have a large scattering length (
going to infinity, i.e. the unitarity limit).
The relevance of this question for atomic gases has been outlined above, 
whereas for nuclear physics it is relevant for instance in neutron matter
where the scattering length in the singlet channel (opposite spin direction) is 
large compared to the range of the nucleon-nucleon interaction. This 
question has sparked immense activity both for large and for small 
systems (a recent review can be found in \cite{giorgini2008}). 
Here we are interested in the connection to nuclei and 
to cold atom experiments with a limited number of particles. We will
therefore focus almost exclusively on small particle number.

Early studies of this interesting question in the homogeneous 
case have already been discussed in Section \ref{pairing4} and 
here we focus on the case where the particles are confined by an 
external harmonic potential. An analogy to nuclei can be made
by considering a mean-field potential where nucleons are 
confined to interact through the residual interaction.

\begin{table}
\caption{\label{tabsmmc} Energies (in units of $\hbar\omega$) calculated with the SMMC method for a trapped fermion gas
with scattering lengths
$a/b=11$ (BEC) and $a/b=-1.0$ (BCS)
for different particle numbers $N$. The statistical uncertainty is given in parenthesis.
HOSD denotes the non-interacting energies. From \protect\cite{zinner2009b}.}
\begin{indented}\item[]
\begin{tabular}{llllllll}
\hline
N  & HOSD & BEC & BCS   &   N & HOSD & BEC & BCS  \\
\hline
2  & 3  & 1.72(3) & 2.49(3)  & 12 & 32 &20.7(2)&27.23(2)    \\
3  & 5.5 & 3.9(2)& 4.84(2)  & 13 & 35.5 & 24.0(2)&30.21(2)    \\
4  & 8    & 4.96(3)& 6.84(4)    & 14 & 39 &26.0(1)&33.16(2) \\
5  & 10.5 & 7.1(3)& 9.14(2)    & 15 & 42.5 &29.1(1)&36.08(5)  \\
6  & 13& 8.11(7)& 11.08(5)      & 16 & 46 &30.9(1)&39.03(2)   \\
7  & 15.5 &10.6(2)&13.28(5)   & 17 & 49.5 &33.6(2)&41.98(2)  \\
8  & 18 &11.58(5) &15.21(4)    & 18 & 53 &35.7(1)&44.89(2)  \\
9  & 21.5 &14.8(2)&18.30(4)   & 19 & 56.5 &39.0(1)&47.85(2)   \\
10 & 25 &16.34(6)&21.27(3)     & 20 & 60 & 40.8(1)&50.75(2)\\
11 & 28.5 & 19.3(2)&24.28(2)    & 21 & 65.4 &44.5(2)  &54.50(2)  \\
\hline
\end{tabular}
\end{indented}
\end{table}

The early seminal study of Busch {\it et al.} \cite{busch1997} demonstrates
that for two trapped fermions in opposite internal states the ground-state
energy is $2\hbar\omega$ (see \cite{zinrev2012} a discussion of this model 
in both two- and three-dimensional traps and for a review of the relevant
theoretical work and experimental support). 
An important benchmark for the two-component fermionic few-body problem in
a harmonic trap is the exact solution for three particles at unitarity
where the ground-state energy is $4.27\hbar\omega$ \cite{werner2006a,werner2006b}.
These results was followed by several numerical studies of three and four 
trapped fermions \cite{stetcu2007,alhassid2008,stecher2007}, and for 
up to 30 fermions \cite{chang2007,blume2007}. All these studies found 
agreement with the exact result for three particles and furthermore 
established that five fermions have an energy of $5\hbar\omega$ 
to within a few percent at unitarity. We note that these nice universal 
results at unitarity break down if the two kinds of fermions have different
masses at which point finite-range corrections must be taken into account \cite{blume2010}.

A number of studies have pursued the energy and structural properties of 
fermionic systems as the scattering length is varied across the 
Feshbach resonance using various numerical methods for three and four
fermions \cite{stetcu2007,stecher2007,blume2009,daily2010} 
and for larger systems as well \cite{stecher2008,zinner2009}. This 
can be viewed as an attempt to address the physics of the BCS-BEC 
crossover from the few-body side. In particular, methods rooted in 
exact diagonalization that have been used abundantly in nuclear physics
can be directly applied to the study of fermionic systems as the
scattering length is varied \cite{stetcu2007,alhassid2008}.

\begin{figure}
\centering
\epsfig{file=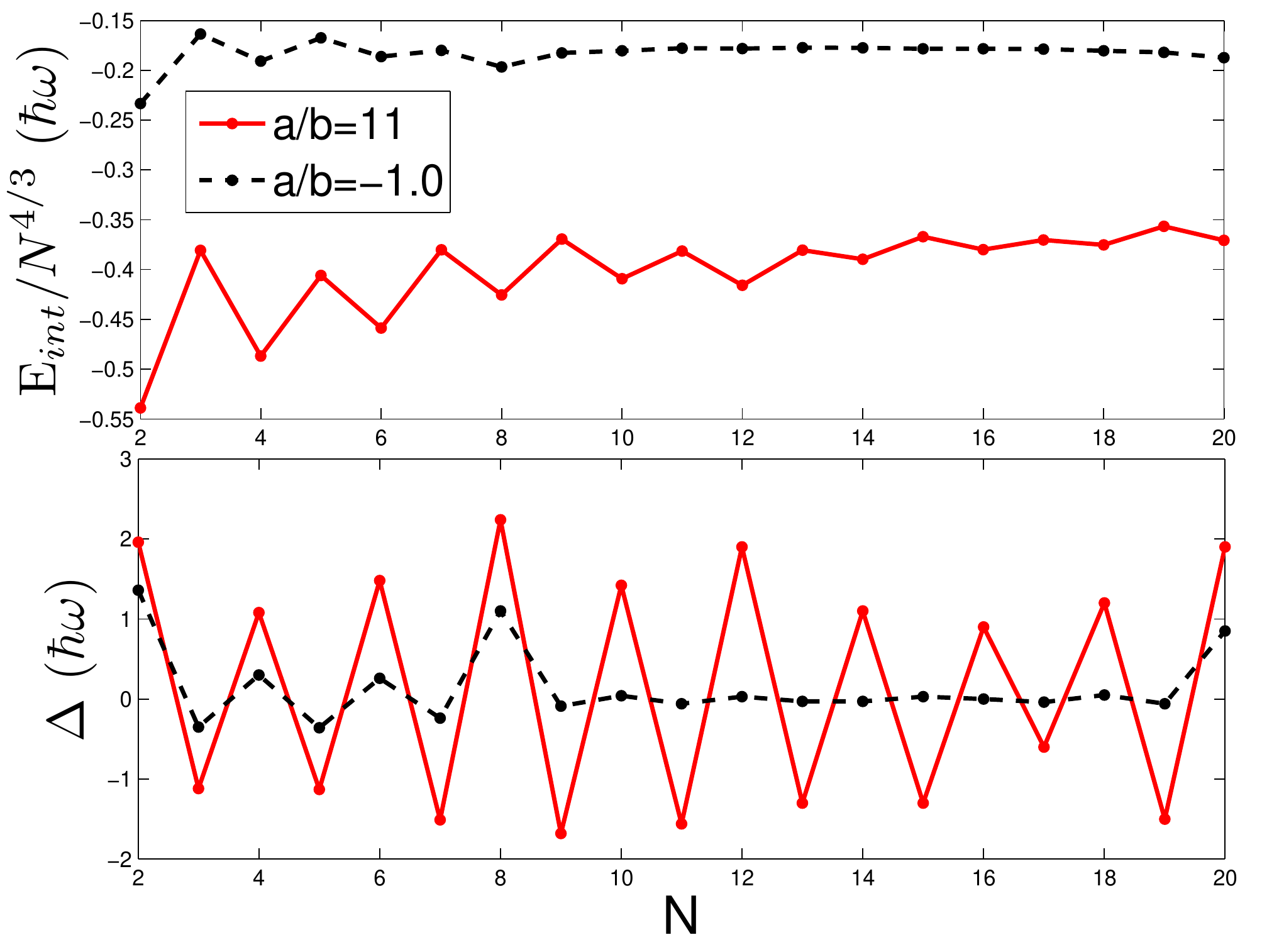,clip=true,scale=0.4}
\caption{\label{figsmmc} Scaled energies 
$E_{int}/N^{4/3}$, where $E_{int}=E-E_{g=0}$ (upper panel), 
and $\Delta(N)=E(N+1)-2E(N)+E(N-1)$ (lower panel) as a function of particle number $N$ for 
$a/b=11$ (solid) and $a/b=-1.0$ (dashed). From \protect\cite{zinner2009b}}
\end{figure}

The shell model Monte Carlo approach \cite{smmc} is a traditional 
nuclear physics method that has been recently applied to study 
trapped few-body systems with fermions as the interaction is varied
\cite{zinner2009}. In Table~\ref{tabsmmc} we show results for 
up to 21 fermions in a harmonic trap for a positive and a negative
scattering length that is close to unitarity. Notice how the results
for two, three, and four particles lie on each side of the energies
at unitarity quoted above. The three-dimensional harmonic oscillator
trap comes with its well-known shell structure. This is illustrated in 
figure~\ref{figsmmc} where the upper panel shows the energy divided by
the scaling expected in a Thomas-Fermi approximation. One clearly 
see an odd-even effect which is most pronounced on the BEC side of the 
resonance. A useful observable for such effects is the pairing gap 
discussed at the end of Section \ref{pairing3}. In the lower panel
of figure~\ref{figsmmc} the gap is depicted as function of particle
number. The shells are seen to matter most on the weaker BCS side of the 
resonance. This is consistent with the finding using other methods 
such as fixed-node diffusion \cite{blume2007} and Green's function 
Monte Carlo \cite{chang2007}, and with density functional theory 
using the local density approximation \cite{bulgac2007}. Similar
gaps are also found in two-dimensional systems \cite{rontani2009}.

We now consider some analogous systems from few-body nuclear physics. 
Nucleons are fermionic spin one-half particles and one could imagine
having a two-neutron or two-proton state, the former would be more 
desirable due to its lack of Coulomb repulsion. However, it is 
well-known that only the neutron-proton system is bound (naturally 
occuring as the atomic deuteron, $^2$D). Considering then three-nucleon structures
the possiblities are three neutrons, $^3$H or $^3$He. We 
are interested in a comparison to the few-body cold atom systems discussed 
above, and will focus on neutrons for the moment since the scattering 
length for neutrons is large.

Since the neutron-neutron scattering length is so large it indicates that 
the system is close to the resonance ($|a|\to \infty$) where a bound 
state appears. It is then reasonable to explore whether the 
additional neutrons in three and four neutron clusters could provide an 
overall bound system. Early experimental studies provide no evidence for 
either system \cite{fairman1973}. A claim of a tetra-neutron 
signal appeared in 2002 \cite{marques2002}, which could, however, not be reproduced
\cite{alek2005}. From a theoretical point of view, it was realized much
earlier that tri- and tetra-neutron cluster are unbound when using
nuclear forces that give reasonable predictions for nearby bound nuclei
such as $^3$H, $^3$He and $^4$He \cite{bevel1980}. Recent calculations
using state-of-the-art modern nucleon-nucleon potentials support 
this conclusion for both the tri- \cite{hemm2002,lazau2005} 
and tetra-neutron system \cite{pieper2003}.

The situation is somewhat different if one starts to consider neutron-rich
few-body nuclei that have one or a couple of protons. A nice example is
$^5$H. It could be the structure of a deuteron (neutron-proton) and a 
tri-neutron but in fact has the structure of triton, $^3$H, plus two 
neutrons \cite{diego2007}.
The latter implies that it could have an interesting three-body 
spectrum related to Efimov states to be discussed in Section \ref{efimov}.
Adding one or two more neutrons would give hope of
realizing a tri- or tetra-neutron plus $^3$H system. However, 
these attempts have also been futile. In the case of 
two protons, i.e. in the Helium isotopic chain, one can play the same 
game and this has been continued all the way up to $^{10}$He. A 
review concerning the limits of stability for different
isotopic chains can be found in \cite{tho2004}.

The nuclear studies of such exotic few-body states with large 
imbalance of neutrons over protons are difficult due to the 
complicated structure of the nuclear interaction. As the 
interactions are well under control in cold atoms in general
and also in experiments with only a few fermionic atoms, one
can hope to gain insights into the general issue of small 
interacting Fermi systems that can be transfered into the 
nuclear physics domain.

\section{Bosonic structures in few-body systems}\label{fewbody}
The BEC structure is defined by mean-field properties. The mean-field
approximation for self-bound finite systems violates translational
invariance and includes a spurious center-of-mass motion. This
approximation easily obscures all finite number effects, as e.g. shell effects. 
Conversely, a proper $N$-body wave function $\Phi$ only depends on
$N-1$ independent relative coordinates in conflict with the mean-field
product structure.  Suitable coordinates in the one-body density
matrix have to be specified to allow computation of the eigenvalues
which in turn decide whether a given structure can be classified as a
BEC.  We believe that only one combination of coordinates and
wave functions is reasonable, i.e. use of initial single-particle
coordinates and a total wave function where the relative wave function 
$\Psi$ is multiplied by a
center-of-mass function adjusted to maximize the largest eigenvalue of
the resulting one-body density matrix.

Before application of BEC definitions it is desirable to know
whether a given system is strongly or weakly correlated as in solids
or liquids, crystal or mean-field structures, localized or delocalized
structures.  Second, constraints from basic symmetry requirements
related to conserved quantum numbers are crucial for finite systems.
After discussions of these general questions we turn to applications
of $\alpha$-clustering and condensation in nuclei.

\subsection{Mean-field versus Localization}\label{fewbody1}

A system of $N$ particles prefer crystallization at low temperature
($T$) if the two-body interaction has a sufficiently deep minimum to
localize the relative wave function around the corresponding distance.
Alternatively, the zero point motion in a weaker attraction could be
larger than the range of the potential and the particles would move
across the potential wells created by neighboring particles. Then
each particle would rather be subject to the average potential from
all particles and the mean-field approximation would be appropriate.

\begin{figure}
\centering
\epsfig{file=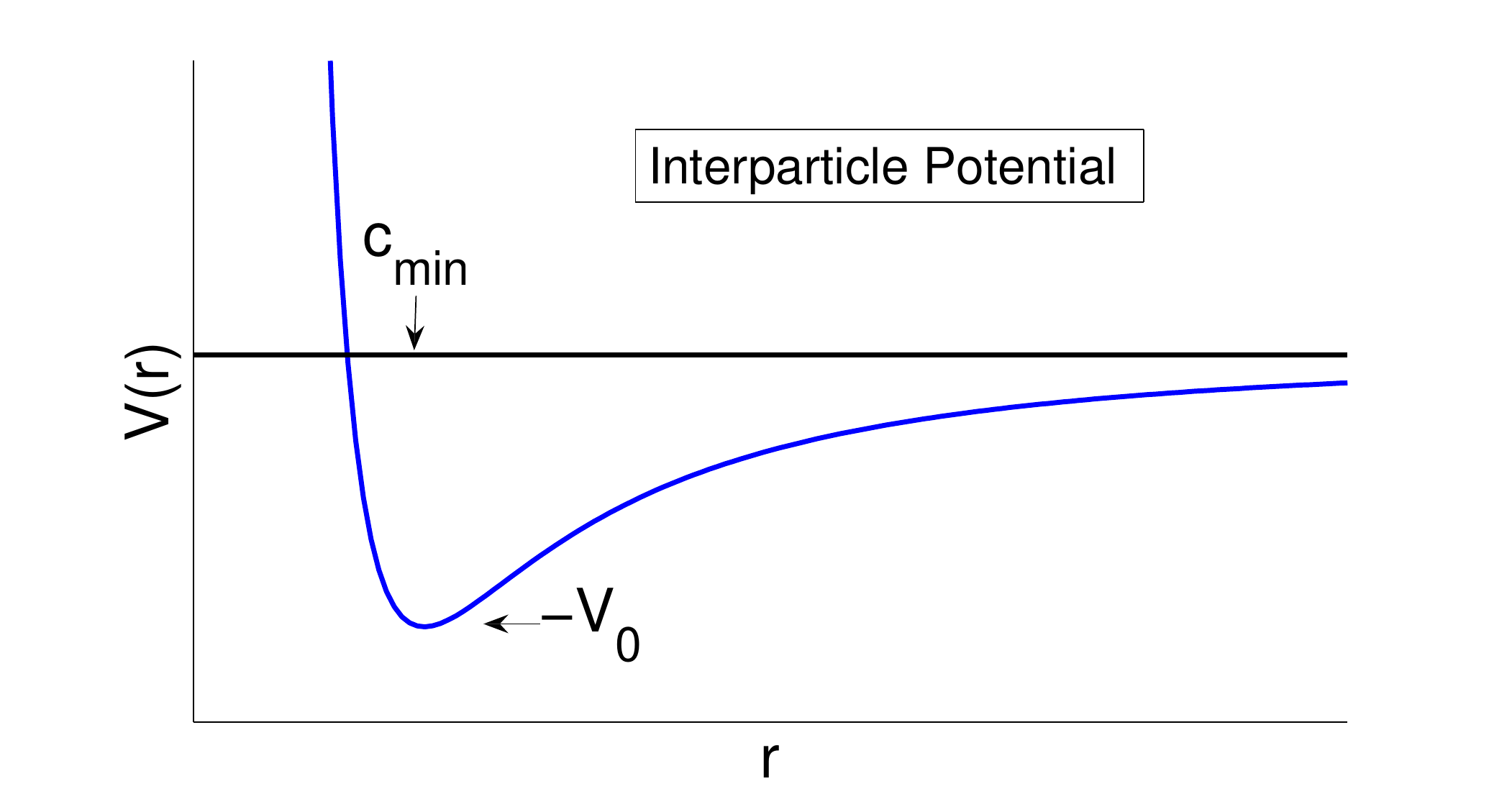,clip=true,scale=0.41,angle=0}
\caption{\label{fig1} Illustrative plot of the inter-particle potential 
as a function of relative distance.}
\end{figure}

To classify the potentials Mottelson, following earlier work by de Boer on the
solid state of noble gases \cite{am1976}, 
introduced in \cite{mot99} the
quantality parameter $\Lambda_{Mot}$ as the ratio between the kinetic
and potential energy of particles located in a relative two-body
potential (see Fig.~\ref{fig1}).  Using the notation of the figure,
this parameter can be written
\begin{equation} \label{mot}
\Lambda_{Mot}=\frac{K}{|V|}= \frac{\hbar^2}{m c^{2}_{min}|V_0|},
\end{equation}
where $m$ is the mass of the particles, with $V_0$ the depth
and $c_{min}$ the position of the assumed potential minimum.  The
kinetic energy $K$ is estimated as $K=\hbar^2/mc_{min}^{2}$ with the
momentum $p\sim \hbar/c_{min}$ and the reduced mass $m/2$.  The
potential energy is estimated as the value $V_0$ at the equilibrium
position. The quantality parameter does not
distinguish between bosons and fermions, implying that the criterion
is a truly profound way to classify physical systems.

Mottelson noticed \cite{mot99} that $\Lambda_{Mot}\sim 0.1$ marks a
transition between solid and liquid (mean-field) behavior. At $T=0$
both atomic $^4$He and $^3$He are liquids with $\Lambda_{Mot}>0.1$
whereas Ne and molecular H$_2$ are solids with
$\Lambda_{Mot}<0.1$.  For nucleons $\Lambda_{Mot}\sim 0.4$ and
therefore nuclei are expected to be described rather well by
mean-field models.  The physical interpretation is
straightforward. When $\Lambda_{Mot}$ is small, the kinetic energy is
small compared to the potential which keeps the particles in their
equilibrium positions and produces a solid.  When $\Lambda_{Mot}$ is
large, the kinetic energy (zero point energy in the well) is larger
than the depth and the particles move to distances greater than
$c_{min}$. We note that the usual characteristic of a solid in condensed-matter
physics which is the breaking of translational symmetry in the crystal lattice
is not obviously applicable to the finite systems we consider here. The total
wavefunction must of course be rotationally invariant, but by looking at 
the wavefunction as a function of the spatial coordinate for fixed center-of-mass
position the localized 'solid' structure will emerge as concentrations of the 
probability density. Fig.~(\ref{fig2}) illustrates this for the $N=3$ case.

When we say that the system exhibits mean-field behavior the
statement is that there are single-particle wave functions extending
over the whole volume occupied by the particles.  However, since we
are addressing interacting systems these mean-field wave functions
include effects from all the other particles. The corresponding
particles are ``effective'' in contrast to physical and the
wave functions are often called quasi-particle orbits.  The statement
that $\Lambda_{Mot}>0.1$ can then be reformulated by saying the
low-energy quasi-particle excitations have large (compared to the 
systems spatial extension) mean-free path
\cite{mot99}, which is the same as total delocalization.

The quantality condition can be related to other criteria from quantum
and statistical mechanics. We first distinguish between systems at
a given temperature determined by an external reservoir 
and finite (nuclear) systems where an unambigous way to assign
a temperature does not exist.  For a specified $T$ the thermal de Broglie
wavelength for ideal gases is given by \cite{fet71,hu87}
\begin{equation}
\lambda_{dB}(T)=\frac{2\pi \hbar}{\sqrt{2 \pi m k_B T}},
\end{equation}
which may be obtained from $\lambda = 2\pi \hbar/p$ and $p^2= 2\pi m k_B
T$, where $p$ is the momentum, $m$ is the particle mass and $k_B$ is
Boltzmann's constant.  This wavelength can be used to characterize
when a quantum gas of either fermions or bosons will display strong
quantum behavior as the temperature is lowered. For a Bose gas the
condition is $1\ll [\lambda_{dB}(T)]^3 n$, where $n$ is the particle
density. We have $d\sim n^{-1/3}$ (with $d$ the inter-particle distance), 
and this criterion can thus be written $1\ll
\lambda_{dB}(T)/d$. For a Fermi gas one has similarly $\lambda_{dB}/d
\propto \sqrt{\epsilon_F/k_B T}$, where $\epsilon_F$ is the Fermi energy. We
thus see that for $k_B T\ll \epsilon_F$, the de Broglie wavelength will be
much greater than the average inter-particle spacing.

We used that the kinetic energy of each particle on average is $\pi
k_B T$ up to factors of order unity.  
We assume that the total energy $E$ of the two-body system is
close to zero, implying that an estimate of the kinetic energy is
$|V_0|$, see Fig.~(\ref{fig1}).  Then the criterion for quantum
degenerate behavior is
\begin{equation} 
1\ll \Lambda_{deg}\equiv \frac{\lambda_{dB}(T)}{d}=
\frac{2\pi \hbar}{d\sqrt{2\pi m |V_{0}|}}= 
 \sqrt{2\pi\Lambda_{Mot}},
\label{cri}
\end{equation}
where we used $d\sim c_{min}$ and $V=V_{0}$ (see Fig.~(\ref{fig1})). We
therefore see how one can relate the quantality condition to the
relation for the breakdown of the classical gas regime in statistical
mechanics. Again there is no distinction between bosons and fermions,
explaining the universality of Mottelson's criterion for such
seemingly different systems.

Alternatively, in a finite isolated system where temperature is not assigned 
the de Broglie wavelength can instead be
defined for the motion of the constituents as
$\lambda_{dB}=2\pi\hbar/\sqrt{2m(E-V)}$.  If we now use $E\sim 0$ and
$V=-V_{0}$, we get the same relation as in Eq.~(\ref{cri}) except for an
unimportant factor of $\sqrt{\pi}$.  Thus the quantality parameter
$\Lambda_{Mot}$ is again related to the de Broglie wavelength of
motion, $\lambda_{dB}$, and the typical inter-particle distance.

The physics interpretation is that the wavelength, both with and
without temperature, compared to the inter-particle distance determines
the character of the structure, i.e. small wavelengths prefer solids
and large wavelengths prefer mean-field structures.  More
quantitatively if the wavelength reaches the order of the inter-particle distance, 
coherent quantum
structure is preferred. The critical value is $\Lambda_{deg} \simeq 1$
or equivalently $\Lambda_{Mot} \simeq 1/(2\pi) = 0.16$ in nice
agreement with the observation that $\Lambda_{Mot} \simeq 0.1$ is the
critical value \cite{mot99}. The competition of mean-field and cluster
states in nuclei has recently been addressed using state-of-the-art
nuclear energy density functionals \cite{ebran2012a,ebran2012b} using
both relativistic and non-relativistic functionals. The results obtained
are consistent with the simpler models discussed above.

\subsection{Symmetries, Localization versus Delocalization}\label{fewbody2}
\begin{figure}
\centering
\epsfig{file=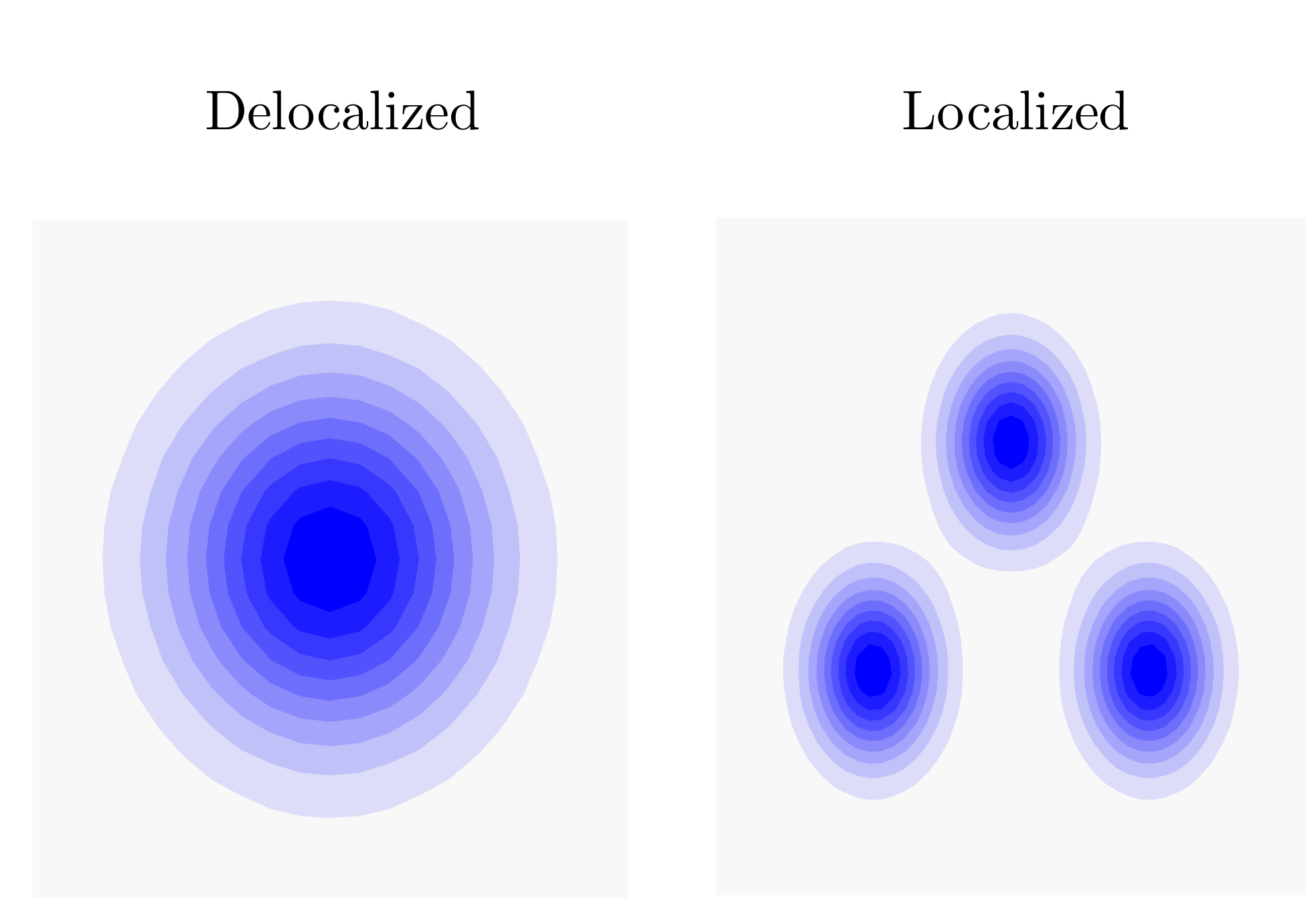,clip=true,scale=0.6}
\caption{\label{fig2} Illustrative plot of the delocalized wave function in Eq.~\eref{e33} and the
localized one in Eq.~\eref{e43} for $N=3$.}
\end{figure}

Precise condensate criteria are derived from the one-body density matrix,
$\rho_1(\vec{r},\vec{r}^{\prime}) $, defined in Eq.~(\ref{e410}).  To
illustrate effects of symmetry and localization we use gaussian
wave functions. This allows analytic derivations yet containing the
generic features of interest. A perfect mean-field product $N$-body wave
function is given by
\begin{equation} \label{e33}
 \Psi(\{\bm{r}_i\}) = (b\sqrt{\pi})^{-3N/2}\exp(- \sum_{i=1}^N r_i^2/(2b^2))\;,
\end{equation}
where $\bm{r_{i}}$ is the $i$th coordinate. This wave function, illustrated in 
Fig.~(\ref{fig2}), as well as all other
mean-field wave functions violate translation invariance, or
equivalently momentum conservation, which is restored by integrating
$\Psi(\{\bm{r}_i-\bm{R}'\}) \exp(i\bm{P}\cdot \bm{R}')$ over all
$\bm{R}'$ \cite{sie87}.  The result, $\Psi_{int}$, of lowest energy has $\bm{P}=0$
which for Eq.~(\ref{e33}) results in Eq.~(\ref{e35}) which also is invariant
under rotations around the center-of-mass $\bm{R}$. 

In contrast to the mean-field solution in Eq.~(\ref{e33}), the particles
can be correlated and the wave function $\Psi_{loc}$ in a body-fixed
coordinate system localized in distributions around preferred points
$\bm{R}_k$, i.e.
\begin{eqnarray} \label{e43}
\Psi_{loc}(\{\bm{r}_i\})  \propto 
 \sum_{\{p\} } \exp(- \sum_{i=1}^N   (\bm{r_{i}} - \bm{R}_{p(i)})^2 /(2B^2)) \;,
\end{eqnarray}
where the normalization is omitted and full symmetry is achieved by
the summation over all permutations $p$ of the set of numbers
$\{1,2,....,N\}$. This is illustrated in Fig.~(\ref{fig2}) for $N=3$.  
The translational invariance is restored precisely
as for Eq.~(\ref{e33}), i.e. the invariant wave function is obtained from
Eq.~(\ref{e43}) by the substitution $\bm{r_{i}} \rightarrow \bm{q_{i}}
= \bm{r_{i}} - \bm{R}$ with a corresponding change of normalization
constant.  The rotational invariance is broken for $\Psi_{loc}$ in
Eq.~(\ref{e43}) but recovered for states of zero angular momentum by
equally weighted linear combinations of all spatial rotations of $\Psi_{loc}$.

The condensate fraction depends strongly on the degree of localization
as we can see explicitly by computing the one-body density matrix for
Eq.~(\ref{e43}).  We assume very narrow non-overlapping gaussians
and obtain
\begin{eqnarray}
 &&\rho_1(\bm{r},\bm{r}') = (B\sqrt{\pi})^{-3/2} \\ \nonumber &&\times
\sum_{k=1}^N \exp(- ((\bm{r} - \bm{R}_k)^2 + (\bm{r'} - \bm{R}_k)^2)
  /(2B^2))\;,
\end{eqnarray}
which has $N$ equally large eigenvalues while all others are zero.
This is a condensate fraction of $1/N$ corresponding to one
single-particle state for each of the $N$ particles.  However, after
restoration of rotational symmetry only eigenvalues zero remain.  If
the widths, $B$, of the gaussians increase and they begin to overlap
with each other one eigenvalue separates out and becomes finite.
Increasing the width leads to increasing overlap with a product wave
function like Eq.~(\ref{e33}).

We can quantify by computing the overlap between the factorized and
localized wave functions in Eqs.~(\ref{e33}) and (\ref{e43}), i.e.
\begin{eqnarray} \label{e47b}
 \langle \Psi | \Psi_{loc} \rangle =  \bigg(\frac{2b B}{b^2+B^2}\bigg)^{3N/2}
 \exp( - \frac{\sum_{k=1}^N {R}_k^2}{2b^2+2B^2})  \;,
\end{eqnarray}
which only is close to unity when $b\sim B$ and either ${R}_i/B \ll 1$
or ${R}_i/b\ll 1$.  Eq.~(\ref{e47b}) is also obtained if one averages
$\Psi_{loc}$ over all possible angles to obtain a rotationally invariant wave function.  
Thus a
substantial condensate fraction requires that the overlap of $\Psi_{loc}$ with
Eq.~(\ref{e33}) is large.

The discussion above is very simple to relate to the Mottelson
quantality. The widths of the gaussians increase with the zero-point
energy in the two-body potentials, which corresponds to increasing
kinetic energy in Eq.~(\ref{mot}).  The states then overlap and one
eigenvalue of the density matrix will grow in magnitude.  On the other
hand, the depth of the potential tends to localize the particles with
decreasing widths for increasingly attractive potentials.  The density
matrix will thus have $N$ eigenvalues of order one.

\subsection{Trapped Few-Boson Systems}\label{fewbody3}
As discussed in Section \ref{condens1}, the characterization of
cold atomic Bose gases from the one-body density matrix has been 
explored in great detail for large samples of atoms. To compare
nuclei with relatively few particles to atomic gases we now
discuss the case of a small number of bosons in a trap.
Monte Carlo studies exist for the
homogeneous \cite{giorgini1999} and the trapped case \cite{dubois2001}
that consider the total energy per particle. However, these results are
for large systems of more than a hundred bosons. Diffusion Monte Carlo was performed
for systems of $N=2-50$ particles \cite{blume2001}, and also variational 
Monte Carlo has been used \cite{han06}. The energy per particle
obtained in the smaller systems are in good agreement with that obtained from 
larger systems.
However, they also show considerable depletion of the condensate fraction
at large scattering lengths. Common to these studies is the use
of repulsive two-body potentials. This has the drawback that the
range of the potential increases essentially linearly with the scattering
length, i.e. at large scattering length the range of the potential can
become as large as the inter-particle spacing and the details of the
potential can no longer be neglected. In this non-universal limit, 
where the range of the potential is comparable to the scattering length, the
energy cannot be expressed in terms of only the scattering length, and 
an effective one-channel model for the effective interaction is no
longer sufficient.

\begin{figure}[htb]
  \centering
  \includegraphics{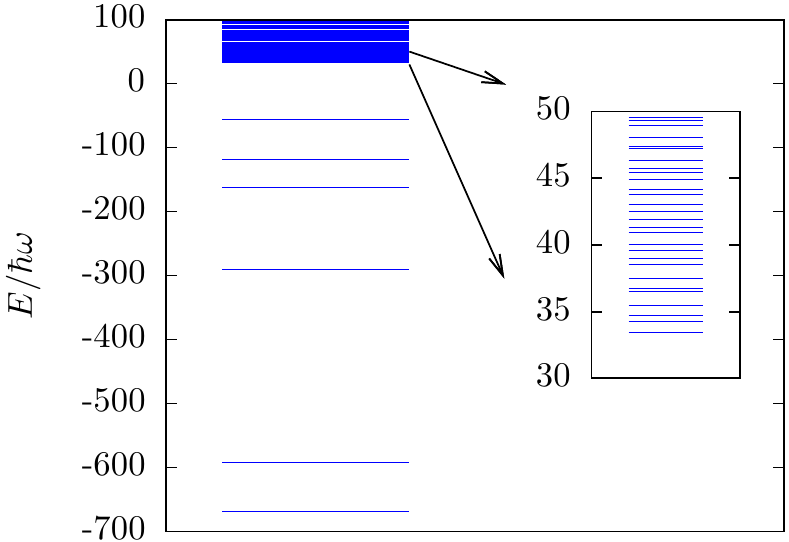}
  \caption{\label{trap-spec} A typical spectrum (in the vicinity of zero energy) of a
    system of $N=20$ bosons in an isotropic harmonic oscillator trap,
    interacting via an attractive gaussian two-body potential
    with one bound state and a
    positive scattering length. The inset shows the beginning of the
    so-called quasi-continuum spectrum. From \protect\cite{tho07}.}
\end{figure}

A solution to this problem is to use an attractive finite-range interaction 
as a model potential. This would also seem more realistic in terms
of providing an effective one-channel model for the Feshbach 
resonances of experiments. It
is well-known that potentials like the attractive gaussian or square
well can attain any value for the scattering length as one increases 
the strength around the threshold for the appearance of a two-body
bound state. This brings in the complication that the many-body problem 
will have many self-bound negative-energy states when bound states are
allowed for a pair of bosons. However, if one uses direct numerical 
diagonalization, then all states are determined and it turns
out that states with the properties expected of condensates appear 
as excited states in a quasi-continuum that are orthogonal to all the 
bound negative-energy states
\cite{tho07}. An example of this is shown 
in Fig.~\ref{trap-spec}.

\begin{figure}[htbp]
  \centering
  \includegraphics{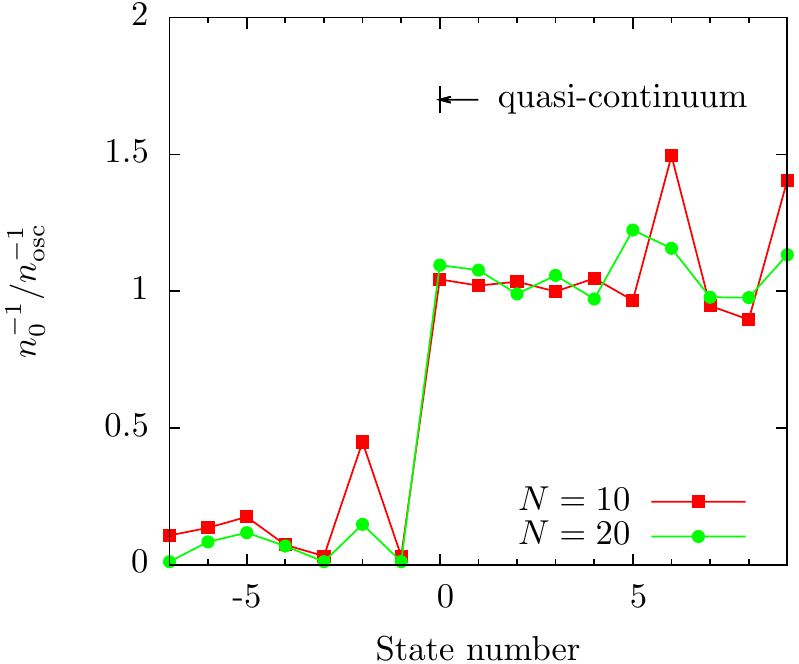}
  \caption{\label{trap-den} The inverse central density $n_0^{-1}$ (in oscillator units
    $n_\mathrm{osc}=\pi^{-3/2}Nb^{-3}$, where $b$ is the 
    oscillator length) for a system of $N$ bosons
    in an isotropic harmonic oscillator trap, interacting via an
    attractive gaussian two-body potential,
    with scattering length $a=119$au as function of the state number.
    The lowest state with positive energy is numbered zero. From \protect\cite{tho07}.}
\end{figure}

The condensate nature of the lowest state in the quasi-continuum can 
be verified by calculating the central density of the cloud. 
For a completely delocalized 
mean-field condensate state one would 
expect the bosons to occupy the entire trap. 
Indeed, as shown in Fig.~\ref{trap-den}, one finds that the first state
of positive energy fulfills this conditions, whereas the negative-energy
self-bound states have a much larger density and thus smaller volume per
particle. This is also reflected in the condensate fraction which 
is small for the negative-energy states, is unity for the lowest 
state in the quasi-continuum, and then decreses slowly for higher-lying
states \cite{tho07}.

\begin{figure}[htbp]
  \centering
  \includegraphics[scale=0.9]{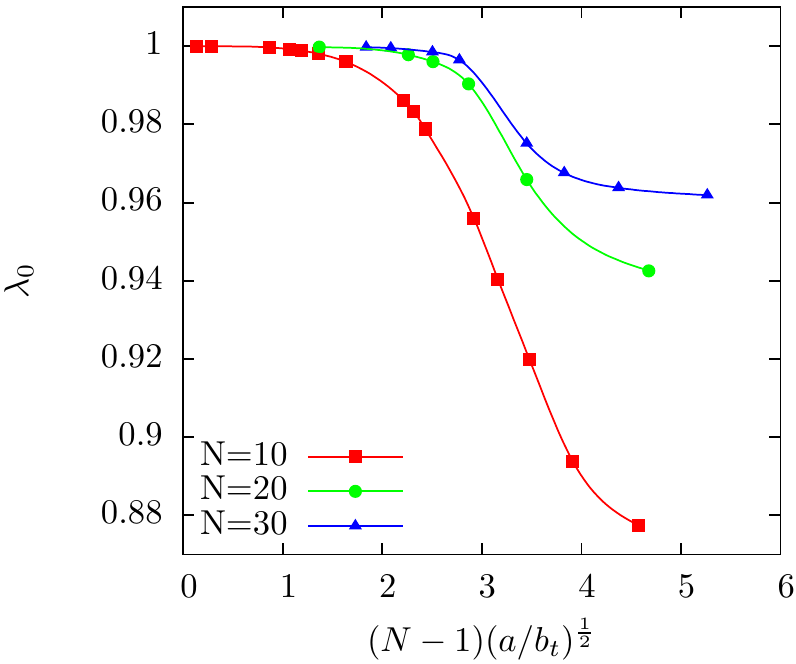}
  \caption{\label{trap-cf} The condensate fraction of a BEC-state of a system of $N$
    identical bosons in an isotropoci harmonic oscillator trap with
    attractive gaussian two-body potential
    as function of $(N-1)(a/b)^{1/2}$ where $b$ is the oscillator length. 
    From \protect\cite{tho07}.}
\end{figure}

Another very important question is the behavior of the condensate fraction as
a function of particle number. In Fig.~\ref{trap-cf} the condensate fraction
is shown as the (positive) scattering length is increased for 10, 20, and 30
bosons in a trap. We clearly see the depletion of the condensate as
the interaction is increased, and that this effect is more severe as the 
particle number decreases. This demonstrates the general trend that small 
condensates are very fragile when interactions, and in turn correlations, are
strong. Similar results can be obtained using harmonic approximation 
methods to the $N$-body problem in one, two and three 
dimensions at the Hamiltonian level \cite{kotur00,yan03,gaj06,armstrong2011,armstrong2012a,armstrong2012b,armstrong2012c,artem2013}
and via path integrals \cite{bro97-1,bro97-2,bro97-3,bro98-1,bro98-2,temp98,temp00,lem99}.
These facts should serve as a warning that systems with very low particle numbers
are hard to meaningfully describe in the same terms as condensates. Below
we will look at an example from nuclear physics and the use of $\alpha$-particles
as a degree of freedom in the nucleus.

\subsection{Nuclear $\alpha$-particle condensates}\label{fewbody4}
The idea of $\alpha$-particles as essential entities in the structure
of nuclei arises from the small $\alpha$-radius, the relatively large binding
and the spin saturation of both neutrons and protons. Attempts were
made in the early days of nuclear physics to construct nuclear
structure from $\alpha$-particles and valence nucleons
\cite{wig37,whe37,wef37}.  When more information became available these
cluster structures turned out to be less than robust as indicated by
the overlapping densities of the different $\alpha$-particles.  Thus
there are no compelling reasons for clusterization of nucleons into
$\alpha$-particles in the bulk at nuclear densities. A review on 
clustering in general and $\alpha$-clusters in particular can be found in 
\cite{freer2007}.

On the other hand, $\alpha$-cluster models are able to explain many
properties of specific light nuclei \cite{bri66}.  The most prominent
examples are the ground state of $^{8}$Be and the so-called
Hoyle-state in $^{12}$C, both $\alpha$-unstable $0^+$-states below all
other particle emission thresholds \cite{hoyle1954,cook1957}.  These
structures were accurately described many years ago \cite{hor70,har72,ueg77}
and later on confirmed in numerous theoretical works, see
e.g. \cite{kan98,des02,nef04}.  Based on calculations that are 
approximations to earlier 
results these states were recently interpreted as
a condensate of respectively two and three $\alpha$-particles
\cite{toh01,toh04,fun02,funaki2008,fun09,fun10}.

Since the $\alpha$-particles have spatial extension and internal structure,
the general condensate condition must be supplemented by the further 
requirements (i) that inert
$\alpha$-particle constituents must be present, and (ii) that the
$\alpha$-cluster wave function must be orthogonal to all other many-body
excited states or equivalently that the internal $\alpha$-particle
degrees of freedom decouple from relative degrees of freedom.

The quantality condition for an effective $\alpha-\alpha$ potential
can be estimated through a common parametrization in terms of an
attractive and a repulsive gaussian \cite{ali66}.  The minimum value
is $V_{0}\approx 5-8$~MeV (including the Coulomb energy of $\approx 2~
$MeV) for $c_{min}\approx 2.5-3.0$~fm and $\Lambda_{Mot} \approx
0.1-0.2$.  Thus the mean-field model is first choice for point-like
$\alpha$-particles but they have a strong tendency to localize.  Thus a
substantial condensate fraction can be expected for such idealized
point-like particles.

However, $\alpha$-particles have internal structure, finite extension,
and interact via both attractive short-range and repulsive long-range forces. They
have essentially to touch each other to form a resonance or bound
state because the Coulomb repulsion otherwise would blow the pieces
apart.  At such relatively small distances the intrinsic nucleonic
many-body degrees of freedom easily get excited. The restricted
Hilbert space describing condensates, i.e. mean-field for
$\alpha$-particles and all nucleonic structures frozen, gets
increasingly inadequate with increasing excitation energy, $E^*$,
simply because the many-body density of states increases
exponentially.  The $\alpha$-cluster structure is most likely to be
prominent at the $\alpha$-disintegration thresholds which increase
with nuclear excitation energy as $E^* \approx 7 (N_{\alpha}-2)$~MeV,
where $N_{\alpha}$ is the number of $\alpha$-particles in the
cluster. Thus only small $N_{\alpha}$ is possible.  Quantitative
estimates can be found in \cite{zin07}.

To be specific we illustrate with details from the most well-known
candidate, i.e. the Hoyle state of three $\alpha$-particles in
$^{12}$C which has about 90\% overlap with $\alpha$-particles in
relative $s$-waves around the center-of-mass
\cite{mat04,suz02,fun03,che07}. This corresponds to a one-body density 
eigenvalue of about $0.7$, computed by assuming the optimum
center-of-mass wave function \cite{mat04,suz02}.  At the same time
$\alpha$-cluster models show $\alpha$-particle density distributions
localized around specific points in space \cite{che07}.  Reconciling
these results, where apparently both localization and large condensate
fraction are present in the same wave function, is only possible with
large widths of the localized wave in Eq.~(\ref{e43}).  This
effectively recovers (most of) the independent particle wave function
in Eq.~(\ref{e33}).  Since the width is large the particles are
sufficiently separated to remove the need for nucleon
antisymmetrization and the $\alpha$-particles are present.

Thus, condition (i) is rather easily fulfilled provided the
wave functions are constrained to describe essentially non-overlapping
$\alpha$-clusters.  However, this structure is rarely an accurate
approximation as conditions (ii) only is expected to hold for $N_\alpha\leq 4$
for the condensate candidates. For low excitation energy, the
$\alpha$-particles would overlap and the nucleon degrees of freedom
wash out the $\alpha$-clusterization. At higher excitation energy, the
condensate wave function constructed in the limited Hilbert space,
would be smeared out over numerous other close-lying states. Collective
states are also very unlikely to display features of a condensate as 
the Coulomb interaction will mix these with other states.
The 
realization of condensates of $\alpha$ particles in nuclei therefore
face severe and likely insurmountable obstacles.

In relation to our discussion of trapped few-body Bose systems in Section \ref{fewbody3},
we note that for three or four bosons the depletion for strong
interaction is severe. The $\alpha$-particle studies use an effective 
$\alpha$-$\alpha$ interaction that is typically a parametrisation as in
\cite{ali66}. Small changes in this interaction can thus cause large 
deviations in the condensate fraction. Describing these $\alpha$-cluster
states in terms of properties applied to condensates can therefore 
quickly be very model-dependent, leaving the interpretation less useful.

The linear chain structures of $\alpha$-particles \cite{zin07} at the
break-up threshold (which can be organized in so-called Ikeda diagrams \cite{ike68}) 
are conceptually
similar to the one-dimensional atomic condensates called
Tonks-Giradeau structures \cite{par04,kin04}.  Such nuclear states have been
searched for and for many years the Hoyle state in $^{12}$C was the
favorite candidate. However, the linear chain suffers from precisely the same
difficulties as the $\alpha$-condensate discussed above \cite{zin07}.

Discussions about possible condensate structure are not constructive
without an experimentally observable distinction from other
structures. First, all one-body properties are excluded as signals.
Condensates are diagnosed through coherence properties of the
wave function, not by density distributions. This immediately implies
that one-body observables like elastic scattering cross sections used
in \cite{che07} cannot be used to distinguish between condensates and
other structures.  Second, the experiments almost inevitably at some
point detect large-distance properties which in the context of
coherence is related to the long-range order parameter. For
$\alpha$-particles no product structure at large distance is possible
due to the Coulomb repulsion.  To see the short-distance bulk
structure responsible for the large one-body density eigenvalue
require detection of properties of structures differing by one or more
particles \cite{gaj06}.  The results should remain unchanged except
for the normalization proportional to particle number.  
Based on these observations, it is unclear what
new physical consequences arise from approximating the fully
correlated cluster states (as in \cite{ueg77}) with less accurate 
$\alpha$-particle condensate
states as  no observable are available that can distinguish between
the two. In any case one must also remember that all cluster states 
are approximations to the real nuclear states.

\section{Efimov Physics}\label{efimov}
\begin{figure}[htb!]
\centering
\epsfig{file=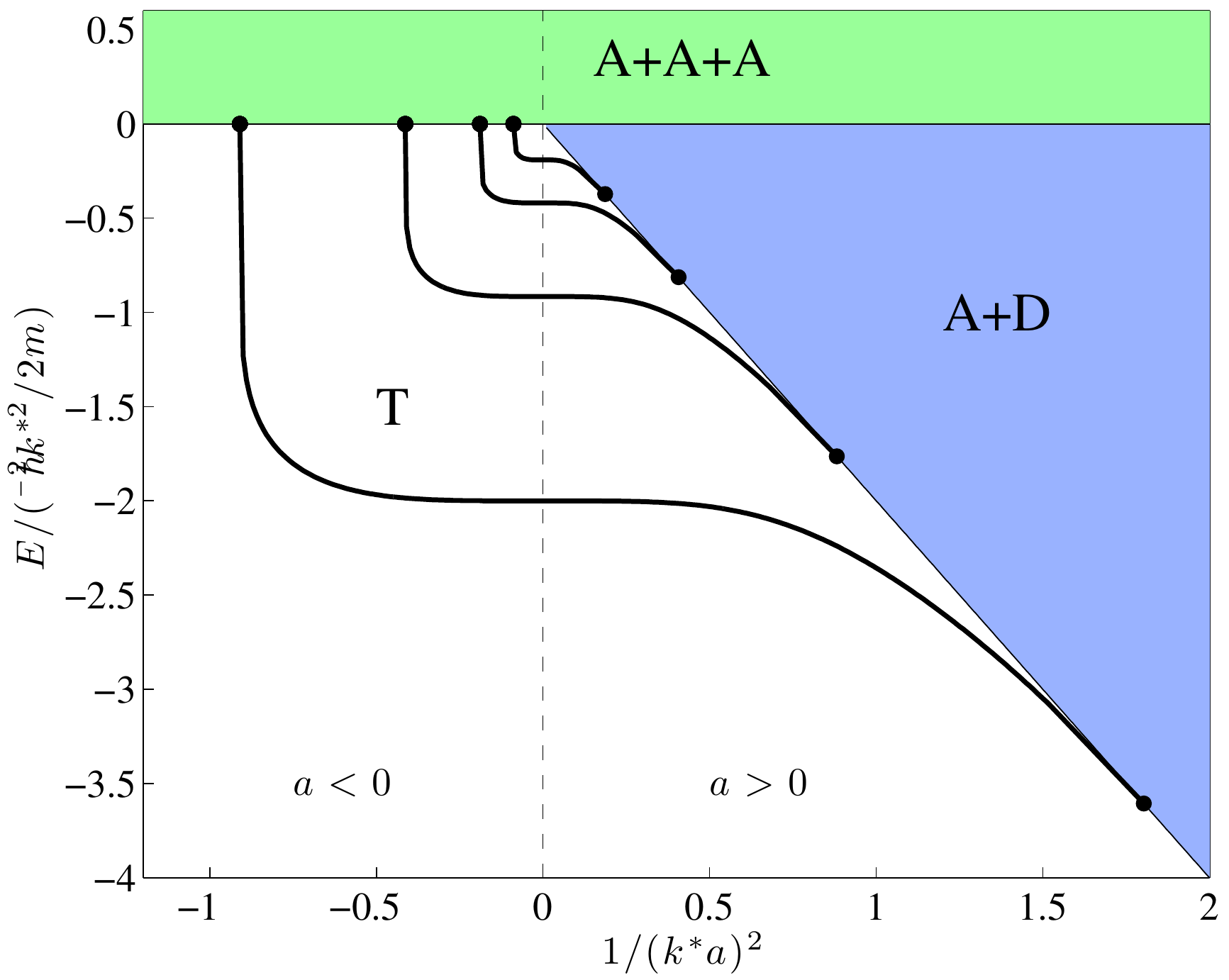,clip=true,scale=0.5}
\caption{\label{fig-efimov} Energy of Efimov States as function of the scattering length. $k^*$ is 
the background scale fixing the Efimov spectrum as explained in the text. On the 
$a<0$ side there are no bound dimers and the trimer states ($T$) terminate in the three-particle
continuum ($A+A+A$), whereas on the $a>0$ side the states terminate at the atom-dimer continuum ($A+D$).
Both axis have been scaled by a power of 1/8 in order to make the similiarities in the spectrum clear.}
\end{figure}

About forty years ago the Efimov effect was suggested as an anomaly
appearing in a three-body system when at least two of the two-body
subsystems simultaneously have short-ranged bound states at zero
energy \cite{efimov1970a,efimov1970b}. 
The three-body system then has infinitely many
bound states with exponentially decreasing binding energies and
correspondingly increasing mean square radii. In figure~\ref{fig-efimov}
we present an example of the three-body spectrum considered by Efimov.
The properties of these
states are independent of the potentials supporting the two-body
states at zero binding energy.  The only requirements are that three
dimensions are available, the potentials are of short-range and
sufficiently attractive to provide borderline binding. We shall in
this section give the simple arguments with corresponding derivations,
and generalize the concept of model independence to a broader class of
phenomena called Efimov physics.  We shall discuss possible occurrence
in nuclear, molecular or atomic systems, and the recent experimental
verification in cold atoms using the technique of Feshbach resonances.

\subsection{Basic derivation and occurrence conditions}\label{efimov1}
Convenient coordinates for three particles are the hyperradius $\rho$
and the five hyperangles $\Omega$. It suffice here to define $\rho$ by
\begin{equation} \label{e333}
m \rho^2 = \frac{1}{m_i+m_k+m_j}  \sum_{i<j} m_i m_j
({\bf r}_{i} -  {\bf r}_{j})^2
\end{equation}
in terms of an arbitrary normalization mass $m$, the three masses
$m_k$, $k=1,2,3$, and the three particle coordinates ${\bf
  r}_{k}$. The pertinent properties of the Schr\"{o}dinger or Faddeev
equations in the present context can be extracted from the lowest effective
(hyper)radial equation \cite{nie01}
\begin{eqnarray} \label{e337}
\left[ -\frac{d^2}{d\rho^2} - \frac{\xi^2 + 1/4}{\rho^2}  + \kappa^2
  \right] f(\rho) = 0 \; ,
\end{eqnarray}
where $\kappa^2 = - 2m E /\hbar^2$, and $\xi$ is a dimensionless
function of $\rho$ determined from the interactions and the angular
equations.  Outside the ranges of the interactions the function $\xi$
turns out to be independent of $\rho$, strongly mass ($m_k$) dependent
\cite{jen03}, and $\xi$ is a real number 
provided two or all three subsystems have
bound states at zero energy.

The solution for the radial wavefunction is then $f \propto
\sqrt{\rho} K_{i\xi}(\kappa\rho)$ where $K_{i\xi}$ is the modified
Bessel function of the second kind.  For small $\kappa$ we get $f
\propto \sqrt{\rho} \sin(\xi\ln(\rho/\rho_0)$ where $\rho_0$ is a
constant defining the arbitrary length scale chosen from
eq.(\ref{e337}) when $E \rightarrow 0$. Then $\rho_0$ is the first
node of the wavefunction.  This approximation for $f$ could be found
directly by testing the solution $f \propto \rho^{1/2 \pm i\xi}$ in
eq.(\ref{e337}) for small $\kappa$.  For large $\rho$ the wavefunction must fall off
exponentially for a bound state, i.e.  $f \propto \exp(-\kappa \rho)$.
The bound state energies are found by the condition
$K_{i\xi}(\kappa\rho_0)=0$, that is from the nodes of the Bessel
function $K_{i\xi}(z_n)=0$ with $\kappa_n=z_n/\rho_0$. For small
$\kappa$ this implies $z_n=z_0\exp(-n\pi/\xi)$ or
\begin{equation} \label{e339}
 E_n = E_0\exp(-2n\pi/\xi) \;\;,\;\; \langle\rho^2\rangle_n =
 \frac{\hbar^2(1+\xi^2)}{3m|E_n|} \;.
\end{equation}
These scaling relations clearly only depend on the value of $\xi$ and
an energy or length scale $E_0$ or $\rho_0$ which are determined by
the finite-range potentials and the boundary conditions at small
distance.  If the potentials are strongly attractive a number of bound
three-body states may appear at energies lower than the scaled
sequence in eq.(\ref{e339}).

This scaling limit is reached when the scattering lengths diverges to
infinity (bound state at zero energy) in comparison with the ranges of
the potential.  The sequence is terminated when the spatial extension
of the state becomes comparable to an average of the scattering
lengths.  Thus we find the Efimov sequence for large scattering
length, as well as for potential ranges approaching zero which is the
Thomas collapse \cite{thomas1935}.  Both effects are therefore described by these
expressions.

The crucial property is the inverse square behavior of the effective
potential in eq.(\ref{e337}) in (at least) a region of space. This
form arises at distances between the effective range $R_e$ of the
two-body potentials and their scattering length $a$.  All Efimov
states are located in this region of space.  By counting the number of
nodes of $f$ in this interval we find the number of bound states to be
$N_E \approx \xi/\pi \ln(a_e/R_e)$.  If the strength of the
$1/\rho^2$ potential is $-1/4$ only an infinitesimal negative or
positive addition would produce either infinitely many bound states or
none at all.

\begin{figure}[ht]
\centering
  \epsfig{file=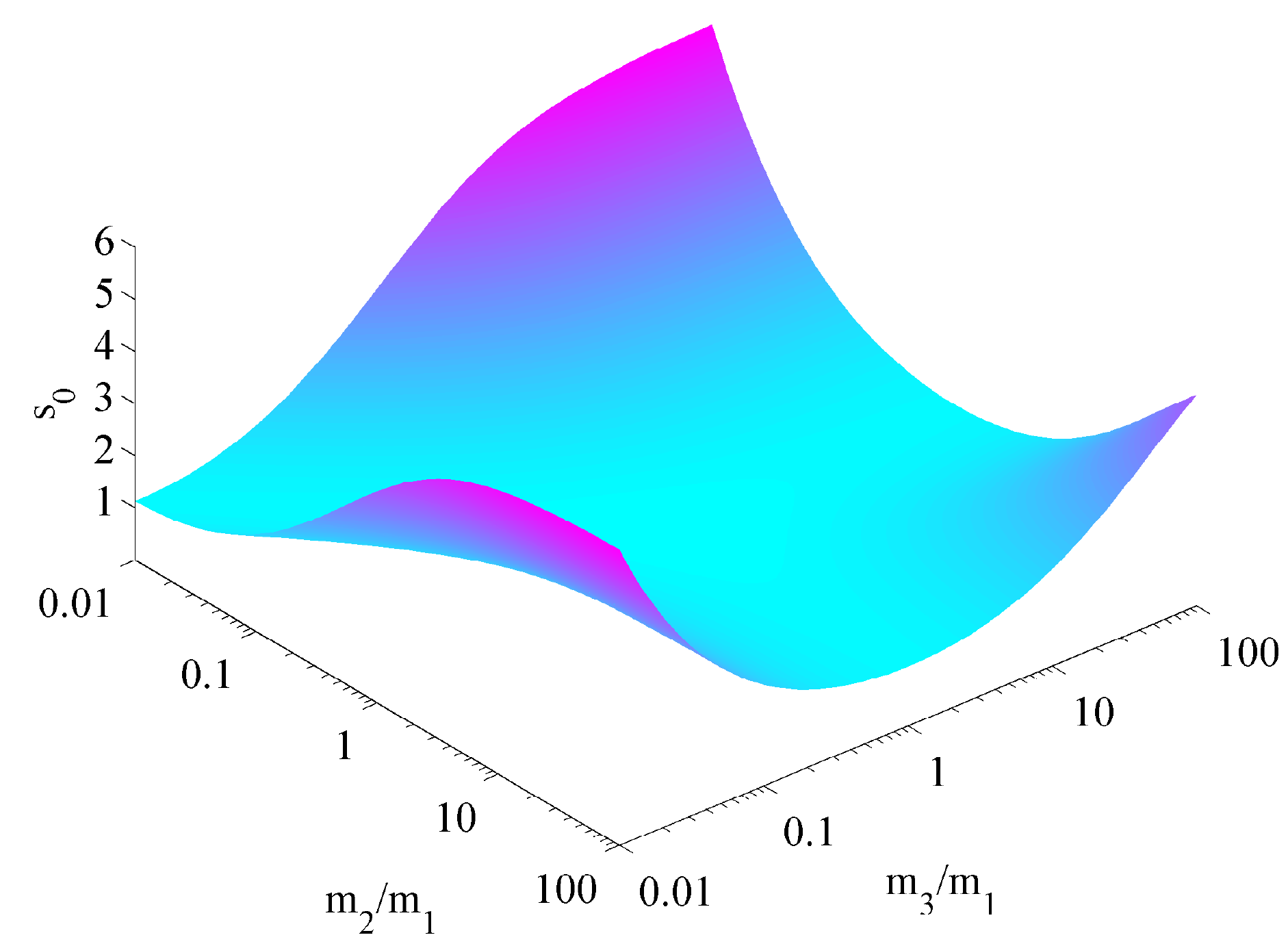,clip=true,scale=0.6}
  \caption{\label{scaling} Surface plot of the scaling factor, $s_0=\xi$, as function of the two mass
  ratios, $m_2/m_1$ and $m_3/m_1$. The mass ratios are on 
  a logarithmic scale.}
\end{figure}

The decisive parameter is then the strength $\xi$ which determines the
character of the potential, that is infinitely many for real $\xi$ or
no bound state for imaginary $\xi$.  The eigenvalue of the
hyper-angular equation determines $\xi$. For identical bosons the
equation is \cite{jen03}
\begin{equation} \label{e441}
 8 \sinh(\xi\pi/6) = \xi \sqrt{3} \cosh(\xi\pi/2) \;.
\end{equation}
with the solution $\xi= 1.00624$. For non-identical bosons where all
scattering lengths still are large the equation is \cite{jen97}
\begin{equation} \label{e443}
 \left(\frac{\xi\cosh(\xi\pi/2)}{2F}\right)^3 - \left(\frac{\xi\cosh(\xi\pi/2)}
{2F}\right)\frac{(f_{1}^{2}+f_{2}^{2}+f_{3}^{2})}{F^{2}} =2 \;,
\end{equation}
where $F=(f_1f_2f_3)^{1/3}$ and
\begin{eqnarray} \label{e445}
f_k&=&\frac{\sinh(\xi(\pi/2-\varphi_k))}{\sin(2\varphi_k)} \\
\varphi_k&=& \arctan\left(\sqrt{m_k(m_i+m_j+m_k)/(m_im_j)}\right) \;.
\end{eqnarray}
The limit of identical bosons is obtained from eq.(\ref{e443}) for
$\varphi_k=\pi/3$.  When only the two scattering lengths between
particle pairs $(j,k)$ and $(i,k)$ are large the $\xi$ equation becomes
\cite{jen03}
\begin{equation} \label{e447}
\xi\cosh(\xi\pi/2) \sin(2\varphi_k) = 2 \sinh(\xi(\pi/2-\varphi_k) \;,
\end{equation}
where $\varphi_k$ is now the only mass dependent parameter. The solution
for $s_0=\xi$ as function of the two mass ratios, $m_2/m_1$ and $m_3/m_1$, 
is shown in figure \ref{scaling}.

The variation of $\xi$ with the mass ratios cover all values from zero
to infinity. Small values imply large distances between the Efimov
states and for realistic finite scattering lengths there is only room
for very few with smaller binding. For identical bosons the
ratio of radii between two neigboring states is $22.7$, that is to
allow three states the scattering length must be about $10.000$ times
larger than the range of the potential.

On the other hand large values of $\xi$ is also possible and many
Efimov states can fit into the allowed interval. This is achieved when
two masses are much larger than the third. The opposite situation with
two light and one heavy mass is the least favorable to find Efimov
states.  Since at least two scattering lengths must be large it is
much simpler to use two identical particles where any required tuning
automatically is done simultaneously for both particles.  However,
this may also reduce the contributions to only these two subsystems
where eq.(\ref{e447}) applies instead of eq.(\ref{e441}). If we
maintain equal masses we find the somewhat less favorable value of
$\xi=0.499$ instead of $\xi=1.00624$.  However, if the mass of
particle $k$ is small compared to the masses of the other particles
the value of $\xi$ can be very large. The mass variation is much more
efficient than increasing from two to three contributing subsystems.

The quantum statistics due to identical particles does not change
anything for bosons. In contrast, none of the derivations are valid
for three identical fermions where the antisymmetry requires either
different $s$-states or at least one $p$-state. For two identical
fermions precisely the same formulations apply since each spatial
state can hold at least two fermions by use of 
different spin projections. The antisymmetry is taken care
of by the spin degree of freedom which effectively is absent for
spin-independent interactions.

Unfortunately the jargon in present days cold atom physics can easily
cause some confusion as we have discuss above.  
The single-particle states are hyperfine
states of given angular momentum projection and a set of other quantum 
numbers related to the trap or momentum vector for a homogeneous system.
The spin projection
is then replaced by the hyperfine quantum numbers.
Thus, since antisymmetry still is required,
$s$-wave contact interactions are forbidden between particles 
in the same hyperfine state. 
Since two
subsystems in $s$-waves are necessary with large scattering lengths 
we must have at least two different hyperfine states. Thus fermions
in one of these states and the third particle (boson or fermion)
in another state leave one $p$-wave interaction and two possible
$s$-wave contact interactions. The total angular momentum is then 
at least 1 unit of $\hbar$. It is of course also possible to have
identical fermions in three different hyperfine states. This allows 
all pairs to interact by $s$-wave contact interactions \cite{wil09,wen09,naka11}.

The question then arises if Efimov states exist for finite three-body angular
momentum $L>0$. The answer is yes \cite{nie01} but with
conditions of sufficiently small mass, $m_k$, for the third $s$-wave interacting particle. 
The
critical mass depends on $L$, which is the same as the orbital angular
momentum of the two non-$s$ wave interacting particles. For $L=1$ the
result was given as
\begin{equation} \label{e449}
 m_k < 0.5 (-m_i -m_j + \sqrt{(m_i+m_j)^2 + 0.60 m_i m_j} \;.
\end{equation}
Again only mass ratios enter the condition. This result is valid
independent of the boson or fermion character of the particles.  In
particular, with all masses equal the requirement cannot be fulfilled
since $1> 0.5(-2+\sqrt{4.6}$. Thus equal masses cannot produce $L=1$
Efimov states. Other studies have addressed the case of three identical
fermions which require higher-partial wave interactions to accomodate the necessary
antisymmetry \cite{suno2003,macek2006,jona2008}. We note that very recent
studies indicate that $p$-wave interactions produce unphysical negative
probabalities at the two-body level and thus also rule out an Efimov effect 
with $p$-wave interactions \cite{nishida2012,braaten2012}.

As a curiosum we mention that the formulations of occurrence
conditions can be investigated in any dimension $d$. The mathematical
equations only leads to potentials of form eq.(\ref{e337}) and real
values of $\xi$ when $2.3 < d < 3.8$. Thus, $d=3$ is the only integer
dimension to exhibit the Efimov effect, and for that matter also the Thomas
effect. However, recent investigations softened these conditions by allowing
different participating particles to move in different dimensions, e.g. 
two particles confined to $d=2$ and a third in $d=3$ (see \cite{nishida2011} 
for a recent review of such possibilities).

In a strict two dimensional setup we do not have the Efimov anomaly with 
infinitely many bound states around threshold but the spectrum of three-body 
states in the zero-range limit still has some rather peculiar features \cite{tjon79,nie97,nie99,borro2012}. 
In fact, for a system of three identical bosons, 
one finds exactly two bound three-body states, a ground state with energy $E_3=16.52E_2$ 
and an excited state with energy $E_3=1.27E_2$. Here $E_2<0$ is the energy
of the two-body bound state of the zero-range potentials which always 
exists \cite{lan81}. Bosonic few-body states in two-dimensional 
setups with three \cite{adhi88,adhi93,brod06,kart06} and four particles \cite{brod06,plat04}, as well 
as larger cluster have been studied \cite{ham04,blume05,lee06}. In addition, the 
scattering \cite{brod06} and recombination \cite{hel11} has also been
discussed. 
Recently, it explored how the number of bound states changes
in system of two identical particles and a third distinct 
particle and for three distinct particles \cite{bellotti2011,bellotti2012a,bellotti2013}.

\subsection{Scaling properties and examples of Efimov physics}\label{efimov2}
The almost complete absence of specific potential dependence of the
derived results strongly suggests that all observables are connected
by universal relations.  Such model independent results are obviously
valuable as predictions virtually without any input, apart from
validity conditions, and as reference units to evaluate similarities
and differences.  A prominent and relevant example in the present
context is early nucleon-nucleon scattering experiments where
disparate potentials were able to reproduce the measured low-energy
cross sections. The explanation was given by the effective range
expansion theory where the phase shift $\delta$ was expanded as function of
energy, i.e.
\begin{equation} \label{e551}
k \cot \delta(k)  = - \frac{1}{a} + \frac{1}{2} k^2 R(k) \;,
\end{equation}
and $E=\hbar^2k^2/(2\mu)$. The scattering length, $a$, is a constant
and $R$ approaches a constant, the effective range, $r_{e0}$, for small $k$.
Since $a$ is a quantity depending on the entire potential similar to a
weighted integral, the same low-energy cross section proportional to
$a^2$ can be found from tremendously different potentials.  Thus, all
low-energy observables can be expressed by $a$ without reference to
specific potentials. It should be emphasized that the scattering
length is periodic as function of potential strength. The same
$a$-dependence emerges independent of the number of deeper lying bound
states.

A related result is the mean square radius of a weakly bound two-body
state, i.e. $<r^2> = \hbar^2/(4 \mu|E|)$, which only depends on the
energy and the reduced mass $\mu$.  
This radius-energy relation suggests a similar correlation
for three-body systems.  The mean square radius is now uniquely
defined by $<\rho^2> = N <r^2> $ but now far from being a uniques
function of the three-body energy.  First of all the quantities are
not dimensionless but this is a problem already for the two-body
system.

The first attempt towards universal curves is to use an appropriate
length $R$ related to the range of the potentials and consider
$<(r/R)^2> = \hbar^2/(4 \mu|E|R^2)$. Then all information about both
radii and energies are maintained and the scale of the system does not
enter.  For short-range potentials a square well radius equivalent to
the range of the true potential would be tempting. Since the spatial
extension usually is measured by the root mean square value we could
choose identical second moments of the potentials as criteron which
for a gaussian of range $b$ implies that $R=b\sqrt{5/2} $.

For three-body systems we still have to choose the scale $\rho_0$ (see
below for some recent developments regarding this cut-off).
One simple option would be to use eq.(\ref{e333}) with
$\rho\rightarrow \rho_0$ and the two-body square well equivalent radii
substituted on the right hand side. The curves of $<(\rho/\rho_0)^2)>$
as function $\mu|E|\rho_0^2/\hbar^2$ is now closer to being universal.
However, two different universal scaling functions appear \cite{jen04,yama2011}.  One
corresponds to the Efimov states with radii and energies related
through eq.(\ref{e339}) either by varying $\xi$ or for fixed $\xi$
following the sequence of excited states. This is for large $|a|$. 
The other universal function
corresponds to size and energy of Borromean states approaching the
threshold for binding which occurs for moderate values of $|a|$.

For non-identical particles the individual two-body interactions can
be varied independently to approach different thresholds for binding.
Completely different behaviors, although still universal, are then
possible for example transitions to two-body scaling behavior must appear in
the limit towards a bound two-body subsystem.

Another established universal curve is the Phillips line, i.e.  the
three-body binding (triton) energy as function of the scattering
length of the particle-dimer (nucleon-deuteron) system. This
universality can be shown to originate from small energies of all
involved subsystems in comparison to the nuclear potentials.
The advantage of both radius and energy information can be traded for
a proper universal curve.  The ratio of two consecutive three-body
binding energies ($|E_{n+1}|/|E_n|$) as function of relative two to
three-body energies, $B_2/|E_n|$, was suggested in \cite{fre99}.
Measured points on such a curve are then signals of structure similar 
to the Efimov states.

The Efimov effect is concerned with the anomaly found for three
particles when two-body subsystems are at the threshold of binding.
The results are universal or model independent by depending only on
the masses.  This window of universality is only open
when the scattering length, $|a|$, is large compared to the effective range of the
interaction.  It is then interesting to extend to other systems with a
similar model independent description.  The broader class of phenomena
is naturally named Efimov physics which could be defined as physics
where Universality and Scale Invariance apply.  To be specific
Universality then means that one global parameter determines all
properties of the $N$-body system.  Scale Invariance means that the
same properties appear for any length scale whether it is in nuclei,
atoms or molecules.

These concepts are easily misinterpreted to interchange meaning,
i.e. scale invariance is taken to mean that the next state in a Efimov
sequence of a given system has the same property as the previous
state, and universality is taken to mean that different systems can be
described with the same theory. We shall use our first definitions
where universality means model independence and scale invariance means
that the theory applies to any length and energy scale.  The two
ingredients begin to be similar when dimensionless quantities are used
to express the physics observables. The scale dependence disappears
and universal features are isolated.

The window of universality is for three particles between the two-body
effective range and the scattering length. In terms of the 
hyperradius we have $r_e<\rho<a$. Viewed from the adiabatic
hyper-radial potential the states are in the attractive region with
positive hyper-radial kinetic energy. However, the size of the system
is large and the two-body subsystems are quickly found far outside the
range of their attraction. This means that classically forbidden
regions are occupied with negative two-body kinetic energy. The origin
of the model-independent features are then simply understood as
properties determined outside the region of non-zero potential where the
kinetic energy term decides the behavior. Due to quantum mechanics,
the boundary conditions still link to the wavefunction behavior at
small disance, but this is then the entire dependence describable by
the model-independent scattering length parameter.
Successful descriptions require spatially extended cluster systems 
which is difficult to achieve for self-bound systems with short-range 
interactions. Only $s$- and $p$-wave systems allow this. The other
direction of more detailed information require short-distance 
properties which essentially excludes zero-range models even when
effective range improvements are employed.

A number of systems are already found to exhibit universal
features. The four-body energy ($\alpha$-particle) is a unique
function (Tjon line) of the three-body energy, i.e.  the three and
four-body energies are always found on the same curve \cite{tjon75}.  
This result
emerge from zero-range models which must be regularized to avoid
Thomas collapse. It was shown in \cite{fed01} that stabilization of
the three-body system automatically ensures finite results for the
$N$-body system.  However, this does not immediately imply that all
$N$-body systems are determined from two and three-body properties.
This is still an open problem \cite{yam10}.

Universal behavior of excited four-boson states was recently found in both
variational \cite{ste09} and zero-range models \cite{ham07}.  Each
three-body state has two related four-body states at larger binding
energy than this three-body state \cite{han06,ste09}.  They have too
little energy to decay into this three-body state.  In contrast, it
was found in \cite{yam06} that the four-body state depends on
interaction details.

The concept of halo nuclei has been rudimentary extended to more than
three particles.  In general fully model independent nuclear $N$-body
structures for $N>3$ cannot be expected since all clusters are charged
and either confined to small distances or repel each other by the
Coulomb interaction.  The line of arguments begins with assumptions of
only $s$-waves and no correlations which prohibits halo existence
\cite{riisager2000,jen04}.  If correlations develop the $N$-body system tends
to form substructures and effective reduction of the number of
clusters.  It is therefore interesting that \cite{yam10} found that
four and five identical bosons converge to a radius larger than the
interaction range as the threshold of binding is approached.
These weakly bound states have universal structures. More generally, 
it may be possible to use Efimov-type structures for $N=3$ to 
build general so-called higher-order Brunnian states \cite{baas2012}.

Another class of universal Efimov-like structures was found in
\cite{sor02,tho08,stecher09} for $N$-body systems with the assumption of only
two-body correlations in the variational wavefunction.  This does not
contradict the conclusion in \cite{ama71a,ama71b} that the Efimov effect does
not exist for $N>3$.  This result is more specifically stating that
the threshold for binding the $N$-body system is not an accumulation
point for infinitely many states.

For dilute systems two-body correlations are expected to be
dominating.  The sequence of excited states found between the
interaction range and the scattering length obey the Efimov scaling
equations in eq.(\ref{e339}).  Whether these states (approximately)
remain after extension of the Hilbert space to allow all correlations
still needs to be tested.  Where this conservation of identity is most
likely to be preserved, for large scattering length or close to
threshold of binding the $N$-body system, is unclear at the moment.
Ground and lowest excited states may be outside the universal region
and even higher excited states are the only hope for finding universal 
behavior.  The scaling
properties are expected to differ in these regions, as e.g. Borromean
three-body states scale differently from Efimov states. In any case,
Coulomb interactions between the particles strongly suppress the
possibilities, see Section \ref{fewbody}.

\subsection{Measurable consequences in physics systems}\label{efimov3}
In nuclei the clusters are all charged except for systems with
neutrons surrounding a core. The Coulomb interaction is long-ranged
and prevent the Efimov effect as it reaches to larger distances than
the $1/\rho^2$ potential. In the pure form we are left with two
neutrons and a core which again is unfavorable since the core
necessarily is heavier than the neutrons.  The well known case is
$^{11}$Li with approximately three scattering lengths each of about
$20$~fm which is about $8$ times the core radius. The $\xi$ value is
$0.074$ and the scale factor on energy and square radius is about
$3\cdot 10^{18}$. One Efimov state could then by chance be seen in the
interval, but not two, and the scaling can thus not be tested.

Realistic two-body interactions for the two neutron-$^{9}$Li system
produce one bound state which moves down to about $1$~MeV below
threshold when the neutron-$^{9}$Li interaction is tuned to have a
bound state at zero energy. However, the many excited Efimov states
can not be seen since already the first is located at an enormous
distance outside the radius of the nucleus \cite{fed94}.  This
calculation assumes that $^{9}$Li has angular momentum zero and
therefore precisely the same interaction with both neutrons. For a
non-zero $^{9}$Li-spin the hyperfine splitting of the strong
interaction prevents simultaneous large scattering lengths for both
neutrons. Thus the Efimov conditions are therefore most easily met in
nuclei for two neutrons and an even-even core-nucleus with zero spin.

It is still possible to see reminiscenses of the Efimov effect in
nuclear decays and possibly in nuclear scattering \cite{gar06}.
Assume for example that $^{11}$Li is excited to a $1^-$ state above
threshold for two-neutron emission.  The momentum distributions of the
decay products, two neutrons and $^{9}$Li, can be measured
simultaneously. Then characteristic peaks should appear in the
probability for emission of both low and high energy neutrons. This
corresponds to one neutron correlated with $^{9}$Li at high energy and
the other then emerges with very low energy. The energy distribution
for $^{9}$Li has a peak at high energy corresponding to correlated
emission of the two neutrons and a peak at intermediate energy
originating from correlated emission of the neutron-$^{9}$Li system.
These structures are found experimentally \cite{nak09}.

An example of an unexpected consequence is in dense plasma of
$\alpha$-particles and electrons \cite{jen95}. The electron screening
reduces the Coulomb repulsion between the $\alpha$-particles, and in
particular the long-range Coulomb tail is removed. This has
first two interesting consequences, i.e. the energy of the $^{8}$Be
nucleus decreases towards zero and the Coulomb tail decreases between
the $\alpha$-particles.  This in turn implies that the Efimov
conditions are approached, and the corresponding three-body system,
$^{12}$C, should have decreasing energies and if the screening is
sufficiently strong more states should appear at the threshold.  This
is the triple $\alpha$ process proceeding via the Hoyle state. The
rate could easily be dramatically changed for high plasma density
and temperature.

An even more speculative example is the suggestion of recombination of
three deuterons where two deuterons are bound in a molecule in a dense
lattice \cite{eng06}.  The third deuteron is injected and,
catalysis-like, via three-body long-range interactions cause fusion of
the molecular deuterons and expulsion of the injected deuteron.  The
energy gain is from formation of the $\alpha$-particle.

The possibilities are enhanced by use of asymmetric systems with two
heavy particles. Three-body combinations of $^{3}$He-atoms and two alkali
atoms could potentially increase the $\xi$-values substantially.
As the scaling is exponential a number of Efimov states could appear
with energy ratios of only $5-10$.  The mass ratio can be made much
more extreme by using an electron as one of the particles and atoms or
molecules for the other two \cite{jen03}.  Then the energy and radius
ratio of consecutive states could be very much more favorable by differing only a few
percent.

Efimov scaling related to $1/r^2$ potentials appear obviously already for a charged particle in a
sufficiently strong dipole field. However, coupling to rotational
degrees of freedom would quickly destroy the simple properties. It was
recently suggested that quantum dots and artificial atoms could
prevent this rotational coupling and thus maintain the Efimov sequence
\cite{sch10}.

Within the last five years a number of papers reported on experimental
results obtained for cold atoms in traps and with Feshbach resonance
techniques \cite{kra06,ott08,zac09,gro09,pol09,bar09,kno09,huc09}.  
The systems always consist of many particles
and the Efimov states are only indirectly observed as enhanced (or
depleted) probabilities for three-body decay of the trapped
particles. The two-body effective interaction is varied through the
magnetic field to pass values where the three-body system has zero
energy. The decay into a deep
dimer and a high energy third particle is enhanced since the relative
and total energy of all three particles in the Efimov state is
identical to their energies in the ultra cold trap. The coupling
between these states is maximized and there is a transition through the
Efimov state at threshold into deeper-lying dimer states accompanied
by the third particle with the surplus energy to ensure energy
conservation.

For very attractive two-body interactions that have positive scattering 
lengths ($a>0$) and where a high-lying or shallow so-called Feshbach dimer
state is present, a somewhat similar process also is enhanced.
Two states have the same energy, that is the well-bound three-body
Efimov state and a bound dimer with zero binding of the third atom.
From an atom-molecular cold gas the coupling between these states is
largest and decay through this Efimov state into deeper-lying dimers
is maximized.  Periodic loss rate minima were
predicted \cite{fedi1996,esr99,nielsen1999,nielsen2002} and recently also
observed for two neighboring structures \cite{zac09}.  These
variations should appear periodically with sizes scaled by the 
Efimov scaling factor of $22.7$ for equal mass bosons.
This three-body recombination process is discussed in details in
\cite{esr07} and subsequently applied in analysis of the experiments.
A number of these features are now confirmed \cite{pol09,zac09,gro09},
including the scale factor. We note that it is now also possible
to study Efimov states in three-component $^{6}$Li gases via
radio-frequency association \cite{lompe2010,naka11}.

The experimental study using $^{39}$K presented in \cite{zac09} reported
some deviations from universality which could potentially come
from finite-range corrections beyond the universal theories that
only take the scattering length into account. Some recent theoretical
works have addressed the corrections coming from the finite range
of inter-atomic potentials in the coordinate-space 
formalism \cite{thoger2008,thoger2009,wang2011,sorensen2012} 
and using momentum-space effective field 
theory \cite{pietro2008,platter2009,ji2010,ueda11}. In particular,
the study of \cite{pietro2008} provide results that are close to the experimental 
data. On the nuclear physics side, finite-range corrections for 
three-nucleon systems have been considered by Efimov himself about
two decades ago \cite{efimov1991}.

\subsection{New directions}\label{efimov4}
Investigations of different masses, dimensionality, 
and quantum statistics are also
producing results both for two \cite{bar09} and three-component
systems \cite{wil09,wen09,naka11}.  Now more than one three-body system can
be tuned to fulfill the Efimov conditions.  Two components can form
light-light-heavy and light-heavy-heavy combinations with very
different scaling properties as described above.
For three different particles the pair interactions involve three
different scattering lengths and a correpondingly more complicated
analysis. To be in the universal window at least two scattering
lengths must be large compared to the effective ranges of the
interactions.

Extension to four-body universal states and corresponding
recombination via four-body universal structures has been predicted
\cite{ham07,ste09} and now also experimentally observed for cold
gasses \cite{pol09,fer09}. The general structures of Brunnian systems
(no bound subsystems) can be anticipated to be universal provided the
states are located outside the effective range. Such investigations
are only barely conceived at the moment in cold gases.  In nuclei such
systems exist but are too spatially confined to be truly universal
\cite{cur08,muta11}. One of the latest experiments and corresponding 
theoretical work even seem to suggest that five-body features could 
be accessible in atomic gases \cite{zene2012}.

A particularly surprising recent development has been the realization 
that the three-body parameter seems to be universal when expressed in 
units of the two-body interaction scale of the inter-atomic potential,
which is the van der Waals length, $r_\textrm{vdW}$. The three-body
parameter was denoted $\rho_0$ in the coordinate-space hyperspherical 
approach discussed above but is also often quoted as a momentum-space
cut-off (typically denoted $\Lambda^*$ or similar). Since $\rho_0$ determines
the lowest-lying Efimov state, the experimental findings can be written
in terms of the threshold for the lowest Efimov state to appear out of 
the three-atom continuum on the $a<0$ side of a Feshbach resonance. We
denote this threshold $a_-$. 
The surprising result is that measurements on different atoms and using
different Feshbach resonances yield $a_-/r_\textrm{vdW}\sim -9.1$ to 
within about 15\% accuracy \cite{ber2011,wild2012,knoop2012}. 
This implies that there is some generic
universality even in the three-body parameter which cannot be captured 
by simple zero-range models that need the $\rho_0$ supplied from elsewhere.
A number of theoretical works have presented various models that explain
the observations \cite{ueda11,chin11,wang12,schmidt12,yujun2012} and it seems 
clear that the two-body inter-atomic
potential is the culprit since it has a large repulsion at short distance
which will naturally provide a three-body cut-off \cite{peder12,peder13}. 
However, there is still a question of how the number of bound two-body
states in the inter-atomic potential influences $a_-$ \cite{wang12,peder12}.

In the case of nuclear physics, it seems clear that the dense environment and 
the complicated nuclear interaction should not allow such an easy 
relation. For the sake of argument, imagine that we find an Efimov state in 
some nuclear system with a corresponding $a_-/r_0$, with $r_0$ the nuclear
interaction range. If this number is of order -10 or so, then this implies that
the nuclear system is universal since $a_-$ is so much larger than $r_0$. However,
as we have discussed at length in the previous sections, this is very unlikely
except for the case of pure neutron matter. Unfortunately, as pointed out 
above, the three neutron system is unbound.

Another interesting aspect in Efimov physics of recent times is the potential
for Efimov states in the presence of long-range dipolar interactions. For 
fixed external polarization of the dipole moments (via applied magnetic or electric
fields), it has been shown that the Efimov effect does occur for bosonic 
particles \cite{wang2012a} while for fermionic atoms universal three-body 
states are possible \cite{wang2012b}. This is interesting in comparison to 
nuclear physics (and condensed-matter physics) since long-range interactions
like Coulomb are generically present. One could thus hope that the analogy 
can be used to learn something about nuclei. These dipolar system have
also been studied in two-dimensional 
geometries \cite{wang2006,wang2007,jeremy2010,volosniev2011a,zinner2012} and one-dimensional 
tubes \cite{klawunn2010,wunsch2011a,wunsch2011b,knap2012,artem2013}, where few-body 
bound states are even more prolific \cite{jeremy2012,vol2012}. 

Efimov bound states also play a role in the recent exploration of the 
unitarity limit, where the so-called Tan relations can be used to 
deduce properties of bulk many-body systems from basic knowledge of 
few-body quantities 
\cite{tan08a,tan08b,tan08c,braaten2008,combescot2009,barth2011,pricoupenko2011,manuel2011,langmack2012,valiente2012,hofmann2012,manuel2012,werner2012a}. 
It turns out that 
macroscopic observables such as the momentum distribution of a two-component
Fermi system with $|a|\to\infty$ are universal and depend on only one
parameter called the contact, $C$ \cite{stewart2010,kuhnle2010}. However, in the case of bosonic systems
that allow Efimov three-body states, there are in fact both a parameter
for two- ($C_2$) and three-body contributions ($C_3$) \cite{braaten2011,castin2011,werner2012b}.
Even the fact that Efimov physics does not occur in two dimensions can be 
observed by measuring the contact parameters \cite{bel2012,werner2012b}. 
As the 
contact paramters are expected to be universal, they should be the same
for a nuclear system in the limit of large scattering length. Measuring
the momentum distribution during nuclear break-up could perhaps be
a way to test predictions from this universal theory of strong interaction 
and compare to the measurements in cold atomic gases. 
This could in turn teach us something
about the strongly-coupled nuclear environment.

\section{BCS-BEC in Nuclear Physics}\label{nucbcs}

The basic physics of the BCS-BEC crossover is contained in the Fermi gas of a homogeneous
system of particles as described in Section \ref{atomgas}.  The parameters are
the Fermi momentum $k_F$, or the density, and the $s$-wave scattering
length $a$.  In the ultracold gases one can fix the density $n$
and vary the interaction parameter $a$ through a Feshbach resonance.
This variation takes the system from the BCS regime of unbound pairs,
through the unitarity limit of $|a|^{-1} = 0$, and into the BEC regime
of strongly bound pairs. 
The unitary regime can also be approached 
from either side by varying $n$ for the Fermi gas since $n|a|^3$ (or $k_F|a|$) 
is the important parameter. 
This observation is the basis for attempts to find
crossover effects in nuclear physics. Note, however, that it is not possible
to cross unitarity by density variation and one will therefore be 
restricted to either BCS or BEC side of the crossover.

\subsection{From infinite matter to neutron halo nuclei}\label{nucbcs1}

As discussed earlier and listed in Tab.~(\ref{tab:overview}), 
the very large bare neutron-neutron scattering length
implies that a large system of neutrons might
reveal universal behavior corresponding to the limit $1/a
=0$.  Such an example is a neutron star, essentially infinite neutron
matter, which perhaps might be described as the
universal point in a BCS-BEC crossover model.

Further investigations require an interaction but as we have 
discussed the bare
nuclear force is only known phenomenologically, and the
bare and effective in-medium nuclear interactions are different.
This was evident in the discussion of polarization effects on 
the gap in Section \ref{pairing}.
The details of the crossover strongly depends on the
residual interaction beyond the mean-field approximation. Therefore to
proceed we should separate mean-field and residual interactions.  To
illustrate, the BCS theory with the bare nuclear interaction produce a
much smaller pairing gap in infinite nuclear matter than measured
\cite{fet71,dickhoff05}.  To improve one can introduce an effective and modified 
nuclear interaction with desired division between mean-field and
residual interaction. However, this is then only applicable for a
specific Hilbert space and a corresponding (BCS) approximation.
The theoretical advantage of infinite nuclear matter is that the
density and the corresponding interaction can be treated as a
parameter.  The crossover physics can then be studied as function of
density.  

We now turn to realistic finite nuclear systems with
wave functions obtained from appropriate nuclear interactions.  We
immediately face several fundamental issues concerned with validity or
transfer of the concepts to such systems.  The interaction is in
principle fixed and at best results for the universal point can be
obtained.  Choice of interaction and separation into mean-field and
residual parts could perhaps allow some freedom but then the chosen
input conditions directly control the output.
The finite systems further introduces conceptual
problems of coordinates, the spurious center-of-mass motion of a
self-bound system, and the definition of condensates in terms of
density matrices as discussed in Section~(\ref{fewbody}).  
We will focus on extracting the universal
features related to the basic crossover concepts. Therefore we first need to address
the seperation of mean-field and residual interaction. The troubles of nuclear physics in this respect are
no less pronounced in nuclear studies of predominantly neutron matter.
Here structures (such as a bound state of two neutrons, see below)
can arise purely out of particular choices of 
residual pairing interaction between the nucleons.
Suitably adjusted zero-range interactions to describe the residual
interaction between neutrons in finite nuclei have been used for
many years \cite{sky56,ber91}. On the surface of it such interactions
appear similar to the zero-range or short-range approximations
of atomic physics. There is, however, problems with this analogy as we
now explain.

The neutron-neutron interaction is often referred to as the
pairing interaction which in nuclear physics is distinctly different
from the full two-body interaction \cite{ber91}.  A commonly
employed neutron-neutron pair residual interaction \cite{doba2001}
has the form
\begin{equation}\label{vnn}
V_{nn}(\bm r,\bm R)= V_0 \delta(\bm r)
\left[1-\eta\left(\frac{\rho(R)}{\rho_0}\right)^{\alpha}\right],
\end{equation}
where $r=|\bm r_1-\bm r_2|$ and $R=|\bm r_1+\bm r_2|/2$ 
are the relative and center-of-mass coordinates of the
neutrons and $\rho(R)$ the nuclear density at $R$ with $\rho_0$
the typical nuclear saturation density \cite{boh69}.  The parameters
$\eta$ and $\alpha$ determine the density-dependence of the
interaction.  Such density-dependent interactions are not used in
atomic physics and any analogy must therefore be carefully
examined. As one can see, for $0<\eta< 1$ and $0 <\alpha<1$, the
pairing strength in Eq.~(\ref{vnn}) will be peaked on the surface of
the nucleus, something we will return to below. Unfortunately it is
not clear what values are consistent with experiments \cite{doba2001},
and the residual interaction is therefore somewhat arbitrary.

Recent studies have made it clear that particularly the $\alpha$
parameter is crucial to get consistent, physically sound results
\cite{doba2001,rot2009}. The problem is that using $\eta=1$ (strong
pairing at the surface) and small $\alpha<1$ can lead to crossover
from BCS to BEC of neutron matter because the interaction unphysically
binds the di-neutron system in the zero-density limit
\cite{baldo1995}. This leads, among other things, to prediction of
neutron-rich halo nuclei that are much too large
\cite{doba2001,rot2009}. In particular, the pair density was found to
have a constant profile out to large radii in Hartree-Fock-Bogoliubov
calculations with $\alpha<0.5$ \cite{rot2009} due to the bound di-neutron.
A recent study using the BCS gap equations \cite{matsuo2006}
recommends potentials with $\alpha<1$ for use in nuclear calculations, 
which is in conflict with the observations that this binds the di-neutron
system. In the latter
study, the features of the neutron wave function were seen
as signatures of BCS-BEC crossover behavior. However, the troublesome
potential means that this conclusion is unlikely to be transferable
to other nuclear or atomic systems.

The BCS-BEC crossover behavior seen in some nuclear studies using these
parameters is then natural as the interaction favors it, but
it might not be physical. However, such density-dependent interaction can
still be used to reliably calculate observables such as energies
\cite{ber91}, and also the pairing gap in nuclear matter
\cite{baldo1995}. Thus whereas average quantities like energies are
accessible, the density-dependent pairing interactions like
Eq.~(\ref{vnn}) can be unphysical in their predictions for the spatial
structure of the neutron pair wave function.

\subsection{Crossover in Finite Halo Nuclei}\label{nucbcs2}
We now discuss recent studies of BCS-BEC crossover physics in finite
nuclei. A bound
state of two neutrons in a potential from an ordinary core-nucleus was
recently introduced as an example of BCS-BEC crossover
\cite{hagino2006,horiuchi2006a,horiuchi2006b}.  Thus a bosonic state of 
two neutrons is suggested to display both BCS and BEC behavior.
This is truly the simplest scenario imaginable.

One of the most prominent examples of a nuclear three-body system is
$^{11}$Li ($^{9}$Li+n+n) where the two neutrons are distributed far
outside the core-radius $R_c$. This is a Borromean halo system where
none of the two-body subsystems are bound.  Valence and core
coordinates separate and the corresponding degrees of freedom
decouple.  The density variation then changes from normal nuclear
density to an exponentially decreasing tail of the two neutrons. This
interesting system was considered within BCS theory more than a 
decade ago \cite{barranco2001}.

The physics of the halo nucleus $^{11}$Li has been
discussed in the context of the BEC and BCS concepts. The analogy is deduced from the
behavior of the three-body wave function, $\Psi(\vec r,\vec R)$,
described in one set of Jacobi coordinates, where  $\vec
r=\vec{r}_1-\vec{r}_2$ connects the neutrons and $\vec
R=(\vec{r}_1+\vec{r}_2)/2$ connect their center-of-mass to the
core-nucleus.  The $s$-wave part can then be projected out of $\Psi$ and
features of the remaining radial wave function $f_0(r,R)$ can be discussed
in detail as function of the two coordinates \cite{hagino2006}.
The probability $|f_0|^2$ exhibits an oscillatory structure as
function of $r$ for fixed small values of $R$ below about 1.5~fm.
These oscillations disappear when $R$ becomes larger than about 4~fm
where only one peak remains.  This is qualitatively the BCS-BEC
crossover behavior as function of the strength of an attractive
residual interaction, i.e. first the BCS oscillations and then the BEC
bound state.  The other potential signature of crossover is the
root mean square radius of $f_0$ with respect to $r$, again as
function of $R$.  This average distance between the two neutrons goes
through a minimum roughly when $R$ is equal to the core-radius. This is
a qualitative feature of crossover since the
extension of a pair-wave function, as in Eq.~(\ref{e330}), is decreasing
from the BCS to the BEC-side because the attraction becomes stronger
and finally binds \cite{eag69,leg80,leg06}.

We now discuss this interpretation of halo nuclei in terms of crossover physics
to isolate the universality of the features of the physics involved and 
illustrate the influence of a particular model of the nuclear system under
study and its interactions.
The oscillatory behavior of the wave function and the minimum of the
radius of the pair wave function must be
generic, universal, and model-independent; unique signals only arising
through the mechanism describing the BCS-BEC crossover.
However, before discussing the question of universality, we caution that 
the suggested signatures involve non-observable quantities for the
particular finite halo nuclei. It is therefore an indirect way to study 
crossover physics.

\begin{figure}[htb!]
\centering
\epsfig{file=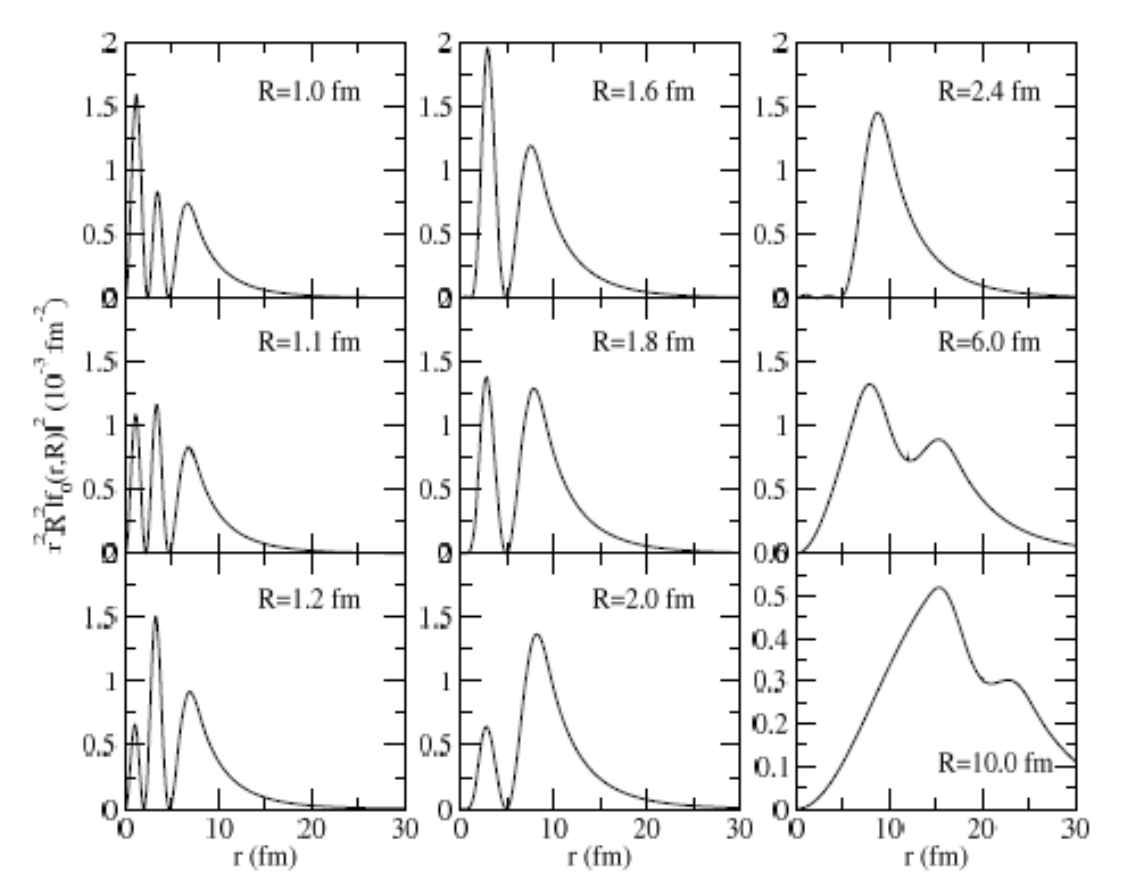,clip=true,scale=1.0}
\caption{\label{fig4} The probabilities as functions of $r$ for different values of $R$
for two non-interacting particles in the mean-field from a square well
potential with depth $108$~MeV and radius $3$~fm. Both particles are in
the second $s$-state each with energy $-0.17$~MeV. Taken from \protect\cite{zinner2008}.}
\end{figure}

The oscillating probability as
function of $r$ is a structure appearing through
correlations of different single-particle orbits in BCS theory \cite{leg06}. 
To assess whether this is a universal feature in a finite nucleus, we take $\Psi$
to be the uncorrelated product of two non-interacting single-particle
neutron-$^{9}$Li wave functions.  We choose the second weakly bound
$s$-wave in a square well potential and show in Fig.~(\ref{fig4}) the
probabilities defined in \cite{hagino2006}.  The resemblance to the
findings in \cite{hagino2006} is striking.  However, the oscillations
at small $R$ are directly due to the nodes of the single-particle
wave functions. They disappear when the lowest $s$-wave without nodes
is used. As $R$ increases these nodes are no longer geometrically
compatible with at least one particle in the attractive part of the
potential.  Then the structure turns into one peak \cite{zinner2008}.
Adding a short-range neutron-neutron attraction of moderate
strength would maintain these features and keep the neutrons a little
closer resulting in a slower increase of the peak
position with $R$ for large $R$.

We are led to conclude that the oscillatory behavior is not connected to
many-body BCS-correlations, they are due to nodes of the
single-particle neutron-core wave functions. Furthermore, the peak for
large $R$ is not a signal of a bound state or of
condensation. This simple model reproduces the features which clearly 
demonstrates how careful one must be when assessing potential generic 
properties. It is always necessary to use more than one particular 
model of the system to check such behavior.

\begin{figure}[htb!]
\centering
\epsfig{file=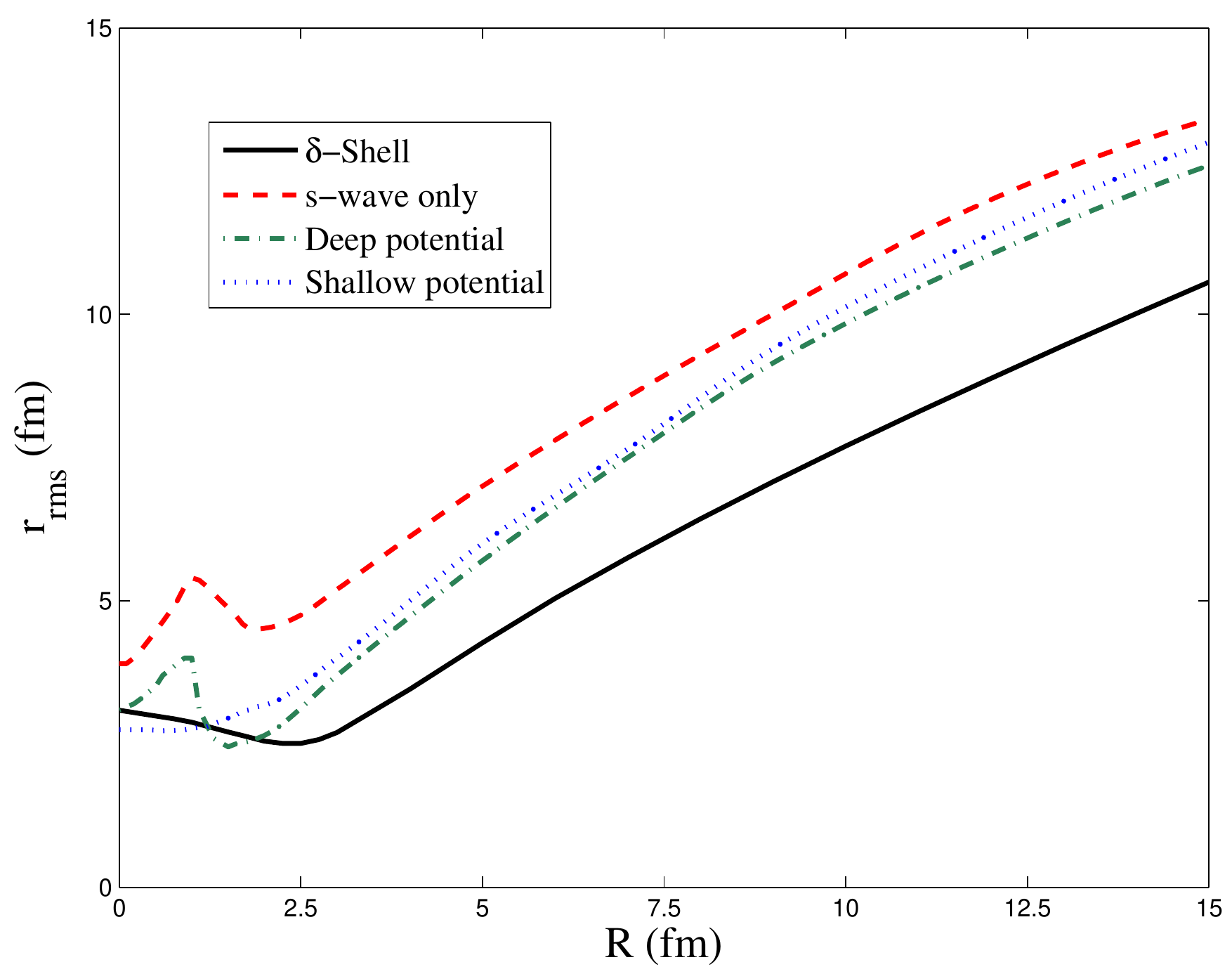,clip=true,scale=0.45}
\caption{\label{fig3} The root mean square distance between the two neutrons
as function of their center-of-mass distance from the core.  Deep (dot-dashed) and
shallow (dotted) gaussian neutron-core potentials with one and with no bound
states are used.  We also show the results when only $s$-waves are
retained for the deep potential (dashed), and for non-interacting neutrons in a
delta-shell potential around the core (full).}
\end{figure}

The second feature, a minimum in the root mean square neutron-neutron
distance at a finite center-of-mass distance, is interpreted as evidence
for change of BCS-correlations into BEC-structure with $R$. The
minimum itself can be related to the choice of interactions in
\cite{hagino2006}, which closely resembles Eq.~(\ref{vnn}) with 
$\eta\lesssim 1$ and $\alpha=1$. The neutron-core interaction is
attractive inside the core radius $R_c$ and very small outside, and
the neutron-neutron interaction is a $\delta$-function in $\vec r$
with an attractive strength which essentially is zero below $R_c$ and
full strength for larger $R$.  When $R=0$ the two neutrons can move
independently inside the core, for $R=R_c$ the two neutrons can
fully exploit their mutual attraction for $r=0$, and for larger $R$
they have to move apart to benefit from the more important
neutron-core attraction.  This produces the minimum in the rms value
for $R \approx R_c$. We illustrate this in Fig.~(\ref{fig3}) by assuming
a $\delta$-shell potential at $R=R_c=3$ fm. 
The key properties are that the neutron-neutron
attraction increases with $R$ while the neutron-core attraction vanish
outside the core.

Since $^{11}$Li is very well studied in other successful three-body
models, see e.g. \cite{garrido1999}, we can compare results to test
the universality requirement.  We show in Fig.~(\ref{fig3})
neutron-neutron distances for a realistic bare neutron-neutron
potential and different neutron-core potentials, i.e. a shallow and a
deep gaussian without or with one bound state, respectively. These
potentials are both able to reproduce the known three-body observables
for $^{11}$Li.  They differ at short distances where the neutron is
inside the core but this inner part is unimportant in descriptions of
spatially extended halo nuclei.  The behavior seen in Fig.~(\ref{fig3})
differ significantly from that of \cite{hagino2006}.  Either no
minimum when the lowest $s$-state is populated or an oscillation with
one maximum and one minimum when the second $s$-state is populated.

To see the effect of mixing partial waves we omitted $p$-waves, i.e.
included only the $s$-waves in all Jacobi systems for the deep
potential.  The rms-curve remains similar but pushed to higher values.
Finally we used a delta-shell neutron-core potential for two
non-interacting neutrons.  For a strongly attractive strength, where
the neutrons independently prefer to stay at $R_c$, the rms-curve
should go from $2R_c$ through a small minimum at $R=R_c$ to a linearly
increasing function at larger $R$.  This only happens for very strong
attraction otherwise the wave function is too smeared out although the
same qualitative behavior remains. The conclusion is that generic pair
correlations are not responsible for the rms minimum, which may appear
for product wave functions and may be absent for correlated wave-mixed
particles.

We have to conclude that features of BCS-BEC crossover in two-neutron halo nuclei
are not generic signals and
the behavior is not universal. The example demonstrates the
troubles that one faces in finite systems when comparing to BCS and
BEC concepts. It also demonstrates some of the inherent dangers with 
specific residual nuclear interactions when they are pushed beyond 
their designated boundaries.

\section{Conclusions and perspective}\label{conc}
The status in the fields of cold atomic gases and nuclear physics
differ at the moment very much from each other.  The cold atomic gases
have well controlled, simple interactions and obey
simultaneously simple dilute limit conditions.  Nuclei are dense
self-bound many-body systems with complicated phenomenological
interactions but well-controlled advanced techniques. Concepts and
techniques from these fields can support each other. Theoretical
methods developed for nuclei are very suitable for investigations of
moderate numbers of bosons and fermions and mixtures of non-identical
particles. Concepts originating from condensed matter and nuclear physics, now
investigated in cold atoms may reappear in nuclei, and
experimental techniques will soon allow investigations of systems
with striking similarity to finite nuclei, but with controlled
variation of the interactions.

We have mainly focused 
on the simplest structures obtained in
mean-field approximations. For bosons this implies that condensation 
could occur while for fermions pair correlations can lead to a 
superfluid state.
In the fermionic case, the BCS-approximation is revealing after
analogous degrees of freedom are identified.  Pairs in time-reversed
states in nuclei correspond to particles in two different internal (hyperfine)
states but the same set of external confinement quantum numbers (usually
parabolic traps).
Relative $s$-waves are by far the dominating contributors to the
interaction in the zero-range approximation.
We consider how to relate BCS and BEC coherent quantum states from
condensed matter and cold atomic gas physics to states in finite
systems of few particles. The goal is to discuss
universal features of these quantum states and phases 
and investigate whether they can be manifest in few-body
systems with nuclei as the example. A particularly 
interesting venture at a time when trapping of small samples
of only a few atoms have been realized experimentally 
\cite{selim2011,zurn2011,bakr2010,sherson2010,weitenberg2011}, implying
that systems resembling few-body nuclei can be produced and 
studied in cold atomic gase experiments.

An important contrast is the fact that cold atomic gases are held together 
by an external field. For
repulsive two-body interactions, an almost perfect BEC is possible
\cite{wieman1995,ketterle1995}.
The influence of correlations are small because the system is
extremely dilute \cite{pet02} (away from two-body scattering resonances, i.e. 
Feshbach resonances). For moderate attractive interactions and relatively
few particles similar stable structures also exist \cite{ruprecht95,baym1996}. However,
attraction destabilizes these systems and collapse occurs into very
dense and highly correlated structures
\cite{bradley1997,roberts2001,donley2001,lahaye2008}.
Nuclear systems become similar to large samples of cold atoms 
only in the dilute limit 
when the density
is about two orders less than that of nuclear saturation. Attempts to
approach the limit of dilute nuclear states are found in the study of
cluster states or weakly bound states close to particle or cluster
thresholds.  The BEC analog has been introduced as a few, or many,
$\alpha$-particles in a dilute state.  However, the Coulomb
repulsion would lead to immediate fragmentation or, if binding is
achieved through the attractive nuclear interaction, the particles
must be close-lying as for ordinary nuclei.
This in turn implies that
the diluteness criteria is violated and the BEC feature destroyed.
Furthermore, detecting a condensate requires knowledge of the 
coherence in the system, for example through a two-particle 
correlation function. In the case of nuclear $\alpha$-condensates
this is impossible to measure and only standard one-body density 
information can be obtained. The notion of condensate in finite nuclei is therefore
ill-defined and should be
regarded as a crude approximation to the cluster structure
of excited states in $\alpha$-nuclei.
Attempts to avoid the Coulomb interaction by using neutrons suffer
from a similar deficiency, i.e. two neutrons can be weakly bound to a
core and form a neutron halo, but adding more than a few extra nucleons
would either make the
system unstable due to the Pauli principle or lead to ordinary bound
nuclear states. Many-body neutron halo states would fall apart or form clusters
of two, three or more pieces. 
The external confining field is not present.
Combinations allowing bosons as deuterons does not seem promising for
the same reasons, i.e. the system becomes either unstable due to lack
of attraction or the spatially extended deuteron structure is lost as
soon as nuclear surface densities are reached.

A promising direction of cross-fertilization between nuclear physics and cold
atomic gases has turned out to be the investigation of three-body bound states
in the universal low-energy regime. The Efimov effect where an infinite number of three-body bound states 
appear near the threshold for two-body binding was initially predicted
in the context of nuclear physics. With its recent experimental observation
in cold atomic gases it has seen a great revival and tremendous activity 
in the area has ensued. Here we have discussed the universal predictions
from the original hyperspherical point of view, and have suggested 
systems which have favourable parameters for observing the effect. The 
flexible tunability of experimental parameters in cold atomic gases 
can help elucidate the features of the Efimov effect, also beyond 
the universal predictions. This information could then 
be transfered to nuclear physics and aid in understanding why
it is so difficult to see the Efimov effect in nuclear systems.

Understanding cold atomic Fermi gases
are to a large extend based on pairing Hamiltonians, and the related
BCS-type approximations.  This is a much better approximation
than for nuclei, but the theoretical models often employed in
atomic physics are not always as accurate as the
impressive experimental precision.  
In nuclear physics, 
improved models had to be developed, and these have now
proven their capability
in the description of many-fermion systems. 
Examples 
are large-scale shell-model methods, hyperspherical 
basis expansions, algebraic methods for pairing Hamiltonians, 
and density functional theory. 
They are based on first principle
and go far beyond mean-field and pairing correlations. They
are highly flexible in connection with choice of interactions and
should thus be able to provide accurate results also
for atomic gases.  
The methods are based on effective interactions adapted to reduced
Hilbert spaces which is a tremendous advantage for the atomic $N$-body
systems. These methods were first of all
developed for handling identical fermions (nucleons), but also simpler
bosonic systems can in most cases be treated with the same numerical
techniques.  Strongly-coupled systems at unitarity and 
the BCS-BEC crossover region are thus obvious fields to explore.
Also finite-range effects and other more complicated corrections are 
naturally handled by nuclear methods.

In summary, we have discussed the physics of BEC, BCS, and BCS-BEC
crossover as studied in cold atomic gases from a nuclear physics perspective. 
We emphasized
similarities, and differences through a detailed
discussion of some relevant methods in wide use. We find
that finite size effects in nuclear systems will most often destroy
signatures of the coherent behavior that is routinely observed in 
cold atomic gases. No current models or
measurements seem to support the interpretation of nuclear states 
as manifestations of BEC, BCS or crossover physics (with one potential
exception being infinite neutron matter).  
At the moment it appears to be difficult to specify
distinguishing measurable consequences or to design test experiments
of the required generic features of universality, a basic first step in any successful
transfer of these concepts to nuclei in particular and few-body
systems in general. However, as cold atomic gas experiments 
with few particles become reality the situation is likely to change.
On the other hand, here 
we have suggested some tempting cold atom systems to investigate with 
nuclear techniques in order to facilitate further exchange between the 
fields.

\paragraph*{{\bf Acknowledgments.}}
Numerous enlightening inputs from H. Fynbo are gratefully acknowledged. We thank
M. Th{\o}gersen, K. Riisager, D.~V. Fedorov, N. Nygaard, H. Fogedby, K. M{\o}lmer,
and T. Neff for useful discussions. We thank E. Garrido for providing numerical results
and N. Nygaard for providing data and invaluable help in making some of the figures.

\end{document}